\documentclass[12pt,a4paper,oneside]{book}

\usepackage{indentfirst}

\usepackage{graphicx}
\usepackage{epstopdf}
\usepackage{color}
\usepackage{setspace}
\usepackage{amssymb}
\usepackage{amsthm}

\usepackage[centertags]{amsmath}
\usepackage{dcolumn}
\usepackage{bm}
\usepackage[mathscr]{eucal}
\usepackage[nooneline]{subfigure}

\usepackage[small,bf]{caption}
 \usepackage[square,numbers,sort&compress]{natbib}
\usepackage{calc} 
\usepackage{psfrag} 
\usepackage[nottoc]{tocbibind} 
\usepackage{titling}
\newcommand{\ket}[1]{\ensuremath{\left|#1\right>}}
\newcommand{\bra}[1]{\ensuremath{\left<#1\right|}}
\newcommand{\braket}[2]{\ensuremath{\left<#1|#2\right>}}

\newcommand{\ketbra}[2]{| #1 \rangle\langle #2 |}

\newcommand{\be}{\begin{equation}}
\newcommand{\ee}{\end{equation}}
\newcommand{\bea}{\begin{eqnarray}}
\newcommand{\eea}{\end{eqnarray}}

\newcommand{\tr}{\textrm{tr}}
\renewcommand{\sp}{\textrm{span}}

\newcommand{\rank}{\textrm{rank}}

\def\opone{\leavevmode\hbox{\small1\kern-3.8pt\normalsize1}}

\theoremstyle{plain}
\newtheorem{theorem}{Theorem}
\newtheorem{lemma}[theorem]{Lemma}
\newtheorem{corollary}[theorem]{Corollary}

\newtheorem{proposition}[theorem]{Proposition}
\newtheorem{fact}[theorem]{Fact}

\theoremstyle{definition}
\newtheorem{definition}{Definition}
\newtheorem{problem}{Problem}

\newcommand{\fig}[1]{Fig.~\ref{#1}}

\newcommand{\poly}{\text{poly}}

\newcommand{\w}{\omega}
\newcommand{\grad}{\nabla}
\newcommand{\hess}{\nabla^2}
\renewcommand{\d}{\mathrm{d}}
\renewcommand{\a}{\mathrm{a}}
\newcommand{\half}{\frac{1}{2}}
\newcommand{\hessfw}{\hess F(\w)}
\newcommand{\hessfx}{\hess F(x)}
\newcommand{\hessfz}{\hess F(z)}

\newcommand{\hessfaw}{\hess F_\a(\w)}
\newcommand{\hessfax}{\hess F_\a(x)}
\newcommand{\hessfaz}{\hess F_\a(z)}

\newcommand{\hessfdw}{\hess F_\d(\w)}

\newcommand{\hessfdz}{\hess F_\d(z)}

\newcommand{\hessfzinv}{(\hess F(z))^{-1}}
\newcommand{\hessfwinv}{(\hess F(\w))^{-1}}
\newcommand{\hessfxinv}{(\hess F(x))^{-1}}

\newcommand{\hessfawinv}{(\hess F_\a(\w))^{-1}}

\newcommand{\hessfazainv}{(\hess F_\a (z_\a))^{-1}}

\newcommand{\hessfzskinv}{(\hess F(z(s_k)))^{-1}}

\newcommand{\hessfawa}{(\hess F_\a (\w_\a))}

\newcommand{\p}{\rho}

\newcommand{\mylemma}[3]{\noindent\textbf{Lemma \ref{#1} (#2).} \emph{#3}}
\newcommand{\mytheorem}[3]{\noindent\textbf{Theorem \ref{#1} (#2).} \emph{#3}}

\newcommand{\dima}{M}
\newcommand{\dimb}{N}

\newcommand{\hermops}{\mathbb{H}_{\dima, \dimb}}
\newcommand{\hermopsa}{\mathbb{H}_{\dima}}
\newcommand{\hermopsb}{\mathbb{H}_{\dimb}}

\newcommand{\densops}{\mathcal{D}_{\dima,\dimb}}
\newcommand{\densopsgen}[1]{\mathcal{D}(#1)}
\newcommand{\sep}{\mathcal{S}_{\dima, \dimb}}
\newcommand{\ent}{\mathcal{E}_{\dima, \dimb}}
\newcommand{\CMotimesCN}{\mathbb{C}^{\dima}\otimes\mathbb{C}^{\dimb}}

\newcommand{\conv}{\mathrm{conv}}

\newcommand{\n}{\dima^2\dimb^2}

\newcommand{\A}{\mathrm{A}}
\newcommand{\B}{\mathrm{B}}

\newcommand{\D}{\mathrm{D}}
\newcommand{\Y}{\mathrm{Y}}

\renewcommand{\P}{\mathrm{P}}
\newcommand{\NP}{\mathrm{NP}}

\newcommand{\NPCK}{\mathrm{NPC_\K}}
\newcommand{\NPCT}{\mathrm{NPC_\T}}

\newcommand{\K}{\mathrm{K}}
\newcommand{\T}{\mathrm{T}}

\newcommand{\col}{\mathrm{col}}

\newcommand{\nchoosek}[2]{\left(\! \begin{array}{c}#1\\#2\end{array}\! \right)}

\renewcommand{\S}{\mathcal{S}}

\newcommand{\Abs}{\mathrm{Abs}}
\newcommand{\M}{\mathrm{Max}}

\newcommand{\OSOPTK}{\mathcal{O}_{\text{SOPT}(K)}}

\newcommand{\OSSEPKprime}{\mathcal{O}_{\text{SSEP}(K')}}
\newcommand{\OSSEPKstar}{\mathcal{O}_{\text{SSEP}(K^\star)}}
\newcommand{\OSSEPQp}{\mathcal{O}_{\text{SSEP}(Q_p)}}

\begin{document}

\frontmatter

\title{Computing finite-dimensional bipartite quantum separability}
\author{Lawrence Mario Ioannou \\
Darwin College\\
University of Cambridge
}
\renewcommand{\maketitlehookc}{\vspace{200pt}}
\date{This dissertation is submitted for the degree of Doctor of
Philosophy.\\
\today}
\begin{titlingpage}


\maketitle
%

\begin{center}
\vspace{5ex}
This dissertation is the result of my own work and includes
nothing which is the outcome of work done in collaboration except
where specifically indicated in the Acknowledgements.\\%
\vspace{5ex}
This dissertation does not exceed 60,000 words.
\end{center}
\end{titlingpage}

\cleardoublepage \addcontentsline{toc}{chapter}{Abstract}
\section*{\huge{Abstract}}

Ever since entanglement was identified as a computational and
cryptographic resource, effort has been made to find an efficient
way to tell whether a given density matrix represents an
unentangled, or \emph{separable}, state.  Essentially, this is the
\emph{quantum separability problem}.

In Chapter \ref{ChapterIntro}, I begin with a brief introduction
to quantum states, entanglement, and a basic formal definition of
the quantum separability problem.  I conclude the first chapter
with a summary of one-sided tests for separability, including
those involving semidefinite programming.

In Chapter \ref{ChapterConvexity}, I apply polyhedral theory to
prove easily that the set of separable states is not a polytope;
for the sake of completeness, I then review the role of polytopes
in nonlocality. Next, I give a novel treatment of entanglement
witnesses and define a new class of entanglement witnesses, which
may prove to be useful beyond the examples given. In the last
section, I briefly review the five basic convex body problems
given in \cite{GLS88}, and their application to the quantum
separability problem.

In Chapter \ref{ChapterSepAsDecisionProblem}, I treat the
separability problem as a computational decision problem and
motivate its approximate formulations. After a review of basic
complexity-theoretic notions, I discuss the computational
complexity of the separability problem: I discuss the issue of
NP-completeness, giving an alternative definition of the
separability problem as an NP-hard problem in NP\@. I finish the
chapter with a comprehensive survey of deterministic algorithmic
solutions to the separability problem, including one that follows
from a second NP formulation.

Chapters 1 to 3 motivate a new interior-point algorithm which,
given the expected values of a subset of an orthogonal basis of
observables of an otherwise unknown quantum state, searches for an
entanglement witness in the span of the subset of observables.
When all the expected values are known, the algorithm solves the
separability problem.  In Chapter
\ref{ChapterReductionToEWSearch}, I give the motivation for the
algorithm and show how it can be used in a particular physical
scenario to detect entanglement (or decide separability) of an
unknown quantum state using as few quantum resources as possible.
I then explain the intuitive idea behind the algorithm and relate
it to the standard algorithms of its kind. I end the chapter with
a comparison of the complexities of the algorithms surveyed in
Chapter 3.  Finally, in Chapter \ref{ChapterAlgorithm}, I present
the details of the algorithm and discuss its performance relative
to standard methods.

\cleardoublepage \addcontentsline{toc}{chapter}{Preface}

\chapter*{Preface}

This work attempts to give a comprehensive treatment of the state
of the art in deterministic algorithms for the quantum
separability problem in the finite-dimensional and bipartite case.
The need for such a treatment stems from the very recent (2003 and
later) proposals for separability algorithms -- all quite
different from one another. It is likely that these recent papers
emerged when they did because of the (disheartening) result of
Gurvits (2001) showing the problem to be computationally
intractable: given that the problem is hard, what is the best we
can do to solve it? Among these proposals is my algorithm (done in
collaboration), which will be shown to compare favorably to the
others, complexity-theoretically.

Gurvits' result, that the separability problem is NP-hard, raised
a question among the quantum information community: ``...but then
isn't it NP-complete?''  After hearing many people ask this
question, I set out to clarify the issue and show that the
separability problem is NP-complete in the usual sense (that is,
with respect to Karp reductions). The latter part of this mission
is as yet unsuccessful, but the partial results are presented,
including a redefining of the separability problem as an NP-hard
problem in NP (previous definitions could not place the problem in
NP, rather only in a modified version of NP).

Entanglement witnesses have been around since 1996, and had been
extensively studied up until recently, especially by the
Innsbruck-Hannover group, which produced interesting
characterisations of entanglement witnesses and showed how to
construct optimal entanglement witnesses.  I approached
entanglement witnesses from the viewpoint of polyhedral theory,
rather than linear-operator theory.  The result was the immediate
solution of an open problem of whether the separable states form a
polytope.  Under a slightly different definition of ``entanglement
witness'',  I discover a new class of entanglement witnesses which
I call ``ambidextrous entanglement witnesses''.  These correspond
to observables whose expected values can indicate that a state is
entangled on opposite sides of the set of separable states.

\cleardoublepage \addcontentsline{toc}{chapter}{Acknowledgements}
\section*{\huge{Acknowledgements}}

I am utterly grateful to my supervisor, Artur Ekert, for his
support and encouragement; his liberal approach to supervision,
which allowed me to pursue my own interests; and for his prophetic
suggestion of my thesis title on the day I arrived in Cambridge.

I am also grateful to the GCHQ for funding this PhD.  Much travel
was also funded by project RESQ (IST-2001-37559) of the IST-FET
programme of the EC.

The main result in this thesis came out of my collaboration with
my main co-authors, Ben Travaglione and Donny Cheung. I would
especially like to thank Ben for essentially co-supervising me
during the first two years of my degree. Discussions with Daniel
Gottesman formed the basis of the NP-formulation of the quantum
separability problem.

Tom Stace has been extremely generous with his time, always
willing to engage in a discussion about the various elements of my
work around which I was having trouble wrapping my head. He was
very helpful during early stages of the development of the
algorithm.

Coralia Cartis introduced me to logarithmic barriers, analytic
centres, and self-concordance; and confirmed my intuition that the
analytic centre in Chapter 5 was indeed a conic combination of the
normal vectors, where I was too inept to calculate $\grad F$
correctly the first time around.

Matthias Christandl taught me about entanglement measures and
pointed me to the work of K{\"{o}}nig and Renner on the finite
quantum de Finetti theorem.  Other colleagues who have been
helpful are Carolina Moura Alves, Garry Bowen, and Daniel Oi.

My examiners made many comments and suggestions which greatly
improved this thesis.

My parents, Art and Josie Ioannou, and brother, John Ioannou, have
been incredibly loving and supportive throughout my studies.

It was absolutely wonderful to marry Sarah Tait in 2005!  She has
been a pillar of support.  She took a risk in coming to Cambridge
with me; I am happy that she is happy here.  I love her.

\cleardoublepage \addcontentsline{toc}{chapter}{List of
Publications}
\section*{\huge{List of Publications}}

The following is a list of papers that have resulted from the work
presented in this thesis.

\begin{enumerate}

\item  L. M. Ioannou and B. C. Travaglione,  A note on quantum
separability, quant-ph/0311184.

\item  L. M. Ioannou, B. C. Travaglione, D. Cheung, A. K. Ekert,
Improved algorithm for quantum separability and entanglement
detection, Physical Review A, \textbf{70} 060303(R) (2004).

\item  L. M. Ioannou, B. C. Travaglione, D. Cheung,  Separation
from optimization using analytic centers, cs.DS/0504110.

\item  L. M. Ioannou and B. C. Travaglione,  Quantum separability
and entanglement detection via entanglement-witness search (in
preparation).

 \end{enumerate}

\listoffigures 

\tableofcontents

\renewcommand{\labelenumi}{(\roman{enumi})}
\mainmatter
\chapter{Introduction}\label{ChapterIntro}

``Just because it's hard, it doesn't mean you don't try.''  When
my mother said these words to me way back when I was a Master's
student, I had no idea they would open my PhD thesis.

Ever since quantum-mechanical phenomena were identified as
computational and cryptographic resources, researchers have become
even more interested in precisely characterising the features of
quantum theory that set it apart from classical physical theory.
Two of these features are nonlocality and entanglement, both of
which are ``provably hard'' to characterise; that is, deciding
whether a quantum state exhibits nonlocality or entanglement is as
hard as some of the hardest and most important problems in
complexity theory.

This thesis concentrates on the latter problem of deciding whether
a quantum state is unentangled, or, separable.  I review all of
the deterministic algorithms proposed for the separability
problem, including two of my own, in an attempt to discover which
has the best asymptotic complexity. Along the way, I look at
entanglement witnesses in a new light and discuss the
computational complexity of the separability problem.

In Section \ref{sec_QuantumPhysics}, I review some elements of
quantum mechanics and define and give the significance of
separable states. The remainder of the chapter discusses partial
solutions to the separability problem.

\section{Quantum physics}\label{sec_QuantumPhysics}

The \emph{pure state} of a $d$-dimensional quantum physical system
is represented mathematically by a complex
unit-vector\footnote{Some conventions do not require the
normalisation constraint; i.e. sometimes it is useful to work
without it and refer to ``unnormalised states''.}
$\ket{\psi}\in\mathbb{C}^d$, where the ``global phase'' of
$\ket{\psi}$ is irrelevant; that is, for any real $\phi$,
$e^{i\phi}\ket{\psi}$ represents the same physical state as
$\ket{\psi}$.  If the system can be physically partitioned into
two subsystems (denoted by superscripts $\A$ and $\B$) of
dimensions $\dima$ and $\dimb$, such that $d=\dima\dimb$, then
$\ket{\psi}$ may be $\emph{separable}$, which means
$\ket{\psi}=\ket{\psi^{\A}}\otimes\ket{\psi^\B}$, for
$\ket{\psi^\A}\in\mathbf{C}^\dima$ and
$\ket{\psi^\B}\in\mathbf{C}^\dimb$ and where ``$\otimes$'' denotes
the Kronecker (tensor) product. Without loss of generality, assume
$\dima\leq \dimb$ unless otherwise stated.  If $\ket{\psi}$ is not
separable, then it is \emph{entangled} (with respect to that
particular partition).

More generally, the state of the system may be a \emph{mixed
state}, which is a statistical distribution of pure states.  A
mixed state $\rho$ is usually represented as the \emph{density
operator} $\rho=\sum_{i=1}^{k} p_i \ket{\psi_i}\bra{\psi_i}$,
where $\ket{\psi_i}\in\mathbf{C}^d$, $\sum_{i=1}^k p_i =1$,
$p_i\geq 0$, and $\bra{\psi_i}$ is the dual vector of
$\ket{\psi_i}$. A mixed state is thus a positive semidefinite (and
hence Hermitian, or self-adjoint) operator with unit
trace\footnote{The previous footnote applies here, too.}: $\rho
\geq 0$ and $\tr(\rho)=1$. Denote the set of all density operators
mapping complex vector space $V$ to itself by $\densopsgen{V}$;
let $\densops:=\densopsgen{\CMotimesCN}$. The \emph{maximally
mixed state} is $I_{\dima,\dimb}:=I/\dima\dimb$, where $I$ denotes
the identity operator.  A density operator $\rho$ satisfies $0\leq
\tr(\rho^2)\leq 1$ and represents a pure state if and only if
$\tr(\rho^2)=1$. A pure state $\ket{\psi}$ is separable if and
only if $\tr_\B(\ketbra{\psi}{\psi})$ is a pure state, where
``$\tr_\B$'' denotes the partial trace with respect to subsystem
$\B$ (e.g. see Exercise 2.78 in \cite{NC00}); a pure state is
called \emph{maximally entangled} if $\tr_\B(\ketbra{\psi}{\psi})$
is the maximally mixed state $I/\dima$ in the space of density
operators on the $\A$-subsystem $\densopsgen{\mathbb{C}^\dima}$.
Thus, the mixedness of $\tr_\B(\ketbra{\psi}{\psi})$ is some
``measure'' of the entanglement of $\ket{\psi}$ (see Section
\ref{subsec_EntanglementMeasures}).


A mixed state $\rho\in\densops$ is \emph{separable} if and only if
it may be written $\rho=\sum_{i=1}^k p_i\rho^\A_i\otimes\rho^\B_i$
with $p_i\geq 0$ and $\sum_ip_i=1$, and where
$\rho^\A_i\in\densopsgen{\mathbb{C}^\dima}$ is a (mixed or pure)
state of the $A$-subsystem (and similarly for
$\rho^\B_i\in\densopsgen{\mathbb{C}^\dimb}$); when $k=1$, $\rho$
is a \emph{product state}. Let $\sep\subset\densops$ denote the
separable states; let $\ent:=\densops\setminus\sep$ denote the
entangled states.  The following fact will be used several times
throughout this thesis:
\begin{fact}[\cite{Hor97}]\label{fact_FiniteDecompOfSepState}  If $\sigma\in\sep$, then $\sigma$ may be written as
a convex combination of $\n$ pure product states, that is,
\begin{eqnarray}
\sigma = \sum_{i=1}^{\n}p_i
\ketbra{\psi^\A_i}{\psi^\A_i}\otimes\ketbra{\psi^\B_i}{\psi^\B_i},
\end{eqnarray}
where $\sum_{i=1}^{\n}p_i=1$ and $0\leq p_i\leq 1$ for all
$i=1,2,\ldots,\n$.
\end{fact}
\noindent Recall that a set of points
$\{x_1,\ldots,x_j\}\subset\mathbb{R}^n$ is \emph{affinely
independent} if and only if the set
$\{x_2-x_1,x_3-x_1,\ldots,x_j-x_1\}$ is linearly independent in
$\mathbb{R}^n$.  Recall also that the \emph{dimension} of
$X\subset\mathbb{R}^n$ is defined as the size of the largest
affinely-independent subset of $X$ minus 1.  Fact
\ref{fact_FiniteDecompOfSepState} is based on the well-known
theorem of Carath{\'{e}}odory that any point in a compact convex
set $X\subset\mathbb{R}^n$ of dimension $k$ can be written as a
convex combination of $k+1$ affinely-independent extreme points of
$X$.


\begin{definition}[Formal quantum separability problem]\label{def_FormalQuSep} Let $\rho\in\densops$ be a mixed state.  Given the
matrix\footnote{We do not yet define how the entries of this
matrix are encoded; at this point, we assume all entries have some
finite representation (e.g. ``$\sqrt{2}$'') and that the
computations on this matrix can be done exactly.} $[\rho]$ (with
respect to the standard basis of
$\mathbb{C}^\dima\otimes\mathbb{C}^\dimb$) representing $\rho$,
decide whether $\rho$ is separable.
\end{definition}


What is the significance of a separable state?  For a pure state
$\ket{\psi}=\ket{\psi^\A}\otimes\ket{\psi^\B}$, we can imagine two
spatially separated people (laboratories) -- called ``Alice''
(``A'') and ``Bob'' (``B'') -- who each have one part of
$\ket{\psi}$: Alice has $\ket{\psi^\A}$ and Bob has
$\ket{\psi^\B}$. We can further imagine that Alice and Bob each
prepared their respective part of the state $\ket{\psi}$; i.e.
Alice prepared a pure state $\ket{\psi^\A}$ and Bob prepared a
pure state $\ket{\psi^\B}$, and $\ket{\psi}$ describes the state
of the union of Alice's system and Bob's system.

In preparing their systems, Alice and Bob could use classical
randomness.  Thus, instead of preparing the pure state
$\ket{\psi^\A}$ with probability 1, Alice prepares the state
$\ket{\psi^\A_i}$ with probability $p^\A_i$.  By imagining
infinitely many repeated trials of this whole scenario, this means
Alice prepares the mixed state $\rho^\A=\sum_i
p^\A_i\ketbra{\psi^\A_i}{\psi^\A_i}$. Similarly, Bob could prepare
his subsystem in the mixed state $\rho^\B$.  The state of the
total system is then represented by $\rho^\A\otimes\rho^\B$.
States of this form can thus be prepared with \emph{local
(randomised) operations}.

Now suppose that Alice and Bob can telephone each other.  Then
they could coordinate their subsystem-preparations: when Alice
(through her local randomness) decides (with probability $p_i$) to
prepare $\ket{\psi^\A_i}$, she tells Bob to prepare
$\ket{\psi^\B_i}$. The state of the total system is now
represented by
\begin{eqnarray}\label{eqn_SEP_LOCC_state}
\rho=\sum_i p_i
\ketbra{\psi^\A_i}{\psi^\A_i}\otimes\ketbra{\psi^\B_i}{\psi^\B_i},
\end{eqnarray}
which may not have a representation of the form
$\rho^\A\otimes\rho^\B$.  States of the form
(\ref{eqn_SEP_LOCC_state}) can thus be prepared with \emph{local
operations and classical communication} (abbreviated ``LOCC'').
These are the separable states.  Instead of a telephone (two-way
classical channel), it suffices that Alice and Bob share a source
of randomness in order to create a separable state.

If Alice and Bob share an entangled state (perhaps Alice prepared
the total system and then sent the B-subsystem to Bob), then they
share something that they could not have made with LOCC\@. Perhaps
unsurprisingly, it turns out that sharing certain types of
entangled states (see Section \ref{sec_PolytopesSep&Nonlocality})
allows Alice and Bob to communicate in ways that they could not
have with just a telephone \cite{Eke91, NC00}.

\section{One-sided tests and restrictions}\label{sec_OneSidedTestsAndRestrictions}

Shortly after the importance of the quantum separability problem
was recognised in the quantum information community, efforts were
made to solve it reasonably efficiently.  In this vein, many
one-sided tests have been discovered. A \emph{one-sided test (for
separability)} is a computational procedure (with input $[\rho]$)
whose output can only every imply \emph{one} of the following
(with certainty):
\begin{itemize}\item $\rho$ is
entangled (in the case of a \emph{necessary test})\item $\rho$ is
separable (in the case of a \emph{sufficient test}).
\end{itemize}

There have been many good articles (e.g. \cite{Bru02, Ter02,
qphSSLS05}) which review the one-sided (necessary) tests. As this
thesis is concerned with algorithms that are both necessary and
sufficient tests for separability for all $\dima$ and $\dimb$ --
and whose computer-implementations have a hope of being useful in
low dimensions -- I only review in detail the one-sided tests
which give rise to such algorithms (see Section
\ref{sec_OneSidedTestsOnSDP}).  But here is a list of popular
conditions on $\rho$ giving rise to efficient one-sided tests for
finite-dimensional bipartite separability:

\vspace{3mm}

\noindent\textbf{Necessary conditions for $\rho$ to be separable}
\begin{itemize}

\item PPT test \cite{Per96}: $\rho^{T_\B}\geq 0$, where ``$T_\B$''
denotes partial transposition

\item Reduction criterion \cite{HH99}: $\rho^\A\otimes I-\rho \geq
0$ and $I\otimes\rho^\B-\rho \geq 0$, where $\rho_\A :=
\tr_\B(\rho)$ and ``$\tr_\B$'' denotes partial trace (and
similarly for $\rho_\B$)

\item Entropic criterion for $\alpha=2$ and in the limit
$\alpha\rightarrow 1$ \cite{HHH96a}: $S_\alpha (\rho)\geq\max
\{S_\alpha (\rho_\A),S_\alpha (\rho_\B)\}$; where, for $\alpha>1$,
$S_\alpha(\rho):=\frac{1}{1-\alpha}\textrm{ln}(\tr(\rho^\alpha))$

\item Majorisation criterion \cite{NK01}: $\lambda_\rho^\downarrow
\prec \lambda_{\rho^\A}^\downarrow$ and $\lambda_\rho^\downarrow
\prec \lambda_{\rho^\B}^\downarrow$, where
$\lambda_\tau^\downarrow$ is the list of eigenvalues of $\tau$ in
nonincreasing order (padded with zeros if necessary), and $x\prec
y$ for two lists of size $s$ if and only if the sum of the first
$k$ elements of list $x$ is less than or equal to that of list $y$
for $k=1,2,...,s$; the majorisation condition implies
$\max\{\rank(\rho^\A), \rank(\rho^\B)\}\leq \rank(\rho)$.

\item Computable cross-norm/reshuffling criterion
\cite{qphRud02,CW03}: $||\mathcal{U}(\rho)||_1\leq 1$, where
$||X||_1:=\tr(\sqrt{X^\dagger X})$ is the trace norm; and
$\mathcal{U}(\rho)$, an $\dima^2\times\dimb^2$ matrix, is defined
on product states as $\mathcal{U}(A\otimes B):=v(A)v(B)^T$, where,
relative to a fixed basis,
$[v(A)]=(\col_1([A])^T,\ldots,\col_\dima([A])^T)^T$ (and similarly
for $v(B)$), where $\col_i([A])$ is the $i$th column of matrix
$[A]$; more generally \cite{qphHHH02}, any linear map
$\mathcal{U}$ that does not increase the trace norm of product
states may be used.

\end{itemize}

\noindent\textbf{Sufficient conditions for $\rho$ to be separable}
\begin{itemize}

\item Distance from maximally mixed state (see also
\cite{BCJLPS99}):

\begin{itemize}\item \cite{GB02}: e.g. $\tr(\rho - I_{\dima,
\dimb})^2\leq1/\dima\dimb(\dima\dimb-1)$\end{itemize}

\begin{itemize}\item \cite{ZHSL98,VT99} $\lambda_{\min}(\rho)\geq
(2+\dima\dimb)^{-1}$, where $\lambda_{\min}(\rho)$ denotes the
smallest eigenvalue of $\rho$\end{itemize}

\item When $\dima=2$ \cite{KCKL00}: $\rho=\rho^{T_A}$.
\end{itemize}

When $\rho$ is of a particular form, the PPT test is necessary and
sufficient for separability. This happens when
\begin{itemize}
\item $\dima\dimb \leq 6$ \cite{HHH96}; or

\item $\rank(\rho)\leq \dimb$ \cite{KCKL00,HLVC00}, see also
\cite{AFG01}.
\end{itemize}

The criteria not based on eigenvalues are obviously efficiently
computed i.e. computing the natural logarithm can be done with a
truncated Taylor series, and the rank can be computed by Gaussian
elimination. That the tests based on the remaining criteria are
efficiently computable follows from the efficiency of algorithms
for calculating the spectrum of a Hermitian
operator.\footnote{Note that $\rho^{T_\B}$ and $\rho^\A$ are
Hermitian.}  The method of choice for computing the entire spectra
is the QR algorithm (see any of \cite{WR71, GL96, SB02}), which
has been shown to have good convergence properties \cite{Wil68}.

In a series of articles (\cite{LS98}, \cite{KCKL00},
\cite{HLVC00}), various conditions for separability were obtained
which involve product vectors in the ranges of $\rho$ and
$\rho^{T_A}$. Any constructive separability checks given therein
involve computing these product vectors, but no general bounds
were obtained by the authors on the complexity of such
computations.

\section{One-sided tests based on semidefinite programming}\label{sec_OneSidedTestsOnSDP}

Let $\hermops$ denote the set of all Hermitian operators mapping
$\mathbb{C}^\dima\otimes\mathbb{C}^\dimb$ to
$\mathbb{C}^\dima\otimes\mathbb{C}^\dimb$; thus,
$\densops\subset\hermops$. This vector space is endowed with the
Hilbert-Schmidt inner product $\langle X, Y \rangle\equiv
\tr(AB)$, which induces the corresponding norm
$||X||\equiv\sqrt{\tr(X^2)}$ and distance measure $||X-Y||$. By
fixing an orthogonal Hermitian basis for $\hermops$, the elements
of $\hermops$ are in one-to-one correspondence with the elements
of the real Euclidean space $\mathbb{R}^{\n}$.  If the Hermitian
basis is orthonormal, then the Hilbert-Schmidt inner product in
$\hermops$ corresponds exactly to the Euclidean dot product in
$\mathbb{R}^{\n}$.

Thus $\densops$ and $\sep$ may be viewed as subsets of the
Euclidean space $\mathbb{R}^{\n}$; actually, because all density
operators have unit trace, $\densops$ and $\sep$ are
full-dimensional subsets of $\mathbb{R}^{\n-1}$.  This observation
aids in solving the quantum separability problem, allowing us to
easily apply well-studied mathematical-programming tools.
Below, I follow the popular
review article of semidefinite programming in \cite{VB96}.

\begin{definition}[Semidefinite program (SDP)] Given the
vector $c\in\mathbb{R}^m$ and Hermitian matrices
$F_i\in\mathbb{C}^{n\times n}$, $i=0,1,\ldots,m$,
\begin{eqnarray}
&\text{minimise} \hspace{2mm}& c^Tx\\
&\text{subject to:}\hspace{2mm}& F(x)\geq 0,
\end{eqnarray}
where $F(x):= F_0 + \sum_{i=1}^{m}x_iF_i$.
\end{definition}
\noindent Call $x$ \emph{(primal) feasible} when $F(x)\geq 0$.
When $c=0$, the SDP reduces to the \emph{semidefinite feasibility
problem}, which is to find an $x$ such that $F(x)\geq 0$ or assert
that no such $x$ exists. Semidefinite programs can be solved
efficiently, in time $O(m^2n^{2.5})$.  Most algorithms are
iterative.  Each iteration can be performed in time $O(m^2n^{2})$.
The number of required iterations has an analytical bound of
$O(\sqrt{n})$, but in practice is more like $O(\log(n))$ or
constant.

%

Let $\hermopsa$ ($\hermopsb$) denote the set of all Hermitian
operators mapping $\mathbb{C}^\dima$ to $\mathbb{C}^\dima$
($\mathbb{C}^\dimb$ to $\mathbb{C}^\dimb$).  The real variables of
the following SDPs will be the real coefficients of some quantum
state with respect to a fixed Hermitian basis of $\hermops$.  The
basis will be separable, that is, made from bases of $\hermopsa$
and $\hermopsb$.  It is usual to take the generators of
$SU(\dima)$ (the generalised Pauli matrices) as a basis for
$\hermopsa$ (see e.g. \cite{TNWM02}).

\subsection{A test based on symmetric extensions}\label{sec_DohertyEtalApproach}

Consider a separable state
$\sigma=\sum_i p_i
\ketbra{\psi^\A_i}{\psi^\A_i}\otimes\ketbra{\psi^\B_i}{\psi^\B_i}$,
and consider the following \emph{symmetric extension of $\sigma$
to $k$ copies of subsystem $\A$} ($k\geq 2$):
\begin{eqnarray}
\tilde{\sigma}_k=\sum_i p_i
(\ketbra{\psi^\A_i}{\psi^\A_i})^{\otimes
k}\otimes\ketbra{\psi^\B_i}{\psi^\B_i}.
\end{eqnarray}
\noindent The state $\tilde{\sigma}_k$ is so called because it
satisfies two properties: (i) it is symmetric (unchanged) under
permutations (swaps) of any two copies of subsystem $\A$; and (ii)
it is an extension of $\sigma$ in that tracing out any of its
$(k-1)$ copies of subsystem $\A$ gives back $\sigma$.  For an
arbitrary density operator $\rho\in\densopsgen{\CMotimesCN}$,
define a \emph{symmetric extension of $\rho$ to $k$ copies of
subsystem $\A$} ($\mathbb{C}^\dima$) as any density operator
$\rho'\in\densopsgen{(\mathbb{C}^\dima)^{\otimes
k}\otimes\mathbb{C}^\dimb}$ that satisfies (i) and (ii) with
$\rho$ in place of $\sigma$. It follows that if an arbitrary state
$\rho$ does not have a symmetric extension to $k_0$ copies of
subsystem $\A$ for some $k_0$, then $\rho\notin\sep$ (else we
could construct $\tilde{\rho}_{k_0}$). Thus a method for searching
for symmetric extensions of $\rho$ to $k$ copies of subsystem $\A$
gives a sufficient test for separability.

Doherty et al.\ \cite{DPS02, DPS04} showed that the search for a
symmetric extension to $k$ copies of $\rho$ (for any fixed $k$)
can be phrased as a SDP\@. This result, combined with the
``quantum de Finetti theorem'' \cite{FLV88, CFS02} that
$\rho\in\sep$ if and only if, for all $k$, $\rho$ has a symmetric
extension to $k$ copies of subsystem $\A$, gives an infinite
hierarchy (indexed by $k=2,3,\ldots$) of SDPs with the property
that, for each entangled state $\rho$, there exists a SDP in the
hierarchy whose solution will imply that $\rho$ is entangled.

Actually, Doherty et al.\ develop a stronger test, inspired by
Peres' PPT test.  The state $\tilde{\sigma}_k$, which is positive
semidefinite, satisfies a third property: (iii) it remains
positive semidefinite under all possible partial transpositions.
Thus $\tilde{\sigma}_k$ is more precisely called a \emph{PPT
symmetric extension}.  The SDP can be easily modified to perform a
search for PPT symmetric extensions without any significant
increase in computational complexity (one just needs to add
constraints that force the partial transpositions to be positive
semidefinite). This strengthens the separability test, because a
given (entangled) state $\rho$ may have a symmetric extension to
$k_0$ copies of subsystem $\A$ but may not have a PPT symmetric
extension to $k_0$ copies of subsystem $\A$ (Doherty et al.\ also
show that the $(k+1)$st test in this stronger hierarchy subsumes
the $k$th test).

The final SDP has the following form:
\begin{equation}\label{prob_DohertyEtalSDP}
\begin{array}{rlrcl}
&\text{minimise} \hspace{2mm} & 0 && \\
&\text{subject to:}\hspace{2mm}& \tilde{X}_k &\geq& 0  \\
&\hspace{2mm}&  (\tilde{X}_k)^{T_j}&\geq&0,\hspace{2mm} j\in J,
\end{array}
\end{equation}
where $\tilde{X}_k$ is a parametrisation of a symmetric extension
of $\rho$ to $k$ copies of subsystem $\A$, and $J$ is the set of
all subsets of the $(k+1)$ subsystems that give rise to
inequivalent partial transposes $(\tilde{X}_k)^{T_j}$ of
$\tilde{X}_k$.  By exploiting the symmetry property, the number of
variables of the SDP is $m=(d_{S_k}^2-\dima^2)\dimb^2$, where
$d_{S_k}=\begin{pmatrix} \dima+k-1 \\ k \end{pmatrix}$ is the
dimension of the symmetric subspace of
$(\mathbb{C}^\dima)^{\otimes k}$.  The size of the matrix
$\tilde{X}_k$ for the first constraint is $d_{S_k}^2\dimb^2$.  The
number of inequivalent partial transpositions is
$|J|=k$.\footnote{Choices are: transpose subsystem $\B$, transpose
1 copy of subsystem $\A$, transpose 2 copies of subsystem $\A$,
..., transpose $k-1$ copies of subsystem $\A$. Transposing all $k$
copies of subsystem $\A$ is equivalent to transposing subsystem
$\B$.  Transposing with respect to both subsystem $\B$ and $l$
copies of subsystem $\A$ is equivalent to transposing with respect
to $k-l$ copies of subsystem $\A$.}  The constraint corresponding
to the transposition of $l$ copies of $\A$, $l=1,2,...,k-1$, has a
matrix of size $d_{S_l}^2d_{S_{(k-l)}}^2\dimb^2$ \cite{DPS04}. I
will estimate the total complexity of this approach to the quantum
separability problem in Section \ref{sec_DohertyEtalKonigRenner}.


\subsection{A test based on semidefinite
relaxations}\label{subsec_EisertsEtalApproach}

Doherty et al.\ formulate a \emph{hierarchy of necessary criteria}
for separability in terms of semidefinite programming -- each
separability criterion in the hierarchy may be checked by a SDP\@.
As it stands, their approach is manifestly a one-sided test for
separability, in that at no point in the hierarchy can one
conclude that the given $[\rho]$ corresponds to a separable state
(happily, recent results show that this is, practically, not the
case; see Section \ref{sec_DohertyEtalKonigRenner}).

Soon after, Eisert et al. \cite{EHGC04} had the idea of
formulating a \emph{necessary and sufficient criterion} for
separability as a hierarchy of SDPs. Define the function
\begin{eqnarray}\label{eqn_DefnOfQuasiRelEntropy}
E_{d^2_2}(\rho) := \min_{x\in\sep} \tr((\rho - x)^2)
\end{eqnarray}
for $\rho\in\densops$.  As $\tr((\rho - x)^2)$ is the square of
the Euclidean distance from $\rho$ to $x$, $\rho$ is separable if
and only if $E_{d^2_2}(\rho)=0$.  The problem of computing
$E_{d^2_2}(\rho)$ (to check whether it is zero) is already
formulated as a constrained optimisation.
\noindent The following observation helps to rewrite these
constraints as low-degree polynomials in the variables of the
problem:\footnote{To see why Fact \ref{fact_EisertPurityFact}
holds, note that in $\mathbb{R}^n$ the surface
$\{(x_1,\ldots,x_n): \hspace{1mm}\sum_{i=1}^n x_i^3=\alpha^3\}$
intersects the hypersphere $\{(x_1,\ldots,x_n):
\hspace{1mm}\sum_{i=1}^n x_i^2=\alpha^2\}$ only at the points
$(\alpha,0,\ldots,0)$, $(0,\alpha,0,\ldots,0)$, ...,
$(0,\ldots,0,\alpha,0,\ldots,0)$, ..., $(0,\ldots,0,\alpha)$.}
\begin{fact}[\cite{EHGC04}]\label{fact_EisertPurityFact}
Let $O$ be a Hermitian operator and let $\alpha\in\mathbb{R}$
satisfy $0<\alpha\leq 1$.  If $\tr(O^2)=\alpha^2$ and
$\tr(O^3)=\alpha^3$, then $\tr(O)=\alpha$ and $\rank(O)=1$ (i.e.
$O$ corresponds to an unnormalised pure state).
\end{fact}
\noindent Combining Fact \ref{fact_EisertPurityFact} with Fact
\ref{fact_FiniteDecompOfSepState}, the problem is equivalent to
\begin{equation}\label{prob_EisertsQSEP}
\begin{array}{rlrcl}
&\text{minimise} \hspace{2mm} & \tr((\rho - \sum_{i=1}^{\n}X_{i})^2) && \\
&\text{subject to:     }\hspace{10mm}& \tr(\sum_{i=1}^{\n}X_{i}) &=& 1  \\
&\hspace{2mm}& \tr((\tr_{j}(X_{i}))^2)&=& (\tr(X_i))^2,\hspace{2mm}\\
&&&&\text{for $i=1,2,\ldots,\n$ and $j\in\{\A,\B\}$}  \\
&\hspace{2mm}& \tr((\tr_{j}(X_{i}))^3)&=& (\tr(X_i))^3,\hspace{2mm}\\
&&&&\text{for $i=1,2,\ldots,\n$ and $j\in\{\A,\B\}$},
\end{array}
\end{equation}
where the new variables are Hermitian matrices $X_{i}$ for
$i=1,2,\ldots,\n$.  The constraints do \emph{not} require $X_{i}$
to be tensor products of \emph{unit-trace} pure density operators,
because the positive coefficients (probabilities summing to 1)
that would normally appear in the expression
$\sum_{i=1}^{\n}X_{i}$ are absorbed into the $X_{i}$, in order to
have fewer variables (i.e. the $X_i$ are constrained to be density
operators corresponding to unnormalised pure product states). Once
an appropriate Hermitian basis is chosen for $\hermops$, the
matrices $X_{i}$ can be parametrised by the real coefficients with
respect to the basis; these coefficients form the real variables
of the feasibility problem. The constraints in
(\ref{prob_EisertsQSEP}) are polynomials in these variables of
degree less than or equal to 3.\footnote{Alternatively, we could
parametrise the pure states (composing $X_i$) in
$\mathbb{C}^\dima$ and $\mathbb{C}^\dimb$ by the real and
imaginary parts of rectangularly-represented complex coefficients
with respect to the standard bases of $\mathbb{C}^\dima$ and
$\mathbb{C}^\dimb$:
\begin{equation}\label{prob_EisertsQSEPStandardBasis}
\begin{array}{rlrcl}
&\text{minimise} \hspace{2mm} & 0 && \\
&\text{subject to:     }\hspace{10mm}& \tr((\rho - \sum_{i=1}^{\n}\ketbra{\psi^{\A}_i}{\psi^{\A}_i}\otimes\ketbra{\psi^{\B}_i}{\psi^{\B}_i})^2)&=& 0  \\
&\hspace{2mm}& \tr\left(\sum_{i=1}^{\n}\ketbra{\psi^{\A}_i}{\psi^{\A}_i}\otimes\ketbra{\psi^{\B}_i}{\psi^{\B}_i}\right)  &=& 1.  \\
\end{array}
\end{equation}
This parametrisation hard-wires the constraint that the
$\ketbra{\psi^{\A}_i}{\psi^{\A}_i}\otimes\ketbra{\psi^{\B}_i}{\psi^{\B}_i}$
are (unnormalised) pure product states, but increases the degree
of the polynomials in the constraint to 4 (for the unit trace
constraint) and 8 (for the distance constraint).}

Polynomially-constrained optimisation problems can be approximated
by, or \emph{relaxed} to, semidefinite programs, via a number of
different approaches (see references in
\cite{EHGC04}).\footnote{For our purposes, the idea of a
relaxation can be briefly described as follows. The given problem
is to solve $\min_{x\in\mathbb{R}^n}\{p(x):\hspace{1mm}g_k(x)\geq
0, k=1,\ldots,m\}$, where
$p(x),g_i(x):\hspace{1mm}\mathbb{R}^n\rightarrow \mathbb{R}$ are
real-valued polynomials in $\mathbb{R}[x_1,\ldots,x_n]$.  By
introducing new variables corresponding to products of the given
variables (the number of these new variables depends on the
maximum degree of the polynomials $p,g_i$), we can make the
objective function linear in the new variables; for example, when
$n=2$ and the maximum degree is 3, if
$p(x)=3x_1+2x_1x_2+4x_1x_2^2$ then the objective function is
$c^Ty$ with $c=(0,3,0,0,2,0,0,0,4,0)\in\mathbb{R}^{10}$ and
$y\in\mathbb{R}^{10}$, where $10$ is the total number of monomials
in $\mathbb{R}[x_1,x_2]$ of degree less than or equal to 3. Each
polynomial defining the feasible set
$G:=\{x\in\mathbb{R}^n:\hspace{1mm}g_k(x)\geq 0, k=1,\ldots,m\}$
can be viewed similarly.  A relaxation of the original problem is
a SDP with objective function $c^Ty$ and with a (convex) feasible
region (in a higher-dimensional space) whose projection onto the
original space $\mathbb{R}^n$ approximates $G$. Better
approximations to $G$ can be obtained by going to higher
dimensions.} Some approaches even give an asymptotically complete
hierarchy of SDPs, indexed on, say, $i=1,2,\ldots$.  The SDP at
level $i+1$ in the hierarchy gives a better approximation to the
original problem than the SDP at level $i$; but, as expected, the
size of the SDPs grows with $i$ so that better approximations are
more costly to compute.  The hierarchy is asymptotically complete
because, under certain conditions, the optimal values of the
relaxations converge to the optimal value of the original problem
as $i\rightarrow\infty$. Of these approaches, the method of
Lasserre \cite{Las01} is appealing because a computational package
\cite{HL03} written in MATLAB is freely available. Moreover, this
package has built into it a method for recognising when the
optimal solution to the original problem has been found (see
\cite{HL03} and references therein). Because of this feature, the
one-sided test becomes, in practice, a full algorithm for the
quantum separability problem.  However, no analytical worst-case
upper bounds on the running time of the algorithm for arbitrary
$\rho\in\densops$ are available.

\subsection{Entanglement
Measures}\label{subsec_EntanglementMeasures}

The function $E_{d^2_2}(\rho)$ defined in Eqn.
(\ref{eqn_DefnOfQuasiRelEntropy}), but first defined in
\cite{VPRK97}, is also known as an \emph{entanglement measure},
which, at the very least, is a nonnegative real function defined
on $\densops$.\footnote{For a comprehensive review of entanglement
measures (and a whole lot more!), see \cite{Chr05}.} If an
entanglement measure $E(\rho)$ satisfies
\begin{eqnarray}\label{crit_VanSepAndPosEnt}
E(\rho)=0\hspace{2mm}\Leftrightarrow\hspace{2mm}\rho\in\sep,
\end{eqnarray}
then, in principle, any algorithm for computing $E(\rho)$ gives an
algorithm for the quantum separability problem. Note that most
entanglement measures $E$ do not satisfy
(\ref{crit_VanSepAndPosEnt}); most just satisfy
$E(\rho)=0\Leftarrow\rho\in\sep$.

A class of entanglement measures that do satisfy
(\ref{crit_VanSepAndPosEnt}) are the so-called ``distance
measures'' $E_d(\rho):= \min_{\sigma\in\sep} d(\rho,\sigma)$, for
any reasonable measure of ``distance'' $d(x,y)$ satisfying
$d(x,y)\geq 0$ and $(d(x,y)=0)\Leftrightarrow (x=y)$. If $d$ is
the square of the Euclidean distance, we get $E_{d^2_2}(\rho)$.
Another ``distance measure'' is the von Neumann relative entropy
$S(x,y):= \tr(x(\log x - \log y))$.

In Eisert et al.'s approach, we could replace $E_{d^2_2}$ by $E_d$
for any ``distance function'' $d(\rho,\sigma)$ that is expressible
as a polynomial in the variables of $\sigma$.  What dominates the
running time of Eisert et al.'s approach is the implicit
minimisation over $\sep$, so using a different ``distance
measure'' (i.e. only changing the first constraint in
(\ref{prob_EisertsQSEP})) like $(\tr(\rho-\sigma))^2$ would not
improve the analytic runtime (because the degree of the polynomial
in the constraint is still 2), but may help in practice.

Another entanglement measure $E$ that satisfies
(\ref{crit_VanSepAndPosEnt}) 
 is the \emph{entanglement of formation} \cite{BDSW96}
\begin{eqnarray}
E_F(\rho):= \min_{\{p_i,\ketbra{\psi_i}{\psi_i}\}_i:\hspace{2mm}
\rho=\sum_ip_i\ketbra{\psi_i}{\psi_i}}\sum_ip_iS(\tr_{\B}(\ketbra{\psi_i}{\psi_i})),
\end{eqnarray}
where $S(\rho):=-\tr(\rho\log(\rho))$ is the von Neumann entropy.
This gives another strategy for a separability algorithm: search
through all decompositions of the given $\rho$ to find one that is
separable.  We can implement this strategy using the same
relaxation technique of Eisert et al., but first we have to
formulate the strategy as a polynomially-constrained optimisation
problem. The role of the function $S$ is to measure the
entanglement of $\ketbra{\psi_i}{\psi_i}$ by measuring the
mixedness of the reduced state
$\tr_{\B}(\ketbra{\psi_i}{\psi_i})$.  For our purposes, we can
replace $S$ with any other function $T$ that measures mixedness
such that, for all $\rho\in\densops$, $T(\rho)\geq 0$ and
$T(\rho)=0$ if and only if $\rho$ is pure. Recalling that, for any
$\rho\in\densops$, $\tr(\rho^2)\leq 1$ with equality if and only
if $\rho$ is pure, the following function $T(\rho):= 1 -
\tr(\rho^2)$ suffices; this function $T$ may be written as a
(finite-degree) polynomial in the real variables of $\rho$,
whereas $S$ could not. Defining
\begin{eqnarray}\label{eqn_DefnMyPurityEntanglementMeasure}
E'_F(\rho):= \min_{\{p_i,\ketbra{\psi_i}{\psi_i}\}_i:\hspace{2mm}
\rho=\sum_ip_i\ketbra{\psi_i}{\psi_i}}\sum_ip_iT(\tr_{\B}(\ketbra{\psi_i}{\psi_i})),
\end{eqnarray}
we have that $E'_F$ satisfies (\ref{crit_VanSepAndPosEnt}). Using
an argument similar to the proof of Lemma 1 in \cite{Uhl98}, we
can show that the minimum in
(\ref{eqn_DefnMyPurityEntanglementMeasure}) is attained by a
\emph{finite} decomposition of $\rho$ into $\n+1$ pure states.
Thus, the following polynomially-constrained optimisation problem
can be approximated by semidefinite relaxations:
\begin{equation}\label{prob_MyEisert-likeQSEP}
\begin{array}{rlrcl}
&\text{minimise} \hspace{2mm} & \sum_{i=1}^{\n+1}\tr(X_{i})T(\tr_{\B}(X_{i})) && \\
&\text{subject to:     }& \tr(\sum_{i=1}^{\n+1}X_{i}-[\rho])^2 &=& 0 \\
&\hspace{2mm}& \tr(\sum_{i=1}^{\n+1}X_{i}) &=& 1  \\
&\hspace{2mm}& \tr(X_{i}^2)&=& (\tr(X_i))^2,\hspace{2mm}\\
&&&&\text{for $i=1,2,\ldots,\n+1$}\\
&\hspace{2mm}& \tr(X_{i}^3)&=& (\tr(X_i))^3,\hspace{2mm}\\
&&&&\text{for $i=1,2,\ldots,\n+1$}.
\end{array}
\end{equation}
The above has about half as many constraints as
(\ref{prob_EisertsQSEP}), so it would be interesting to compare
the performance of the two approaches.


\subsection{Other tests}

There are several one-sided tests which do not lead to full
algorithms for the quantum separability problem for $\sep$.
 Brand{\~{a}}o and Vianna \cite{BV04} have a set of one-sided
necessary tests based on \emph{deterministic} relaxations of a
robust semidefinite program, but this set is not an asymptotically
complete hierarchy.  The same authors also have a related
\emph{randomised} quantum separability algorithm which uses
probabilistic relaxations of the same robust semidefinite program
\cite{BV04a}.  Randomised algorithms for the quantum separability
problem are outside the scope of this thesis.

Woerdeman \cite{Woe03} has a set of one-sided tests for the case
where $\dima=2$.  His approach might be described as the
mirror-image of Doherty et al.'s: Instead of using an infinite
hierarchy of necessary criteria for separability, he uses an
infinite hierarchy of sufficient criteria.  Each criterion in the
hierarchy can be checked with a SDP.


\chapter{Convexity}\label{ChapterConvexity}

The set of bipartite separable quantum states $\sep$ in $\hermops$
is defined as the closed convex hull of the separable pure states:
\begin{eqnarray}
\sep :=\conv \lbrace
\ketbra{\psi^\A_i}{\psi^\A_i}\otimes\ketbra{\psi^\B_i}{\psi^\B_i}\in
\hermops\rbrace.
\end{eqnarray}
$\sep$ is also compact (see e.g. \cite{Hor97}). Since the
separable states form a convex and compact subset of
$\mathbb{R}^{\n}$, a plethora of well-studied mathematical and
computational tools are available for the separability problem, as
we shall see.

First, I apply polyhedral theory to show that $\sep$ is not a
polytope, easily settling an open problem.  I then review the
concept of an entanglement witness and define a new class of
entanglement witnesses which have some advantage over conventional
entanglement witnesses in the detection of entanglement.  I finish
the chapter with a review of the five basic convex body problems
and their relation to the separability problem.

\section{Polyhedra and $\sep$}\label{sec_PolyhedraAndSep}

The following definitions may be found in \cite{NW88} (but I use
operator notation in keeping with the spirit of quantum physics).
If $A\in \hermops$ and $A\neq 0$ and $a\in\mathbb{R}$, then
$\{x\in \hermops:\hspace{2mm} \tr(Ax)\leq a\}$ is called the
\emph{halfspace} $H_{A,a}$.  The boundary $\{x\in
\hermops:\hspace{2mm} \tr(Ax)=a\}$ of $H_{A,a}$ is the
\emph{hyperplane} $\pi_{A,a}$ with \emph{normal} $A$.  Call two
hyperplanes \emph{parallel} if they share the same normal. Let
$H^\circ_{A,a}$ denote the interior $H_{A,a}\setminus \pi_{A,a}$
of $H_{A,a}$.  Note that $H^\circ_{-A,-a}$ is just the complement
of $H_{A,a}$. The density operators of an $M$ by $N$ quantum
system lie on the hyperplane $\pi_{I,1}$:
$\mathcal{D}_{M,N}=\{\rho\in \hermops:\hspace{2mm} \rho\geq
0\}\cap \pi_{I,1}\subset\mathbb{R}^{\n-1}$.

The intersection of finitely many halfspaces is called a
\emph{polyhedron}.  Every polyhedron is a convex set.
 Let $D$ be a polyhedron. A set $F\subseteq D$ is a \emph{face} of
$D$ if there exists a halfspace $H_{A,a}$ containing $D$ such that
$F=D\cap \pi_{A,a}$. If $v$ is a point in $D$ such that the set
$\{v\}$ is a face of $D$, then $v$ is called a \emph{vertex} of
$D$.  A \emph{facet} of $D$ is a nonempty face of $D$ having
dimension one less than the dimension of $D$.  A polyhedron that
is contained in a hypersphere $\{x\in \hermops:\hspace{2mm}
\tr(x^2)= r^2\}$ of finite radius $r$ is called a \emph{polytope}.

What is the shape of $\sep$ in $\mathbb{R}^{\n-1}$ (with respect
to the Euclidean norm)?  Is it a polytope?  This is an interesting
question which arises when considering separability in an
experimental setting and comparing it to nonlocality (Section
\ref{sec_EWs}).

Minkowski's theorem \cite{NW88} says that every polytope in
$\mathbb{R}^{n}$ is the convex hull of its \emph{finitely many}
vertices (extreme points).  Recall that an extreme point of a
convex set is one that cannot be written as a nontrivial convex
combination of other elements of the set.  To show that $\sep$ is
not a polytope, it suffices to show that it has infinitely many
extreme points.  The extreme points of $\sep$ are precisely the
product states, as we now show (see also \cite{Hor97}).  A mixed
state is not extreme, by definition.  Conversely, we have that
\begin{eqnarray}
\ketbra{\psi}{\psi}=\sum_ip_i\ketbra{\psi_i}{\psi_i}
\end{eqnarray}
implies
\begin{eqnarray}
1=\sum_ip_i\bra{\psi}\ketbra{\psi_i}{\psi_i}\ket{\psi}=\sum_ip_i
|\braket{\psi_i}{\psi}|^2,
\end{eqnarray}
which implies that $|\braket{\psi_i}{\psi}|=1$ for all $i$; thus,
a pure state is extreme.  Since $\sep$ has infinitely many pure
product states, we have the following fact, which settles an open
problem posed in \cite{Bru02}.
\begin{fact}
$\sep$ is not a polytope in $\mathbb{R}^{\n-1}$.
\end{fact}

\section{Entanglement witnesses}\label{sec_EWs}

The compactness of $\sep$ and the fact that any point not in a
convex set in $\mathbb{R}^n$ can be separated from the set by a
hyperplane imply that for each entangled state $\rho$ there exists
a halfspace $H_{A,a}$ whose interior $H^\circ_{A,a}$ contains
$\rho$ but contains no member of $\sep$ \cite{HHH96}. Call $A\in
\hermops$ an \emph{entanglement witness} \cite{Ter00} if for some
$a\in\mathbb{R}$
\begin{eqnarray}\label{Def_LeftEW}
\sep\cap
H^\circ_{A,a}=\varnothing\hspace{3mm}\text{and}\hspace{3mm}\ent\cap
H^\circ_{A,a}\neq\varnothing.
\end{eqnarray}
Entanglement witnesses $A$ with $a=0$ in (\ref{Def_LeftEW})
correspond to the conventional definition of ``entanglement
witness'' found in the literature, e.g. \cite{GHBELMS02}.

\subsection{Experimental separability}

Suppose that a physical property $A$ of a state $\rho$ may be
measured or \emph{observed}.  The result of such a measurement is
a real number (in practice having finite representation dictated
by the precision of the measurement apparatus).  An axiom of
quantum mechanics is that all possible real outcomes of measuring
property $A$ form the spectrum of a Hermitian operator (which we
also denote by ``$A$'').  We assume that in principle all such
physical properties $A$ are in one-to-one correspondence with the
Hermitian operators acting on the Hilbert space, so that any
Hermitian operator defines a physical property that can be
measured. When property $A$ of $\rho$ is measured in the
laboratory, the \emph{measurement axiom} dictates that the
\emph{expected value} of the measurement is
\begin{eqnarray}\nonumber
\langle A\rangle_\rho :=\tr(A\rho).
\end{eqnarray}
Such physical properties or Hermitian operators, $A$, are also
called $\emph{observables}$.

Entanglement witnesses can be used to determine that a physical
quantum state is entangled. Suppose $A$ is an EW as in
(\ref{Def_LeftEW}) and that a state $\rho$ that is produced in the
lab is not known to be separable. If sufficiently many copies of
$\rho$ may be produced, then measuring the observable $A$ (once)
on each copy of $\rho$ gives a good estimate of $\langle
A\rangle_\rho$ which, if less than $a$, indicates that $\rho\in
H^\circ_{A,a}$ and hence that $\rho$ is entangled. Otherwise, if
$\langle A\rangle_\rho\geq a$, then $\rho$ may be entangled or
separable. The best value of $a$ to use in (\ref{Def_LeftEW}) is
$a^* = \min_{\ketbra{\psi}{\psi}\in \sep}
\{\bra{\psi}A\ket{\psi}\}$ since, with this value of $a$, the
hyperplane $\pi_{A,a}$ is tangent to $\sep$ and thus the volume of
entangled states that can be detected by measuring observable $A$
is maximised.  With this in mind, define
\begin{eqnarray}\nonumber
a^*(A):= \min_{\ketbra{\psi}{\psi}\in \sep}
\{\bra{\psi}A\ket{\psi}\}
\end{eqnarray}
if $A$ is an entanglement witness.

Much work has been done on entanglement witnesses and their
utility in investigating the separability of quantum states, e.g.
\cite{LKCH00,LKHC01}. Entanglement witnesses have been found to be
particularly useful for experimentally detecting the entanglement
of states of the particular form
$p\ketbra{\psi}{\psi}+(1-p)\sigma$, where $\ket{\psi}$ is an
entangled state and $\sigma$ is a mixed state close to the
maximally mixed state and $0\leq p\leq 1$
\cite{GHBELMS02,qphBMNMDM03}.

\subsection{Polytopes in separability and nonlocality}\label{sec_PolytopesSep&Nonlocality}

Detection of the entanglement of reproducible physical states in
the lab would be straightforward if there were a relatively small
number $K$ of entanglement witnesses $A_i$ such that $\ent$ is
contained in
\begin{eqnarray}\label{ENTComplementPolytope}\nonumber
\bigcup_{i=1}^{K} H_{A_i,a_i},
\end{eqnarray}
where $a_i:=a^*(A_i)$.  This would imply that $\sep$ is
\begin{eqnarray}\label{SEPAPolytope}\nonumber
\bigcap_{i=1}^{K} H_{-A_i,-a_i},
\end{eqnarray}
that is, that $\sep$ is a polytope.  Alas, it is not (see Section
\ref{sec_PolyhedraAndSep}).  But this raises an interesting
question:

\begin{problem}\label{prob_BestPolytopForS}
Given $k\geq \n$, find the $k$-facet polytope $\Pi$ containing
$\sep$ such that the volume of $\Pi\setminus\sep$ is minimal.
\end{problem}


Polytope enthusiasts will be happy to know, however, that their
favorite convex set plays a role in the confounding issue of
\emph{nonlocality}, which I now explain. We know that for any
entangled state there is always an observable (entanglement
witness) acting on the total system whose statistics will imply
that the state of the system is entangled. We also noted earlier
that entangled states could not be prepared by Alice and Bob with
just LOCC\@.  It turns out that the \emph{total} statistics of
some set of \emph{local} observables on an entangled state can
also imply that the state is entangled, by revealing the
inconsistency with LOCC.

Alice and Bob share the bipartite system and want to probe its
properties by each performing some local tests independently of
each other (for a statistical interpretation, we again assume that
Alice and Bob will repeat this procedure with identically prepared
systems infinitely many times).  After performing the tests, they
will communicate their results to a common location to be
analysed.  They will want to see if the results of their tests
violate an assumption that their subsystems are correlated in a
way no stronger that what is allowed by LOCC\@.  Suppose Alice
will choose one of $N^\A$ tests (labelled by $A_i$) to perform,
with the $i^\A$th test having one of $N^\A_i$ mutually exclusive
outcomes (labelled by $A_i(j)$). If Alice's subsystem were totally
independent of Bob's, then the outcomes of her tests may be
thought to be governed by a local variable $\lambda^\A$ which --
while possibly uncontrollable or inaccessible -- may indeed exist
(local realism assumption); the possible values that $\lambda^\A$
may assume are in one-to-one correspondence with the possible
states of Alice's subsystem.  A particular setting of $\lambda^\A$
dictates which outcome each test will have.  Thus, for a given set
of tests, we can view each $\lambda^\A$ as a Boolean vector of
length $\sum_i N^\A_i$ that is the concatenation of $N^\A$ Boolean
vectors each of length $N^\A_i$ and each having exactly 1 nonzero
entry. For example, for $N^\A=2$ and $N^\A_1=2$ and $N^\A_2=3$, a
possible $\lambda^\A$ is $ \lambda^\A = (0,1;0,1,0)$, which says
that test $A_1$ will have outcome $A_1(2)$ and test $A_2$ will
have outcome $A_2(2)$.  We assume a similar setup on Bob's side.
The total hidden variable is then
$\lambda=(\lambda^\A,\lambda^\B)$ which dictates Alice's and Bob's
results.  Now $B_\lambda:=\lambda^\A\otimes\lambda^\B$ is the
vector whose entries are probabilities of getting pairs of
outcomes (conditioned on performing the tests which can give rise
to such outcomes).

Suppose Alice and Bob carry out their experiment which consists of
repeated trials, the measurements in each trial done
simultaneously\footnote{It follows from the postulates of the
theory of relativity that physical influences cannot propagate
faster than light.  More precisely, using the terminology of
relativity, we want the measurements to be done in a causally
disconnected manner.} to prevent Alice's outcome from influencing
Bob's and vice versa. Let $P$ be the vector of measured
(conditional) probabilities of pairs of outcomes. Then the
statistics are consistent with a LOCC state if and only if
\begin{eqnarray}
P\in\conv(\{B_\lambda\}_\lambda),
\end{eqnarray}
where $\conv(\{B_\lambda\}_\lambda)$ is called the
\emph{correlation polytope}.  Note that there is a different
correlation polytope for every different experimental
setup.\footnote{I have followed the formulation of Peres
\cite{Per99}, which is tailored to the nonlocality problem.
Pitowsky's very general formulation \cite{Pit91} has application
beyond the nonlocality problem; however, it is well suited to
tests with two outcomes (Boolean tests), as in photon detectors
(which either ``click'' or do not ``click''), where it gives a
polytope in lower dimension than Peres' construction, e.g. compare
the treatments of \cite{CH74} in \cite{Pit91} and \cite{Per99}.
For tests with more than two outcomes, Pitowsky's correlation
polytope contains ``local junk'' -- product-vectors (e.g.
$(1,1,\dots,1)$) which are not valid statistical vectors $P$ (an
artifact of the generality of the construction which allows for
not necessarily distinct events).}

A hyperplane which separates $P$ from the correlation polytope
(corresponding to some experimental setup) corresponds to a
``violation of a generalised Bell inequality'' \cite{Bel64, Bel66,
CHSH69}, which indicates that the state of the system is not
separable. However, to show that a state is consistent with a
local hidden variables theory would require examining all possible
correlation polytopes and corresponding statistical vectors $P$
i.e. all possible experiments.  Experiments can also be done on
pairs (or triples, etc.) of subsystems at a time, or Alice and Bob
could perform sequences of tests rather than just single tests. In
the case of some ``Werner states'' \cite{Wer89}, this more general
type of experimental setup gives rise to a violation of a Bell
inequality, where the simple setup above does not \cite{Pop95}.
The strange thing about quantum mechanics is that there may exist
states whose statistics are consistent with LOCC but which cannot
be prepared with LOCC; entangled states which pass the PPT test
are conjectured to be such states.

\subsection{Ambidextrous entanglement witnesses}\label{sec_AEW}

Suppose that $A$ is not an entanglement witness but that $-A$ is.
In this case, an estimate of $\tr(A\rho)$ is just as useful in
testing whether $\rho$ is entangled.  We extend the definition of
``entanglement witness'' to reflect this fact:  Call $A\in
\hermops$ a \emph{left (entanglement) witness} if
(\ref{Def_LeftEW}) holds for some $a\in\mathbb{R}$, and a
\emph{right (entanglement) witness} if
\begin{eqnarray}\label{Def_RightEW}
\sep\cap
H^\circ_{-A,-b}=\varnothing\hspace{3mm}\text{and}\hspace{3mm}
\ent\cap H^\circ_{-A,-b}\neq\varnothing
\end{eqnarray}
for some $b\in\mathbb{R}$.  As well, for $A$ a right witness,
define
\begin{eqnarray}\nonumber
b^*(A):= \max_{\ketbra{\psi}{\psi}\in \sep}
\{\bra{\psi}A\ket{\psi}\}.
\end{eqnarray}
Note that $A$ is a left witness if and only if $-A$ is a right
witness.

The operator $A\in \hermops$ defines the family
$\{\pi_{A,a}\}_{a\in\mathbb{R}}$ of parallel hyperplanes in
$\mathbb{R}^{\n}$.  Consider the hyperplane
$\pi_A:=\pi_{A,\frac{\tr(A)}{MN}}$ which cuts through $\sep$ at
the maximally mixed state $I_{MN}$.  When can $\pi_A$ be shifted
parallel to its normal so that it separates $\sep$ from some
entangled states? If $A$ is \emph{both} a left and right witness,
then $\pi_A$ can be shifted either in the positive or negative
directions of the normal.  In this case, the two parallel
hyperplanes $\pi_{A,a^*(A)}$ and $\pi_{A,b^*(A)}$ sandwich $\sep$
with some entangled states outside of the \emph{sandwich}, which
we will denote by $W(A):=H_{-A,-a^*(A)}\cap H_{-A,-b^*(A)}$.


\begin{definition}[Ambidextrous entanglement witness]
An operator $A\in \hermops$ is an \emph{ambidextrous
(entanglement) witness} if it is both a left witness and a right
witness.
\end{definition}
If $A$ is an ambidextrous witness, then $\rho$ is entangled if
$\langle A\rangle_\rho <a^*(A)$ \emph{or} if $\langle
A\rangle_\rho
>b^*(A)$.  We can further define a \emph{left-handed} witness to
be an entanglement witness that is left but not right. Say that
two entangled states $\rho_1$ and $\rho_2$ are \emph{on opposite
sides of $\sep$} if there does not exist a halfspace $H_{A,a}$
such that $H^\circ_{A,a}$ contains $\rho_1$ and $\rho_2$ but
contains no separable states.  Ambidextrous witnesses have the
potential advantage over conventional (left-handed) entanglement
witnesses that they can detect entangled states on opposite sides
of $\sep$ with the \emph{same} physical measurement.

Entanglement witnesses can be simply characterised by their
spectral decomposition. In the following, suppose $A\in \hermops$
has spectral decomposition $A=\sum_{i=0}^{MN-1} \lambda_i
\ketbra{\lambda_i}{\lambda_i}$ with
$\lambda_0\leq\lambda_1\leq\ldots\leq\lambda_{MN-1}$.
\begin{fact}\label{Fact_CharShiftLeft}
The operator $A$ is a left witness if and only if there
 exists $k\in [0,1,\ldots,MN-2]$ such that
 $\sp(\{\ket{\lambda_0},\ket{\lambda_1},\ldots,\ket{\lambda_k}\})$
 contains no separable pure states and $\lambda_{k+1}>\lambda_k$.
\end{fact}

\begin{proof}Suppose first that there exists no such $k$. Then
$\ket{\lambda_0}$ is, without loss of generality, a separable pure
state (because the eigenspace corresponding to $\lambda_0$ must
contain a product state),
so $A$ cannot be a left witness. To prove the converse, suppose
that such a $k$ does exist and that $\lambda_{k+1}>\lambda_k$.
Define the real function $f(\sigma):=\tr(A\sigma)$ on $\sep$.
Since
$\sp(\{\ket{\lambda_0},\ket{\lambda_1},\ldots,\ket{\lambda_k}\})$
contains no separable states and $\lambda_{k+1}>\lambda_k$, the
function satisfies $f(\sigma)>\lambda_0$. Since the set of
separable states is compact, there exists a separable state
$\sigma'$ that minimises $f(\sigma)$. Thus, setting
$a:=f(\sigma')$ gives $\sep\cap H^\circ_{A,a}=\varnothing$. As
well, $\ent\cap H^\circ_{A,a}\neq\varnothing$ since
$\tr(A\ketbra{\lambda_0}{\lambda_0})=\lambda_0<a$, and so $A$ is a
left witness.
\end{proof}

\begin{theorem}\label{Theorem_CharShift}
The operator
 $A$ is a left or right entanglement witness if and only if (i) there
 exists $k\in [0,1,\ldots,MN-2]$ such that
 $\sp\{\ket{\lambda_0},\ket{\lambda_1},\ldots,\ket{\lambda_k}\}$
 contains no separable pure states and $\lambda_{k+1}>\lambda_k$, or (ii) there
 exists $l\in [1,2,\ldots,MN-1]$ such that\\
 $\sp\{\ket{\lambda_l},\ket{\lambda_{l+1}},\ldots,\ket{\lambda_{MN-1}}\}$
 contains no separable pure states and $\lambda_{l}>\lambda_{l-1}$.
\end{theorem}

Theorem \ref{Theorem_CharShift} immediately gives a method for
identifying and constructing entanglement witnesses.
\begin{definition}[Partial Product Basis, Unextendible Product Basis \cite{Ter01}] A \emph{partial product
basis} of $\mathbb{C}^\dima\otimes\mathbb{C}^\dimb$ is a set $S$
of mutually orthonormal pure product states spanning a proper
subspace of $\mathbb{C}^\dima\otimes\mathbb{C}^\dimb$.  An
\emph{unextendible product basis} of
$\mathbb{C}^\dima\otimes\mathbb{C}^\dimb$ is a partial product
basis $S$ of $\mathbb{C}^\dima\otimes\mathbb{C}^\dimb$ whose
complementary subspace $(\sp S)^\perp$ contains no product state.
\end{definition}
\noindent We can use unextendible product bases to construct
ambidextrous witnesses.  Suppose $B$ is an unextendible product
basis of $\mathbb{C}^\dima\otimes\mathbb{C}^\dimb$, and let $B'$
be disjoint from $B$ such that $B\cup B'$ is an orthonormal basis
of $\mathbb{C}^\dima\otimes\mathbb{C}^\dimb$. One possibility is
the left witness defined by $A'$ as
\begin{eqnarray}
A' = -\sum_{\ket{\lambda}\in B'}\ketbra{\lambda}{\lambda}
\end{eqnarray}
As well, we could split $B'$ into $B'_L$ and $B'_R$ and define an
ambidextrous witness $A''$ as
\begin{eqnarray}
A'' = -\sum_{\ket{\lambda_L}\in B'_L}\ketbra{\lambda_L}{\lambda_L}
+\sum_{\ket{\lambda_R}\in B'_R}\ketbra{\lambda_R}{\lambda_R}.
\end{eqnarray}
\noindent Another thing to realise is that $\sp B$ may contain an
entangled pure state, which can be pulled out and put into a
$(+1)$-eigenvalue eigenspace of $A'$. Depending on $B$ (and the
dimensions $\dima$, $\dimb$), there may be several mutually
orthogonal pure entangled states in $\sp B$ whose span contains no
product state; let $B''$ be a set of such pure states.  Define the
ambidextrous witness as
\begin{eqnarray}
A''' = -\sum_{\ket{\lambda}\in B'}\ketbra{\lambda}{\lambda} +
\sum_{\ket{\lambda}\in B''}\ketbra{\lambda}{\lambda}.
\end{eqnarray}
\noindent This suggests the following problem, related to the
combinatorial \cite{AL01} problem of finding unextendible product
bases:
\begin{problem} Given $\dima$ and $\dimb$, find all orthonormal
bases $B$ for $\mathbb{C}^\dima\otimes\mathbb{C}^\dimb$ such that
\begin{itemize}
\item $B$ is the disjoint union of $\Lambda_L$, $\tilde{B}$,
$\Lambda_R$, \item $\sp\Lambda_L$ and $\sp\Lambda_R$ contain no
product state, \item $\sp( \Lambda_L\cup\Lambda_R)$ contains a
product state, and \item $\min\{|\Lambda_L|,|\Lambda_R|\}$ is
maximal.
\end{itemize}
\end{problem}
\noindent Such bases may give ``optimal'' ambidextrous witnesses,
which detect the largest volume of entangled states on opposite
sides of $\sep$.

We will see in Chapter \ref{ChapterReductionToEWSearch} that the
functions $a^*(A)$ and $b^*(A)$ are difficult (NP-hard) to
compute.  Thus a criticism of constructing witnesses via the
spectral decomposition is that even if you can construct the
corresponding observable, you still have to perform a difficult
computation to make them useful. However, most experimental
applications of entanglement witnesses are in very low dimensions,
where computing $a^*(A)$ and $b^*(A)$ deterministically is not a
problem -- it may even be done analytically, as in the example
below.

\subsubsection{Example: Noisy Bell states}\label{sec_NoisyBell}

A simple illustration of how AEWs may be used involves detecting
and distinguishing noisy Bell states. Define the four Bell states
in $\mathbb{C}^2\otimes\mathbb{C}^2$:
\begin{eqnarray}\nonumber
\ket{\psi^\pm}&:=&\left(\ket{00}\pm\ket{11}\right)/\sqrt{2}\\\nonumber
\ket{\phi^\pm}&:=&\left(\ket{01}\pm\ket{10}\right)/\sqrt{2}.
\end{eqnarray}
It is straightforward to show that the Bell states are, pairwise,
on opposite sides of $\mathcal{S}_{2,2}$.\footnote{Suppose a left
entanglement witness $W$, with $a^*(W)=0$, detects $\ket{\psi^+}$
and $\ket{\phi^+}$.  Without loss of generality, $W$ can be
written in the Bell basis $\{  \ket{\psi^+}, \ket{\phi^+}, \ldots
\}$ as
\begin{eqnarray}W=\begin{bmatrix}-\epsilon_1 & a+bi & \times & \times \\
a-bi & -\epsilon_2 & \times & \times
\\ \times & \times & \times & \times \\ \times & \times & \times & \times \\ \end{bmatrix},\end{eqnarray}
for $\epsilon_1$ and $\epsilon_2$ both positive.  But the states
$\ket{s^\pm}\equiv\frac{1}{\sqrt{2}}(\ket{\psi^+}\pm\ket{\phi^+})$
are separable.  Requiring $\bra{s^+}W\ket{s^+}\geq 0$ gives
$2a\geq\epsilon_1+\epsilon_2$ and requiring
$\bra{s^-}W\ket{s^-}\geq 0$ gives $2a\leq -\epsilon_1-\epsilon_2$,
which, together, give a contradiction.  Similar arguments hold for
the other pairs of Bell states.} Define the operators
\begin{eqnarray}\nonumber
A_\psi&:=& -\ketbra{\psi^-}{\psi^-} +
\ketbra{\psi^+}{\psi^+}\\\nonumber A_\phi&:=&
-\ketbra{\phi^-}{\phi^-} + \ketbra{\phi^+}{\phi^+}.
\end{eqnarray}
Both $A_\psi$ and $A_\phi$ are easily seen to be AEWs.  It is also
straightforward to compute the values
\begin{eqnarray}\nonumber
a^*(A_\psi)=a^*(A_\phi)= -1/2
\end{eqnarray}
and
\begin{eqnarray}\nonumber
 b^*(A_\psi)=b^*(A_\phi)=
+1/2.
\end{eqnarray}
Suppose that there is a source that repeatedly emits the same
noisy Bell state $\rho$ and that we want to decide whether $\rho$
is entangled.  Define the Pauli operators:
\begin{equation}\label{eqn_PauliOperators}
\begin{array}{ccrccrl}
\nonumber \sigma_0 &:=&   \frac{1}{\sqrt{2}} (\ketbra{0}{0} +
\ketbra{1}{1})&\\
\nonumber \sigma_1 &:=&   \frac{1}{\sqrt{2}}(\ketbra{0}{1}+ \ketbra{1}{0})&\\
\nonumber \sigma_2 &:=&- \frac{i}{\sqrt{2}}(\ketbra{0}{1}- \ketbra{1}{0})&\\
\nonumber \sigma_3 &:=&   \frac{1}{\sqrt{2}}(\ketbra{0}{0} -
\ketbra{1}{1})&,
\end{array}
\end{equation}
where $\{\ket{0},\ket{1}\}$ is the standard orthonormal basis for
$\mathbb{C}^2$.  Noting that
\begin{eqnarray}\nonumber
A_\psi&=& \sigma_1\otimes\sigma_1 -
\sigma_2\otimes\sigma_2\\
\nonumber A_\phi&=&\sigma_1\otimes\sigma_1 +
\sigma_2\otimes\sigma_2,
\end{eqnarray}
measuring the expected value of the two observables
$\sigma_1\otimes\sigma_1$ and $\sigma_2\otimes\sigma_2$ may be
sufficient to decide that $\rho$ is entangled because $\rho\in
\mathcal{E}_{2,2}$ if one of the following four inequalities is
true:
\begin{eqnarray}\label{ineq_SuffEnt}
\langle \sigma_1\otimes\sigma_1 \rangle_\rho\pm
\langle\sigma_2\otimes\sigma_2\rangle_\rho&>&1/2 \\\nonumber
\langle \sigma_1\otimes\sigma_1 \rangle_\rho\pm
\langle\sigma_2\otimes\sigma_2\rangle_\rho&<&-1/2 .
\end{eqnarray}
If the noise is known to be of a particular form, then we can also
determine \emph{which} noisy Bell state was being produced. Let
$\ket{B}$ be a Bell state. Suppose $\rho$ is known to be of the
form $p\ketbra{B}{B}+(1-p)\sigma$ for some $\sigma$ inside both
sandwiches $W(A_\psi)$ and $W(A_\phi)$. With $\sigma$ so defined,
one of the four inequalities (\ref{ineq_SuffEnt}) holds only if
exactly one of them holds, so that $\ket{B}$ is determined by
which inequality is satisfied. We remark that, if $\sigma$ and
$\ket{B}$ are known, knowledge of the expected value of any single
observable $A$ may allow one to compute $p$ and hence an upper
bound on the $l_2$ distance between $\rho$ and the maximally mixed
state $I/4$. This distance may be enough information to conclude
that $\rho$ is separable by checking if $\rho$ is inside the
largest separable ball centered at $I/4$ \cite{GB02}.

\section{Convex body problems}\label{sec_ConvexBodyProblems}

I end this chapter with a brief review of some basic problems for
a convex subset $K$ of $\mathbb{R}^n$ and their meaning in terms
of the separability problem when $K=\sep$. In Chapter
\ref{ChapterReductionToEWSearch}, the relationship among these
problems will be exploited to solve the quantum separability
problem.

We have already noted that $\sep$ may be viewed as a subset of
$\mathbb{R}^{\n-1}$.  Let us be more precise.  Let
$\mathcal{B}=\{X_i:i=0,1,\ldots,\n-1\}$ be an orthonormal,
Hermitian basis for $\mathbb{H}_{M,N}$, where
$X_0\equiv\frac{1}{\sqrt{MN}}I$. For concreteness, we can assume
that the elements of $\mathcal{B}$ are tensor-products of the
(suitably normalised) canonical generators of SU(M) and SU(N),
given e.g. in \cite{TNWM02}. Note $\tr(X_i)=0$ for all $i>0$.
Define $v:\hermops\rightarrow \mathbb{R}^{\dima^2\dimb^2-1}$ as
\begin{eqnarray}\label{eqn_MappingFromHermopsToRealVecs}
v(A):=\begin{bmatrix} \tr(X_1 A) \\ \tr(X_2 A) \\
\vdots \\ \tr(X_{\dima^2\dimb^2-1} A)\end{bmatrix}.
\end{eqnarray}
Via the mapping $v$, the set of separable states $\sep$ can be
viewed as a full-dimensional convex subset of
$\mathbb{R}^{\dima^2\dimb^2-1}$
\begin{eqnarray}
\{v(\sigma)\in \mathbb{R}^{\dima^2\dimb^2-1}: \sigma\in\sep\},
\end{eqnarray}
which properly contains the origin
$v(I_{\dima,\dimb})=\overline{0}\in\mathbb{R}^{\dima^2\dimb^2-1}$
(recall that there is a ball of separable states of nonzero radius
centred at the maximally mixed state $I_{\dima,\dimb}$). For
traceless $A_1,A_2\in\hermops$, we clearly have $\tr(A_1A_2)\equiv
v(A_1)^Tv(A_2)$.  For $A\in\hermops$ and $\rho\in\densops$, where
$A:=\sum_{i=0}^{\n-1}\alpha_iX_i$ and
$\rho:=\sum_{i=0}^{\n-1}\rho_iX_i$, we have
$\tr(A\rho)=\alpha_0\rho_0 + v(A)^Tv(\rho)$.  But $\rho_0$ is
fixed at $1/\sqrt{\dima\dimb}$ for all $\rho\in\densops$.  Thus,
in terms of entanglement witnesses $A$, we might as well restrict
to those $A$ that have $\alpha_0=0$; that is, we may restrict to
traceless entanglement witnesses without loss of generality.  In
the definitions below, the vector $c$ corresponds to a traceless
right entanglement witness when $K=\sep$.

The following definitions can be found in \cite{GLS88}.

\begin{definition}
[Strong Membership Problem (SMEM)]\label{def_SMEM} Given a point
$p\in\mathbb{R}^n$, decide whether $p\in K$.
\end{definition}

\begin{definition}
[Strong Separation Problem (SSEP)]\label{def_SSEP} Given a point
$p\in\mathbb{R}^n$, either assert that $p\in K$, or find a vector
$c\in\mathbb{R}^n$ such that $c^Tp>\max\{c^Tx|x\in K\}$.
\end{definition}
\noindent For $K=\sep$, SMEM corresponds exactly to the formal
quantum separability problem in Definition \ref{def_FormalQuSep}.
SSEP also solves SMEM, but, in the case where $p$ represents an
entangled state, also provides a right entanglement witness (note
how the unconventional definition of ``entanglement witness'' fits
nicely here).

\begin{definition}
[Strong Optimisation Problem (SOPT)]\label{def_SOPT} Given a
vector $c\in\mathbb{R}^n$, either find a point $k\in K$ that
maximises $c^Tx$ on $K$, or assert that $K$ is empty.
\end{definition}
\noindent SOPT corresponds to the problem of calculating $b^*(A)$
for a potential right entanglement witness $A$.  The optimisation
problem over $\sep$ will continue to play a major role throughout
this thesis.

\begin{definition}
[Strong Validity Problem (SVAL)]\label{def_SVAL} Given a vector
$c\in\mathbb{R}^n$ and a number $\gamma\in\mathbb{R}$, decide
whether $c^Tx\leq\gamma$ holds for all $x\in K$.
\end{definition} \noindent For $K=\sep$, SVAL asks,
``Given a potential right entanglement witness $A$ and a number
$b$, is $b^*(A)\leq b$?''

Let $K'$ be a convex subset of $\mathbb{R}^n$.

\begin{definition}
[Strong Violation Problem (SVIOL)]\label{SVIOL} Given a vector
$d\in\mathbb{R}^n$ and a number $\gamma\in\mathbb{R}$, decide
whether $d^Tx\leq\gamma$ holds for all $x\in K'$, and, if not,
find a vector $y\in K'$ with $d^Ty>\gamma$.
\end{definition}
\noindent Note that taking $d=0$ and $\gamma=-1$, the strong
violation problem reduces to the problem of checking whether $K'$
is empty, and if not, finding a point in $K'$.  This problem is
called the \textbf{Feasibility Problem} and will arise in Chapters
\ref{ChapterReductionToEWSearch} and \ref{ChapterAlgorithm} (but
\emph{not} for $K'$ equal to $\sep$, which is why I switched
notation from ``$K$'' to ``$K'$'' to define this problem).

\chapter{Separability as a Computable Decision Problem}\label{ChapterSepAsDecisionProblem}

Definition \ref{def_FormalQuSep} gave us a concrete definition of
the quantum separability problem that we could use to explore some
important results.  Now we step back from that definition and
consider more carefully how we might define the quantum
separability problem for the purposes of computing it.

For a number of reasons, we settle on approximate formulations of
the problem and give a few examples that are, in a sense,
equivalent. I then formulate the quantum separability problem as
an NP-hard problem in NP. I end the chapter with a survey of
algorithms for the approximate quantum separability problem; one
of the algorithms comes directly from a second NP-formulation and
can be considered as the weakening of a recent algorithm by Hulpke
and Bru{\ss} \cite{qphHB04}.

\section{Formulating the quantum separability problem}

The nature of the quantum separability problem and the possibility
for quantum computers allows a number of approaches, depending on
whether the input to the problem is classical (a matrix
representing $\rho$) or quantum ($T$ copies of a physical system
prepared in state $\rho$) and whether the processing of the input
will be done on a classical computer or on a quantum computer.  In
Chapter \ref{ChapterConvexity}, we dealt with the case of a
quantum input and very limited quantum processing in the form of
measurement of each copy of $\rho$; we will deal with this case in
more detail in Chapter \ref{ChapterReductionToEWSearch}.  The case
of more-sophisticated quantum processing on either a quantum or
classical input is not well studied (see \cite{HE02} for an
instance of more-sophisticated quantum processing on a quantum
input). For the remainder of this chapter, I focus on the case
where input and processing are classical.

\subsection{Exact formulations}

Let us examine Definition \ref{def_FormalQuSep} (or, equivalently,
Definition \ref{def_SMEM}) from a computational viewpoint.  The
matrix $[\rho]$ is allowed to have real entries.  Certainly there
are real numbers that are uncomputable (e.g. a number whose $n$th
binary digit is 1 if and only if the $n$th Turing machine halts on
input $n$); we disallow such inputs.  However, the real numbers
$e$, $\pi$, and $\sqrt{2}$ are computable to any degree of
approximation, so in principle they should be allowed to appear in
$[\rho]$.  In general, we should allow any real number that can be
approximated arbitrarily well by a computer subroutine. If
$[\rho]$ consists of such real numbers (subroutines), say that
``$\rho$ is given as an approximation algorithm for $[\rho]$.'' In
this case, we have a procedure to which we can give an accuracy
parameter $\delta>0$ and out of which will be returned a matrix
$[\rho]_\delta$ that is (in some norm) at most $\delta$ away from
$[\rho]$.  Because $\sep$ is closed, the sequence
$([\rho]_{1/n})_{n=1,2,\ldots}$ may converge to a point on the
boundary of $\sep$ (when $\rho$ is on the boundary of $\sep$). For
such $\rho$, the formal quantum separability problem may be
``undecidable'' because the $\delta$-radius ball centred at
$[\rho]_\delta$ may contain both separable and entangled states
for all $\delta>0$ \cite{Myr97} (more generally, see ``Type II
computability'' in \cite{Wei87}).

If we really want to determine the complexity of deciding
membership in $\sep$, it makes sense not to confuse this with the
complexity of specifying the input.  To give the computer a
fighting chance, it makes more sense to restrict to inputs that
have finite exact representations that can be readily subjected to
elementary arithmetic operations begetting exact answers. For this
reason, we might restrict the formal quantum separability problem
to instances where $[\rho]$ consists of rational entries:

\begin{definition}[Rational quantum separability problem (EXACT QSEP)]\label{def_RationalQuSep}
Let $\rho\in\densops$ be a mixed state such that the matrix
$[\rho]$ (with respect to the standard basis of
$\mathbb{C}^\dima\otimes\mathbb{C}^\dimb$) representing $\rho$
consists of rational entries.  Given $[\rho]$, is $\rho$
separable?
\end{definition}

As pointed out in \cite{DPS04}, Tarski's
algorithm\footnote{Tarski's result is often called the
``Tarski-Seidenberg'' theorem, after Seidenberg, who found a
slightly better algorithm \cite{Sei54} (and elaborated on its
generality) in 1954, shortly after Tarski managed to publish his;
but Tarski discovered his own result in 1930 (the war prevented
him from publishing before 1948).} \cite{Tar51} can be used to
solve EXACT QSEP exactly.  The Tarski-approach is as follows. Note
that the following first-order logical formula\footnote{Recall the
logical connectives: $\vee$ (``OR''), $\wedge$ (``AND''), $\neg$
(``NOT''); the symbol $\rightarrow$ (``IMPLIES''), in
``$x\rightarrow y$'', is a shorthand, as ``$x\rightarrow y$'' is
equivalent to ``$(\neg x) \vee y $''; as well, we can consider
``$x\vee y$'' shorthand for ``$\neg((\neg x)\wedge (\neg y))$''.
Also recall the existential and universal quantifiers $\exists$
(``THERE EXISTS'') and $\forall$ (``FOR ALL''); note that the
universal quantifier $\forall$ is redundant as ``$\forall x
\phi(x)$'' is equivalent to ``$\neg\exists x\neg\phi(x)$''.} is
true if and only if $\rho$ is separable:
\begin{eqnarray}\label{eqn_FOLogicalStatementOfSep}
\forall A[(\forall \Psi (\tr(A\Psi)\geq 0))\rightarrow (\tr
A\rho\geq 0)],
\end{eqnarray}
where $A\in\hermops$ and $\Psi$ is a pure product state. To see
this, note that the subformula enclosed in square brackets means
``$A$ is not a (left) entanglement witness for $\rho$'', so that
if this statement is true for all $A$ then there exists no
entanglement witness detecting $\rho$. When $[\rho]$ is rational,
our experience in Section \ref{subsec_EisertsEtalApproach} with
polynomial constraints tells us that the formula in
(\ref{eqn_FOLogicalStatementOfSep}) can be written in terms of
``quantified polynomial inequalities'' with rational coefficients:
\begin{eqnarray}\label{eqn_FOLS}
\forall X   \lbrace (\forall Y \left[ Q(Y)\rightarrow (r(X,Y)\geq
0)\right])\rightarrow (s(X)\geq 0) \rbrace,
\end{eqnarray}
where
\begin{itemize}
\item $X$ is a block of real variables parametrising the matrix
$A\in\hermops$ (with respect to an orthogonal rational Hermitian
basis of $\hermops$); the ``Hermiticity'' of $X$ is hard-wired by
the parametrisation;

\item $Y$ is a block of real variables parametrising the matrix
$\Psi$;

\item $Q(Y)$ is a conjunction of four polynomial equations that
are equivalent to the four constraints $\tr((\tr_{j}(\Psi))^2)=1$
and $\tr((\tr_{j}(\Psi))^3)=1$ for $j\in\{\A,\B\}$;

\item $r(X,Y)$ is a polynomial representing the expression
$\tr(A\Psi)$;\footnote{To ensure the Hermitian basis is rational,
we do not insist that each of its elements has unit Euclidean
norm. If the basis is $\{X_i\}_{i=0,1,\ldots,\dima^2\dimb^2}$,
where $X_0$ is proportional to the identity operator, then we can
ignore the $X_0$ components write
$A=\sum_{i=1}^{\dima^2\dimb^2}A_iX_i$ and
$\Psi=\sum_{i=1}^{\dima^2\dimb^2}\Psi_iX_i$.  An expression for
$\tr(A\Psi)$ in terms of the real variables $A_i$ and $\Psi_i$ may
then look like $\sum_{i=1}^{\dima^2\dimb^2}A_i\Psi_i\tr(X_i^2)$.}

\item $s(X)$ is a polynomial representing the expression
$\tr(A[\rho])$.
\end{itemize}
The main point of Tarski's result is that the quantifiers (and
variables) in the above sentence can be eliminated so that what is
left is just a formula of elementary algebra involving Boolean
connections of atomic formula of the form $(\alpha \diamond 0)$
involving terms $\alpha$ consisting of rational numbers, where
$\diamond$ stands for any of $<, >, =, \neq$; the truth of the
remaining (very long) formula can be computed in a straightforward
manner.  The best algorithms for deciding (\ref{eqn_FOLS}) require
a number of arithmetic operations roughly equal to
$(PD)^{O(|X|)\times O(|Y|)}$, where $P$ is the number of
polynomials in the input, $D$ is the maximum degree of the
polynomials, and $|X|$ ($|Y|$) denotes the number of variables in
block $X$ ($Y$) \cite{BPR96}\footnote{Ironically, due to some
computer font incompatibility, my copy of this paper, entitled
``On the computational and algebraic complexity of quantifier
elimination,'' did not display any of the quantifiers.}. Since
$P=6$ and $D=3$, the running time is roughly
$18^{O(\dima^2\dimb^2)\times O(\dima^2\dimb^2)}$ (times the length
of the encoding of the rational inputs).

\subsection{Approximate formulations}

The benefit of EXACT QSEP is that, compared to Definition
\ref{def_FormalQuSep}, it eliminated any uncertainty in the input
by disallowing irrational matrix entries. Consider the following
motivation for an alternative to EXACT QSEP, where, roughly, we
only ask whether the input $[\rho]$ corresponds to something
\emph{close to} separable:
\begin{itemize}
\item Suppose we really want to determine the separability of a
density operator $\rho$ such that $[\rho]$ has irrational entries.
If we use the EXACT QSEP formulation (so far, we have no decidable
alternative), we must first find a rational approximation to
$[\rho]$.  Suppose the (Euclidean) distance from $[\rho]$ to the
approximation is $\delta$.  The answer that the Tarski-style
algorithm gives us might be wrong, if $\rho$ is not more than
$\delta$ away from the boundary of $\sep$.

\item Suppose the input matrix came from measurements of many
copies of a physical state $\rho$.  Then we only know $[\rho]$ to
some degree of approximation.

\item The best known Tarski-style algorithms for EXACT QSEP
have gigantic running times. 
 Surely, we can achieve better asymptotic running
times if use an approximate formulation.
\end{itemize}
\noindent Thus, in many cases of interest, insisting that an
algorithm says exactly whether the input matrix corresponds to a
separable state is a waste of time. In Section
\ref{sec_DefnInNPProper}, we will see that there is another reason
to use an approximate formulation, if we would like the problem to
fit nicely in the theory of NP-completeness.

Gurvits was the first to use the weak membership formulation of
the quantum separability problem \cite{GLS88, Gur03}.  For
$x\in\mathbb{R}^n$ and $\delta>0$, let $B(x,\delta):=
\{y\in\mathbb{R}^n: ||x-y||\leq\delta\}$.  For a convex subset
$K\subset\mathbb{R}^n$, let $S(K,\delta):=\cup_{x\in K}
B(x,\delta)$ and $S(K,-\delta):=\{x: B(x,\delta)\subseteq K\}$.
\begin{definition}[Weak membership problem (WMEM)]\label{def_WMEM}
Given a rational vector $p\in\mathbb{R}^n$ and rational
$\delta>0$, assert either that
\begin{eqnarray}
p&\in& S(K,\delta), \hspace{2mm}\text{or}\label{eqn_WMEMSepAssertion}\\
p&\notin& S(K,-\delta)\label{eqn_WMEMEntAssertion}.
\end{eqnarray}
\end{definition}
\noindent Denote by WMEM($\sep$) the quantum separability problem
formulated as the weak membership problem.  An algorithm solving
WMEM($\sep$) is a separability test with two-sided
``error''\footnote{Of course, relative to the problem definition,
there is no error.} in the sense that it may assert
(\ref{eqn_WMEMSepAssertion}) when $p$ represents an entangled
state and may assert (\ref{eqn_WMEMEntAssertion}) when $p$
represents a separable state.  Any formulation of the quantum
separability problem will have (at least) two possible answers --
one corresponding to ``$p$ approximately represents a separable
state'' and the other corresponding to ``$p$ approximately
represents an entangled state''. Like in WMEM($\sep$), there may
be a region of $p$ where both answers are valid. We can use a
different formulation where this region is shifted to be either
completely outside $\sep$ or completely inside $\sep$:
\begin{definition}[In-biased
weak membership problem (WMEM$_\text{In}$)]\label{def_WMEMS} Given
a rational vector $p\in\mathbb{R}^n$ and rational $\delta>0$,
assert either that
\begin{eqnarray}
p&\in& S(K,\delta), \hspace{2mm}\text{or}\label{eqn_WMEMSSepAssertion}\\
p&\notin& K\label{eqn_WMEMSEntAssertion}.
\end{eqnarray}
\end{definition}
\begin{definition}[Out-biased weak membership problem (WMEM$_\text{Out}$)]\label{def_WMEME}
Given a rational vector $p\in\mathbb{R}^n$ and rational
$\delta>0$, assert either that
\begin{eqnarray}
p&\in& K, \hspace{2mm}\text{or}\label{eqn_WMEMESepAssertion}\\
p&\notin& S(K,-\delta)\label{eqn_WMEMEEntAssertion}.
\end{eqnarray}
\end{definition}
\noindent We can also formulate a ``zero-error'' version such that
when $p$ is in such a region, then any algorithm for the problem
has the option of saying so, but otherwise must answer exactly:
\begin{definition}[Zero-error weak membership problem (WMEM$^0$)]\label{def_WMEM0}
Given a rational vector $p\in\mathbb{R}^n$ and rational
$\delta>0$, assert either that
\begin{eqnarray}
p&\in& K, \hspace{2mm}\text{or}\label{eqn_WMEM0SepAssertion}\\
p&\notin& K, \hspace{2mm}\text{or}\label{eqn_WMEM0EntAssertion}\\
p&\in& S(K,\delta)\setminus
S(K,-\delta)\label{eqn_WMEM0BoundaryAssertion}
\end{eqnarray}
\end{definition}

All the above formulations of the quantum separability problem are
based on the Euclidean norm and use the isomorphism between
$\hermops$ and $\mathbb{R}^{\n}$. We could also make similar
formulations based on other operator norms in $\hermops$.  In the
next section, we will see yet another formulation of an entirely
different flavour.  While each formulation is slightly different,
they all have the property that in the limit as the error
parameter approaches 0, the problem coincides with EXACT QSEP.
Thus, despite the apparent inequivalence of these formulations, we
recognise that they all basically do the same job. In fact,
WMEM$(\sep)$, WMEM$_\text{In}(\sep)$, WMEM$_\text{Out}(\sep)$, and
WMEM$(\sep)^0$ are equivalent: given an algorithm for one of the
problems, one can solve an instance $(\rho,\delta)$ of any of the
other three problems by just calling the given algorithm at most
twice (with various parameters).\footnote{To show this
equivalence, it suffices to show that given an algorithm for
WMEM$(\sep)$, one can solve WMEM$_\text{Out}(\sep)$ with one call
to the given algorithm (the converse is trivial); a similar proof
shows that one can solve WMEM$_\text{In}(\sep)$ with one call to
the algorithm for WMEM$(\sep)$. The other relationships follow
immediately.  Let $(\rho,\delta)$ be the given instance of
WMEM$_\text{Out}(\sep)$. Define
$\rho_0:=\rho+\delta(\rho-I_{\dima,\dimb})/2$ and
$\delta_0:=\delta/(2\sqrt{\dima\dimb(\dima\dimb-1)})$. Call the
algorithm for WMEM$(\sep)$ with input $(\rho_0,\delta_0)$. Suppose
the algorithm asserts $\rho_0\notin S(\sep,-\delta_0)$. Then,
because $||\rho-\rho_0||=\frac{\delta}{2}||\rho-I_{\dima,\dimb}||$
and $||\rho-I_{\dima,\dimb}||\leq 1$, we have $\rho\notin
S(\sep,-(\delta_0+\delta/2))$ hence $\rho\notin S(\sep,-\delta)$.
Otherwise, suppose the algorithm asserts $\rho_0\in
S(\sep,\delta_0)$. By way of contradiction, assume that $\rho$ is
entangled.  But then, by convexity of $\sep$ and the fact that
$\sep$ contains the ball
$B(I_{\dima,\dimb},{1}/{\sqrt{\dima\dimb(\dima\dimb-1)}})$, we can
derive that the ball $B(\rho_0,\delta_0)$ does not intersect
$\sep$.  But this implies $\rho_0\notin S(\sep,\delta_0)$ -- a
contradiction.  Thus, $\rho\in\sep$. This proof is a slight
modification of the argument given in \cite{Lut05}.}

\section{Computational complexity}

This section addresses how the quantum separability problem fits
into the framework of complexity theory. I assume the reader is
familiar with concepts such as \emph{problem}, \emph{instance} (of
a problem), \emph{(reasonable, binary) encodings},
\emph{polynomially relatedness}, \emph{size} (of an instance),
\emph{(deterministic and nondeterministic) Turing machine}, and
\emph{polynomial-time algorithm}; all of which can be found in any
of \cite{GJ79, Pap94, NC00}.

Generally, the weak membership problem is defined for a class
$\mathcal{K}$ of convex sets.  For example, in the case of
WMEM($\sep$), this class is $\{\sep\}_{M,N}$ for all integers
$\dima$ and $\dimb$ such that $2\leq\dima\leq\dimb$.  An instance
of WMEM thus includes the specification of a member $K$ of
$\mathcal{K}$.  The size of an instance must take into account the
size $\langle K \rangle$ of the encoding of $K$.  It is reasonable
that $\langle K \rangle\geq n$ when $K\in\mathbb{R}^n$, because an
algorithm for the problem should be able to work
efficiently\footnote{Recall that ``efficiently'' means ``in time
that is upper-bounded by a polynomial in the size of an instance''
(the same polynomial for all instances).} with points in
$\mathbb{R}^n$. But the complexity of $K$ matters, too. For
example, if $K$ extends (doubly-exponentially) far from the origin
(but contains the origin) then $K$ may contain points that require
large amounts of precision to represent; again, an algorithm for
the problem should be able to work with such points efficiently
(for example, it should be able to add such a point and a point
close to the origin, and store the result efficiently).  In the
case of WMEM($\sep$), the size of the encoding of $\sep$ may be
taken as $\dimb$ (assuming $\dima\leq\dimb$), as $\sep$ is not
unreasonably long or unreasonably thin: it is contained in the
unit sphere in $\mathbb{R}^{\n-1}$ and contains a ball of
separable states of radius $\Omega(1/\poly(\dimb))$ (see Section
\ref{sec_OneSidedTestsAndRestrictions}).  Thus, the total size of
an instance of WMEM($\sep$), or any formulation of the quantum
separability problem, may also be taken to be $\dimb$ plus the
size of the encoding of $(\rho, \delta)$.

\subsection{Review of NP-completeness}\label{sec_ReviewNPCness}

Complexity theory, and, particularly, the theory of
NP-completeness, pertains to \emph{decision problems} -- problems
that pose a yes/no question.  Let $\Pi$ be a decision problem.
Denote by $\D_\Pi$ the set of instances of $\Pi$, and denote the
yes-instances of $\Pi$ by $\Y_\Pi$. Recall that the complexity
class P (respectively, NP) is the set of all problems the can be
decided by a deterministic Turing machine (respectively,
nondeterministic Turing machine) in polynomial time. The following
equivalent definition of NP is perhaps more intuitive:
\begin{definition}[NP] A decision problem $\Pi$ is in NP if there exists
a deterministic Turing machine $T_\Pi$ such that for every
instance $I\in \Y_\Pi$ there exists a string $C_I$ of length
$|C_I|\in O(\poly(|I|))$ such that $T_\Pi$, with input $C_I$, can
check that $I$ is in $\Y_\Pi$ in time $O(\poly(|I|))$.
\end{definition}
\noindent  The string $C_I$ is called a \emph{(succinct)
certificate}.  Let $\Pi^c$ be the complementary problem of $\Pi$,
i.e. $\D_{\Pi^c}\equiv \D_{\Pi}$ and $\Y_{\Pi^c}:=\D_\Pi \setminus
\Y_\Pi$. The class co-NP is thus defined as $\{\Pi^c:
\hspace{2mm}\Pi\in\NP\}$.

Let us briefly review the different notions of ``polynomial-time
reduction'' from one problem $\Pi'$ to another $\Pi$. Let
$\mathcal{O}_\Pi$ be an oracle, or black-boxed subroutine, for
solving $\Pi$, to which we assign unit complexity cost. A
\emph{(polynomial-time) Turing reduction} from $\Pi'$ to $\Pi$ is
any polynomial-time algorithm for $\Pi'$ that makes calls to
$\mathcal{O}_\Pi$.  Write $\Pi'\leq_\T\Pi$ if $\Pi'$ is
Turing-reducible to $\Pi$. A \emph{polynomial-time
transformation}, or \emph{Karp reduction}, from $\Pi'$ to $\Pi$ is
a Turing reduction from $\Pi'$ to $\Pi$ in which $\mathcal{O}_\Pi$
is called at most once and at the end of the reduction algorithm,
so that the answer given by $\mathcal{O}_\Pi$ is the answer to the
given instance of $\Pi'$.\footnote{In other words, a Karp
reduction from $\Pi'$ to $\Pi$ is a polynomial-time algorithm that
(under a reasonable encoding) takes as input an (encoding of an)
instance $I'$ of $\Pi'$ and outputs an (encoding of an) instance
$I$ of $\Pi$ such that $I'\in \Y_{\Pi'}\Leftrightarrow I\in
\Y_{\Pi}$.} Write $\Pi'\leq_\K\Pi$ if $\Pi'$ is Karp-reducible to
$\Pi$.  Karp and Turing reductions are on the extreme ends of a
spectrum of polynomial-time reductions; see \cite{LLS75} for a
comparison of several of them.

Reductions between problems are a way of determining how hard one
problem is relative to another.  The notion of NP-completeness is
meant to define the hardest problems in NP\@.  We can define
NP-completeness with respect to any polynomial-time reduction; we
define \emph{Karp-NP-completeness} and
\emph{Turing-NP-completeness}:
\begin{eqnarray}
\NPCK &:=& \{\Pi\in\NP:\hspace{2mm} \Pi'\leq_\K\Pi \text{ for all
$\Pi'\in\NP$ }\}\\
\NPCT &:=& \{\Pi\in\NP:\hspace{2mm} \Pi'\leq_\T\Pi \text{ for all
$\Pi'\in\NP$ }\}.
\end{eqnarray}
We have $\NPCK\subseteq\NPCT$.  Let $\Pi$, $\Pi'$, and $\Pi''$ be
problems in NP, and, furthermore, suppose $\Pi'$ is in $\NPCK$. If
$\Pi'\leq_\T \Pi$, then, in a sense, $\Pi$ is at least as hard as
$\Pi'$ (which gives an interpretation of the symbol
``$\leq_\T$''). Suppose $\Pi'\leq_\T \Pi$ but suppose also that
$\Pi'\nleq_\K \Pi$. If $\Pi'\leq_\K \Pi''$, then we can say that
``$\Pi''$ is at least as hard as $\Pi$'', because, to solve $\Pi'$
(and thus any other problem in NP), $\mathcal{O}_\Pi$ has to be
used at least as many times as $\mathcal{O}_{\Pi''}$; if any
Turing reduction proving $\Pi'\leq_\T \Pi$ requires more than one
call to $\mathcal{O}_\Pi$, then we can say ``$\Pi''$ is harder
than $\Pi$''. Therefore, if $\NPCK\neq\NPCT$, then the problems in
$\NPCK$ are harder than the problems in $\NPCT\setminus\NPCK$;
thus $\NPCK$ are the hardest problems in NP (with respect to
polynomial-time reductions).

A problem $\Pi$ is \emph{NP-hard} when $\Pi'\leq_T\Pi$ for some
Karp-NP-complete problem $\Pi'\in\NPCK$.  The term ``NP-hard'' is
also used for problems other than decision problems. For example,
let $\Pi'\in\NPCK$; then WMEM($\sep$) is NP-hard if there exists a
polynomial-time algorithm for $\Pi'$ that calls
$\mathcal{O}_{\mathrm{WMEM}(\sep)}$.

\subsection{Quantum separability problem in NP}\label{sec_DefnInNPProper}

Fact \ref{fact_FiniteDecompOfSepState} suggests that the quantum
separability problem is ostensibly in NP: a nondeterministic
Turing machine guesses $\{(p_i, [\ket{\psi^\A_i}],
[\ket{\psi^\B_i}])\}_{i=1}^{\n}$,\footnote{As usual, I use square
brackets to denote a matrix with respect to the standard basis.}
and then easily checks that
\begin{eqnarray}\label{eqn_NPCheck}
[\rho]=\sum_{i=1}^{\n}p_i
[\ket{\psi^\A_i}][\bra{\psi^\A_i}]\otimes
[\ket{\psi^\B_i}][\bra{\psi^\B_i}].
\end{eqnarray}
Hulpke and Bru{\ss} \cite{qphHB04} have demonstrated another
hypothetical guess-and-check procedure that does not involve the
numbers $p_i$.  They noticed that, given the vectors
$\{[\ket{\psi^\A_i}], [\ket{\psi^\B_i}]\}_{i=1}^{\n}$, one can
check that
\begin{eqnarray}\label{eqn_NPCheckHulpkeBruss1}
\text{$\{[\ket{\psi^\A_i}][\bra{\psi^\A_i}]\otimes
[\ket{\psi^\B_i}][\bra{\psi^\B_i}]\}_{i=1}^{\n}$ is affinely
independent; and}
\end{eqnarray}
\begin{eqnarray}\label{eqn_NPCheckHulpkeBruss2}
[\rho]\in\conv \{[\ket{\psi^\A_i}][\bra{\psi^\A_i}]\otimes
[\ket{\psi^\B_i}][\bra{\psi^\B_i}]\}_{i=1}^{\n}
\end{eqnarray}
in polynomially many arithmetic operations.

\begin{figure}[ht]
\centering \resizebox{100mm}{!}{\includegraphics{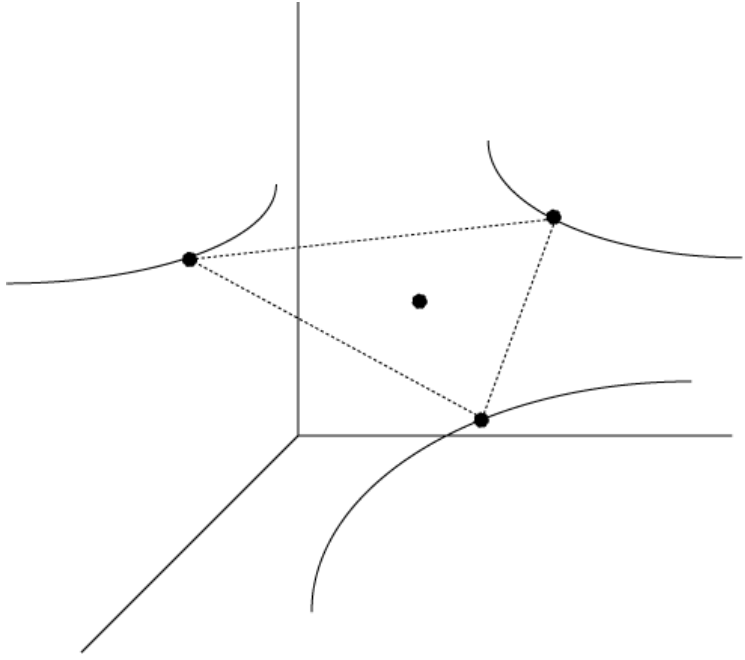}}
\caption[Surface on boundary of $\sep$]{The dashed triangle
outlines the convex hull of $x_1$, $x_2$, and $x_3$, shown as dots
at the triangle's vertices.  This convex hull contains $\rho$,
shown as a dot inside the triangle, and forms a (schematic) facet
of $\sep$. The curves represent the allowable choices for the
$\tilde{x}_i$. Sliding any of the $x_i$ takes
$\conv\{x_1,x_2,x_3\}$ outside of the facet.} \label{QSEPNotInNP}
\end{figure}

Membership in NP is only defined for decision problems.  Since
none of the weak membership formulations of the quantum
separability problem can be rephrased as decision problems
(because problem instances corresponding to states near the
boundary of $\sep$ can satisfy both possible answers), we cannot
consider their membership in NP\@.  However, EXACT QSEP \emph{is}
a decision problem.
\begin{problem}
Is EXACT QSEP in NP?
\end{problem}
\noindent Hulpke and Bru{\ss} have formalised some important
notions related to this problem.  They show that if $\rho\in
S(\sep,-\delta)$, for some $\delta>0$,
then each of the extreme points $x_i\in\sep$ in the expression
$\rho=\sum_{i=1}^{\dima^2\dimb^2}p_ix_i $ can be replaced by
$\tilde{x}_i$, where $[\tilde{x}_i]$ has rational entries.  This
is possible because the extreme points (pure product states) of
$\sep$ with rational entries are dense in the set of all extreme
points of $\sep$. However, when $\rho\notin S(\sep,-\delta)$, then
this argument breaks down. For example, when $\rho$ has full rank
and is on the boundary of $\sep$, then ``sliding'' $x_i$ to a
rational position $\tilde{x}_i$ might cause $\tilde{x}_i$ to be
outside of the affine space generated by $\{x_i\}_{i=1,\ldots,k}$.
Figure \ref{QSEPNotInNP} illustrates this in $\mathbb{R}^3$.
\noindent Furthermore, even if $x_i$ can be nudged comfortably to
a rational $\tilde{x}_i$, one would have to prove that
$<\tilde{x}_i>\in O(\poly(<[\rho]>))$, where $<X>$ is the size of
the encoding of $X$.

So, either the definition of NP does not apply (for weak
membership formulations), or we possibly run into problems near
the boundary of $\sep$ (for exact formulations).  Below we give an
alternative formulation that is in NP; we will refer to this
problem as QSEP.  The definition of QSEP is just a precise
formulation of the question ``Given a density operator $\rho$,
does there exist a separable density operator $\hat{\sigma}$ that
is close to $\rho$?''  We must choose a guess-and-check procedure
on which to base QSEP\@.  Because I want to prove that QSEP is
NP-hard, it is easier to choose the procedure which has the less
complex check (but the larger guess).

\begin{definition}[QSEP] Given a rational density matrix $[\rho]$ of dimension
$MN$-by-$MN$, and positive rational numbers $\delta_p$,
$\epsilon'$ and $\delta'$; does there exist a distribution
$\{(\tilde{p}_i;
\tilde{\alpha}_i,\tilde{\beta}_i)\}_{i=1,2,...,\n}$ of
unnormalised pure states $\tilde{\alpha}_i\in\mathbb{C}^M$,
$\tilde{\beta}_i\in\mathbb{C}^N$ where $\tilde{p}_i\geq 0$, and
$\tilde{p}_i$ and all elements of $\tilde{\alpha}_i$ and
$\tilde{\beta}_i$ are $\lceil \log_2(1/\delta_p)\rceil$-bit
numbers (complex elements are $x+iy$, $x,y\in\mathbb{R}$; where
$x$ and $y$ are $\lceil \log_2(1/\delta_p)\rceil$-bit numbers)
such that
\begin{eqnarray}\label{QSEP_Requirement1}
|1- ||\tilde{\alpha}_i||^2||\tilde{\beta}_i||^2\sum_{j=1}^{\n}
\tilde{p}_j| < \epsilon'\hspace{5mm}\textrm{for all $i$}
\end{eqnarray}
and
\begin{eqnarray}\label{QSEP_Requirement2}
||[\rho] -\tilde{\sigma}||^2_2:= \tr(([\rho] -\tilde{\sigma})^2) <
\delta'^2,
\end{eqnarray}
where $\tilde{\sigma}:= \sum_{i=1}^{\n} \tilde{p}_i
\tilde{\alpha}_i\tilde{\alpha}_i^\dagger\otimes
\tilde{\beta}_i\tilde{\beta}_i^\dagger $?
\end{definition}
\noindent Note that these checks can be done exactly in
polynomial-time, as they only involve elementary arithmetic
operations on rational numbers. To reconcile this definition with
the above intuition, we define $\hat{\sigma}$ as the separable
density matrix that is the ``normalised version'' of
$\tilde{\sigma}$:
\begin{eqnarray}\label{eqn_DefnOfSigmaHat}
\hat{\sigma} := \sum_{i=1}^{\n} \hat{p}_i
\hat{\alpha}_i\hat{\alpha}_i^\dagger\otimes
\hat{\beta}_i\hat{\beta}_i^\dagger,
\end{eqnarray}
where $\hat{p}_i:=\tilde{p}_i/\sum_i \tilde{p}_i$,
$\hat{\alpha}_i:=\tilde{\alpha}_i/||\tilde{\alpha}_i||$, and
$\hat{\beta}_i:=\tilde{\beta}_i/||\tilde{\beta}_i||$.  Using the
triangle inequality, we can derive that
\begin{eqnarray}
||\hat{\sigma}-\tilde{\sigma}||_2 \leq \sum_{i}\hat{p}_i |1-
||\tilde{\alpha}_i||^2||\tilde{\beta}_i||^2\sum_j \tilde{p}_j|,
\end{eqnarray}
where the righthand side is less than $\epsilon'$ when
(\ref{QSEP_Requirement1}) is satisfied.  If
(\ref{QSEP_Requirement2}) is also satisfied, then we have
\begin{eqnarray}
||[\rho]-\hat{\sigma}||_2 \leq ||[\rho] -\tilde{\sigma}||_2+
||\hat{\sigma}-\tilde{\sigma}||_2 \leq \delta' + \epsilon',
\end{eqnarray}
which says that the given $[\rho]$ is no further than $\delta' +
\epsilon'$ away from a separable density matrix (in Euclidean
norm).\footnote{I have formulated these checks to avoid division;
this makes the error analysis of the next section simpler.}

The decision problem QSEP is trivially in NP, as a
nondeterministic Turing machine need only guess the $\lceil\log_2(
1/\delta_p)\rceil$-bit distribution $\{(\tilde{p}_i;
\tilde{\alpha}_i,\tilde{\beta}_i)\}_{i=1,2,...,\n}$ and verify (in
polytime) that (\ref{QSEP_Requirement1}) and
(\ref{QSEP_Requirement2}) are satisfied.

\subsection{NP-Hardness}

Gurvits \cite{Gur03} has shown the weak membership problem for
$\sep$ to be NP-hard with respect to the complexity-measure $(N +
<[\rho]> + <\delta>)$. He demonstrates a Turing-reduction from
PARTITION and makes use of the very powerful Yudin-Nemirovskii
theorem (Theorem 4.3.2 in \cite{GLS88}).


We check now that QSEP is NP-hard, by way of a Karp-reduction from
WMEM($\sep$).  We assume we are given an instance
$I:=([\rho],\delta)$ of WMEM($\sep$) and we seek an instance
$I':=([\rho'],\delta_p,\epsilon',\delta')$ of QSEP such that if
$I'$ is a ``yes''-instance of QSEP, then $I$ satisfies
(\ref{eqn_WMEMSepAssertion}); otherwise $I$ satisfies
(\ref{eqn_WMEMEntAssertion}). It suffices to use $[\rho']=[\rho]$.
It is clear that if $\delta'$ and $\epsilon'$ are chosen such that
$\delta \geq \delta'+\epsilon'$, then $I'$ is a ``yes''-instance
only if $I$ satisfies (\ref{eqn_WMEMSepAssertion}).  For the other
implication, we need to bound the propagation of some
truncation-errors.  Let $p:=\lceil\log_2( 1/\delta_p)\rceil$.

Recall how absolute errors accumulate when multiplying and adding
numbers.  Let $x=\tilde{x}+\Delta_x$ and $y=\tilde{y}+\Delta_y$
where $x$, $y$, $\tilde{x}$, $\tilde{y}$, $\Delta_x$, and
$\Delta_y$ are all real numbers.  Then we have
\begin{eqnarray}\label{eqn_errorMult}
xy &=& \tilde{x}\tilde{y} + \tilde{x}\Delta_y+ \tilde{y}\Delta_x +
\Delta_x\Delta_y\\
x+y &=& \tilde{x}+\tilde{y}+\Delta_x+\Delta_y.
\end{eqnarray}
For $|\tilde{x}|,|\tilde{y}|<1$, because we will be dealing with
summations of products with errors, it is sometimes convenient
just to use
\begin{eqnarray}\label{eqn_errorMult}
|xy -\tilde{x}\tilde{y}| &\leq&   |\Delta_y|+ |\Delta_x| +
\textrm{max}\{|\Delta_x|,|\Delta_y|\}
\end{eqnarray}
to obtain our cumulative errors (which do not need to be tight to
show NP-hardness).  For example, if $\tilde{x}$ and $\tilde{y}$
are the $p$-bit truncations of $x$ and $y$, where $|x|,|y|<1$,
then $|\Delta_x|, |\Delta_y|<2^{-p}$; thus a conservative bound on
the error of $\tilde{x}\tilde{y}$ is
\begin{eqnarray}\nonumber
|xy- \tilde{x}\tilde{y}| < |\Delta_y|+|\Delta_x|+|\Delta_x| =
3|\Delta_x|<2^2|\Delta_x|= 2^{-(p-2)}.
\end{eqnarray}

\begin{proposition}\label{Prop_BoundDist__Sigma_SigmaTilde} Let $\sigma\in \sep$ be such that
$\sigma=\sum_{i=1}^{\n}p_i\alpha_i\alpha_i^\dagger\otimes\beta_i\beta_i^\dagger$,
and let \\ $\{(\tilde{p}_i;
\tilde{\alpha}_i,\tilde{\beta}_i)\}_{i=1,2,...,\n}$ be the $p$-bit
truncation of $\{({p}_i;
{\alpha}_i,{\beta}_i)\}_{i=1,2,...,\n}$.\\
Then $||\sigma-\tilde{\sigma}||_2< M^3N^32^{-(p-7.5)}$, where
\begin{eqnarray}
\tilde{\sigma} :=
\sum_{i=1}^{\n}\tilde{p}_i\tilde{\alpha}_i\tilde{\alpha}_i^\dagger\otimes\tilde{\beta}_i\tilde{\beta}_i^\dagger.
\end{eqnarray}
\end{proposition}
\begin{proof}
Letting
$\gamma_i:=p_i\alpha_i\alpha_i^\dagger\otimes\beta_i\beta_i^\dagger
-
\tilde{p}_i\tilde{\alpha}_i\tilde{\alpha}_i^\dagger\otimes\tilde{\beta}_i\tilde{\beta}_i^\dagger$,
we use the triangle inequality to get
\begin{eqnarray}
||\sigma  - \tilde{\sigma}||_2 &\leq& \sum_i ||\gamma_i||_2
=\sum_i \sqrt{\tr(\gamma_i^2)}.
\end{eqnarray}
It suffices to bound the absolute error on the elements of
$[\tilde{p}_i\tilde{\alpha}_i\tilde{\alpha}_i^\dagger\otimes\tilde{\beta}_i\tilde{\beta}_i^\dagger]$;
using our conservative rule (\ref{eqn_errorMult}), these elements
have absolute error less than $2^{-(p-7)}$.  Thus $[\gamma_i]$ is
an $MN$-by-$MN$ matrix with elements no larger than $2^{-(p-7)}$
in absolute value.  It follows that $(\tr(\gamma_i^2))^{1/2}$ is
no larger than $\sqrt{MN}2^{-(p-7.5)}$ in absolute value. Finally,
we get
\begin{eqnarray}
||\sigma - \tilde{\sigma}||_2 \leq\sum_i
\sqrt{\tr(\gamma_i^2)}\leq M^{3}N^{3}2^{-(p-7.5)}.
\end{eqnarray}
\end{proof}

\begin{proposition}\label{Prop_BoundOnNormalisedSigma}
Let $\tilde{\sigma}$ be as in Proposition
\ref{Prop_BoundDist__Sigma_SigmaTilde}.  Then for all
$i=1,2,\ldots \n$
\begin{eqnarray}
|1- ||\tilde{\alpha}_i||^2||\tilde{\beta}_i||^2\sum_{j=1}^{\n}
\tilde{p}_j| < \dima^3\dimb^3 2^{-(p-5)}.
\end{eqnarray}
\end{proposition}
\begin{proof}
The absolute error on $\sum_j \tilde{p}_j$ is $\n2^{-p}$. The
absolute error on $||\tilde{\alpha}_i||^2$ (resp.
$||\tilde{\beta}_i||^2$) is no more than $M2^{-(p-3)}$ (resp.
$N2^{-(p-3)}$).  This gives total absolute error of
\begin{eqnarray}
|1-{||\tilde{\alpha}_i||^2||\tilde{\beta}_i||^2}\sum_j
\tilde{p}_j| < \dima^3\dimb^3 2^{-(p-5)}.
\end{eqnarray}
\end{proof}

Let $\delta':= M^{3}N^{3}2^{-(p-8)}$ and
$\epsilon':=\dima^3\dimb^3 2^{-(p-5)}$ and set $p$ such that
$\epsilon'+\delta'\leq\delta$. Suppose there exists a separable
density matrix $\sigma$ such that $||[\rho]-\sigma||_2=0$.  Then
Propositions \ref{Prop_BoundDist__Sigma_SigmaTilde} and
\ref{Prop_BoundOnNormalisedSigma} say that there exists a
certificate $\tilde{\sigma}$ such that (\ref{QSEP_Requirement1})
and (\ref{QSEP_Requirement2}) are satisfied. Therefore, if $I'$ is
a ``no''-instance, then for all separable density matrices
$\sigma$, $||[\rho]-\sigma||_2>0$; which implies that $I$
satisfies (\ref{eqn_WMEMEntAssertion}).  I have exhibited a
polytime Karp-reduction from WMEM($\sep$) to QSEP (actually, from
$\text{WMEM}_{\text{In}}(\sep)$ to QSEP).
\begin{fact}\label{Prop_QSEPisNPHard}
\emph{QSEP} is in $\NPCT$.
\end{fact}

\subsection{Towards a Karp Reduction}

To date, every decision problem (except for QSEP) that is in
$\NPCT$ is also known to be in $\NPCK$ \cite{PS01}. While it is
strongly suspected that Karp and Turing reductions are
inequivalent within NP, it would be very strange if QSEP, or some
other formulation of the quantum separability problem,\footnote{By
``formulation of the quantum separability problem'', I mean an
approximate formulation that tends to EXACT QSEP as the accuracy
parameters of the problem tend to zero.} is the first example that
proves this inequivalence.  We have an interesting open problem:
\begin{problem}\label{prob_QSEPinNPCK} Is QSEP
in $\NPCK$?
\end{problem}
\noindent Note that, because of Fact \ref{Prop_QSEPisNPHard}, a
negative answer to this problem implies that $\P\neq\NP$. Thus it
might be safer to work under the assumption that the answer is
positive, and look for a Karp reduction from some Karp-NP-complete
problem to some formulation $\Pi_{\mathrm{QSEP}}$ of the quantum
separability problem.

Technically, WMEM($\sep$) is not in NP because it is not a
decision problem. But the definition of ``NP'' can be modified to
accommodate such weakened problems having overlapping decisions
\cite{GLS88}. According to this different definition, WMEM($\sep$)
is in ``NP''.\footnote{For the weak membership problem, WMEM$(K)$
is in ``NP'' if and only if for all points $p\in S(K,-\delta)$
there exists a succinct certificate of the fact that $p\in
S(K,\delta)$. According to \cite{qphHB04}, any $\rho\in
S(\sep,-\delta)$ is in the convex hull of $\dima^2\dimb^2$
affinely independent elements of a dense set of pure product
states generated by rationals.  By possibly tweaking each element,
we can choose the rational numbers to have denominators no bigger
than $\poly(\dima,\dimb)/\delta$, so we can perform the checks in
(\ref{eqn_NPCheckHulpkeBruss1}) and
(\ref{eqn_NPCheckHulpkeBruss2}) efficiently, to conclude that
$p\in S(\sep,\delta)$.} We can pose the following open problem,
related to the one above.
\begin{problem}
Does there exist a Karp reduction from some Karp-NP-complete
problem to WMEM($\sep$)?
\end{problem}
\noindent Finding a positive answer to this problem implies a
positive answer for Problem \ref{prob_QSEPinNPCK}. Alternatively,
finding a negative answer to this problem does not, technically,
imply that $\P\neq\NP$, so may not win the million-dollar prize.

\subsection{Nonmembership in co-NP}

Is either EXACT QSEP or QSEP in co-NP?   To avoid possible
technicalities, we might first consider the presumably easier
question of whether WMEM($\sep$) is in ``co-NP'': Does every
entangled state $\rho\notin S(\sep,\delta)$ have a succinct
certificate of not being in $S(\sep,-\delta)$?  It may or may not
be the case that P equals NP$\cap$co-NP, but a problem's
membership in NP$\cap$co-NP can be ``regarded as suggesting'' that
the problem is in P \cite{GJ79}. Thus, we might believe that
WMEM($\sep$) is not in ``co-NP'' (since WMEM($\sep$) is NP-hard).

Let us consider this with regard to entanglement witnesses (which
are candidates for succinct certificates of entanglement). We know
that every entangled state has a (right) entanglement witness
$A\in\hermops$ that detects it. However, it follows from the
NP-hardness of WMEM($\sep$) and Theorem 4.4.4 in \cite{GLS88} that
the weak validity problem for $K=\sep$ (WVAL($\sep$)) is
NP-hard:\footnote{Theorem 4.4.4 in \cite{GLS88}, applied to
$\sep$, states that there exists an oracle-polynomial-time
algorithm that solves the WSEP($\sep$) given an oracle for
WVAL($\sep$).}
\begin{definition}[Weak validity problem (WVAL)]\label{def_WVAL}
Given a rational vector $c\in\mathbb{R}^n$, a rational number
$\gamma$, and rational $\epsilon>0$, assert either that
\begin{eqnarray}
c^Tx &\leq& \gamma+\epsilon\text{ for all }x\in K, \hspace{2mm}\text{or}\label{WVAL_AssertIsASepPlane}\\
c^Tx &\geq& \gamma-\epsilon \text{ for some }x\in K.
\end{eqnarray}
\end{definition}
\noindent So there is no known way to check efficiently that a
hyperplane $\pi_{A,b}$ separates $\rho$ from $\sep$ (given just
the hyperplane); thus, an entanglement witness alone does not
serve as a succinct certificate of a state's entanglement unless
WVAL($\sep$) is in P\@.  However, one could imagine that there is
a succinct certificate of the fact that a hyperplane $\pi_{A,b}$
separates $\rho$ from $\sep$. If such a certificate exists, then
WVAL($\sep$) is in ``NP'' and WMEM($\sep$) is in
``co-NP''.\footnote{WVAL(K) is in ``NP'' means that for any $c$,
$\gamma$, $\epsilon$ satisfying $c^Tx\leq\gamma-\epsilon\text{ for
all }x\in K$, there exists a succinct certificate of the fact that
(\ref{WVAL_AssertIsASepPlane}) holds.}

With regard to QSEP, we can prove the following:
\begin{fact}\label{Fact_QSEPincoNPmeansNPiscoNP} QSEP is not in co-NP, unless NP equals co-NP.
\end{fact}
\noindent This fact follows from the general theorem below
\cite{Buh05}:
\begin{theorem}  If $\Pi$ is in $\NPCT$ and $\Pi$ is in co-NP, then NP
equals co-NP.
\end{theorem}
\begin{proof}
Since $\Pi$ is in co-NP, $\Pi^c$ is in NP.  Let $\Pi'$ be any
problem in co-NP.  To show that co-NP equals NP, it suffices to
show that co-NP is contained in NP; thus, it suffices to show that
$\Pi'$ is in NP.  The following reduction chain holds, since
$\Pi'^c$ is in NP: $\Pi' \leq_\T \Pi'^c \leq_\T \Pi$.  Because
both $\Pi$ and $\Pi^c$ are in NP, the reduction $\Pi'\leq_\T \Pi$
can be carried out by a polytime nondeterministic Turing machine,
which can ``solve'' any query to $\mathcal{O}_\Pi$ by
nondeterministically guessing and checking in polynomial-time the
``yes''-certificate (if the query is a ``yes''-instance of $\Pi$)
or the ``no''-certificate (if the query is a ``no''-instance of
$\Pi)$.  Thus $\Pi'$ is in NP.
\end{proof}
\noindent It is strongly conjectured that NP and co-NP are
different \cite{Pap94}, thus we might believe that QSEP is not in
co-NP. \footnote{We would like to be able to use Fact
\ref{Fact_QSEPincoNPmeansNPiscoNP} to show that WVAL($\sep$) is
not in ``NP'' unless NP equals co-NP.  However, for this, we would
require that ``WVAL($\sep$) is in NP only if QSEP is in co-NP'';
but this is not the case (only the converse holds).}

\section{Survey of algorithms for the quantum separability
problem}\label{sec_SurveyOfAlgorithms}

I concentrate on proposed algorithms that solve an approximate
formulation of the quantum separability problem and have
(currently known) asymptotic analytic bounds on their running
times.  For this reason, the SDP relaxation algorithm of Eisert et
al.\ is not mentioned here (see Section
\ref{subsec_EisertsEtalApproach}); though, I do not mean to
suggest that in practice it could not outperform the following
algorithms on typical instances.  As well, I do not analyse the
complexity of the naive implementation of every necessary and
sufficient criterion for separability, as it is assumed that this
would yield algorithms of higher complexity than the following
algorithms.\footnote{For an exhaustive list of all such criteria,
see the forthcoming book by Bengtsson and Zyczkowski \cite{BZ}.}

The main purpose below is to get a time-complexity estimate in
terms of the parameters $\dima$, $\dimb$, and $\delta$, where
$\delta$ is the accuracy parameter in WMEM($\sep$). In the
following, the only way precision and error are dealt with is
similar to the above discussion, where we have a truncation-error
resulting from approximating the continuum of pure product states
by a finite set of finitely precise product vectors.  The
running-time estimates are based on the number of elementary
arithmetic operations and do not attempt to deal with computer
round-off error; I do not give estimates on the total amount of
machine precision required.  Instead, where rounding is necessary
in order to avoid exponential blow-up of the representation of
numbers during the computation, I assume that the working
precision\footnote{``Working precision'' is defined as the number
of significant digits the computer uses to represent numbers
during the computation.} can be set large enough that the overall
effect of the round-off error on the final answer is either much
smaller than $\delta$ or no larger than, say, $\delta/2$ (so that
doubling $\delta$ takes care of the error due to round-off).

\subsection{Search for separable decompositions}\label{sec_BasicAlgorithm}

The most naive algorithm for any problem in NP consists of a
search through all potential succinct certificates that the given
problem instance is a ``yes''-instance.  Thus QSEP immediately
gives an algorithm for the quantum separability problem.  However,
we can, in principle, reformulate QSEP to incorporate the ideas of
Hulpke and Bru{\ss} \cite{qphHB04} in order to get a better
algorithm.

\subsubsection{The algorithm of Hulpke and Bru{\ss}}

First, let us see how to perform the checks in lines
(\ref{eqn_NPCheckHulpkeBruss1}) and
(\ref{eqn_NPCheckHulpkeBruss2}).  Using simpler notation, suppose
we are given
$\{x_i:\hspace{2mm}i=1,2,\ldots,k\}\subset\mathbb{R}^n$.  This set
is \emph{affinely independent} if and only if
$\{x_i-x_1:\hspace{2mm}i=2,\ldots,k\}$ is linearly independent.
Thus Gaussian elimination can be used to test for affine
independence. Suppose $\{x_i:\hspace{2mm}i=1,2,\ldots,n+1\}$ is
affinely independent. Then the $x_i$ form the extreme points of
the polytope $\conv \{x_i:\hspace{2mm}i=1,2,\ldots,n+1\}$.
Consider the facet of this polytope that does not contain $x_j$,
and choose some $x_l\neq x_j$ in the facet. The normal $\nu_j$ to
this facet is orthogonal to $x_i-x_l$, for all $i\neq j,l$, and is
thus the generator of the nullspace of the matrix whose $n-1$ rows
are the vectors $x_i-x_l$.  Again, Gaussian elimination can be
used to solve for $\nu_j$.  A point $\rho$ is in the polytope if
and only if, for all $j=1,2,\ldots,n+1$, the halfspace $\{x:
\nu_j^Tx\leq \nu_j^Tx_l\}$ contains both or neither of $\rho$ and
$x_j$; that is, both $\rho$ and $x_j$ are on the ``same side'' of
the hyperplane $\{x: \nu_j^Tx = \nu_j^Tx_l\}$ corresponding to the
facet not containing $x_j$.

The algorithm of Hulpke and Bru{\ss} is basically a loop through
all possible affinely independent sets $X$ of pure product states,
with the check for whether $\conv X$ contains the given state
$\rho$. However, the algorithm uses unbounded precision and
performs its calculations to arbitrarily high precision so that it
attempts to find such (arbitrarily precise) $X$ for $\rho\in\sep$
that are arbitrarily close to the boundary of $\sep$; it may even
find such $X$ for $\rho\in\sep$ that are on the boundary of the
``cone'' of positive Hermitian operators and hence on the boundary
of $\sep$.  The algorithm only relaxes and solves the weak
membership problem for states $\rho\in\sep$ that are on the
boundary between separable and entangled states. As argued at the
beginning of this chapter, we are satisfied with an algorithm for
the weak membership problem for \emph{all} states.  Thus we will
formulate an approximate version of this algorithm whose precision
requirements for the $X$ are bounded by $\dima$, $\dimb$, and
$\delta$.\footnote{The full algorithm of Hulpke and Bru{\ss} is
the parallel combination of the algorithm of Doherty et al.\ and
this search for an $X$, along with a check for the case when
$\rho$ is $\eta$-close to the boundary between separable and
entangled states.}

\subsubsection{Reformulation of QSEP}\label{subsubsec_ReformulationQSEP}

Recall the mapping $v:\hermops\rightarrow
\mathbb{R}^{\dima^2\dimb^2-1}$ defined in
(\ref{eqn_MappingFromHermopsToRealVecs}) on page
\pageref{eqn_MappingFromHermopsToRealVecs}.

\begin{definition}[QSEP']
Given a rational density matrix $[\rho]$ of dimension
$MN$-by-$MN$, and positive rational numbers $\delta_p$ and
$\epsilon'$; does there exist a set $\{(
\tilde{\alpha}_i,\tilde{\beta}_i)\}_{i=1,2,...,\n}$ of
unnormalised pure states $\tilde{\alpha}_i\in\mathbb{C}^M$,
$\tilde{\beta}_i\in\mathbb{C}^N$ where all elements of
$\tilde{\alpha}_i$ and $\tilde{\beta}_i$ are $\lceil
\log_2(1/\delta_p)\rceil$-bit numbers (complex elements are
$x+iy$, $x,y\in\mathbb{R}$; where $x$ and $y$ are $\lceil
\log_2(1/\delta_p)\rceil$-bit numbers) such that\footnote{Because
I am ignoring round-off error, I assume that the function $v$ can
be computed exactly, even though the elements $X_i$ of
$\mathcal{B}$ have square-root symbols appearing in them. (Because
the computations required for the check are relatively simple, it
might be possible to carry these irrationals symbolically through
most of the computation, only requiring an approximation of them
near the end when computing the normal to a hyperplane and
checking the distance from various points to a hyperplane.) I
wanted to avoid such an assumption in the proof of NP-hardness of
QSEP\@. It will be become clear, though, that QSEP' -- with the
$v(\tilde{\alpha}_i\tilde{\alpha}_i^\dagger\otimes\tilde{\beta}_i\tilde{\beta}_i^\dagger)$
truncated -- could also be shown to be NP-hard with a suitable
truncation-error analysis.}
\begin{eqnarray}\label{QSEP'_Requirement1}
|1- ||\tilde{\alpha}_i||^2||\tilde{\beta}_i||^2| <
\epsilon'\hspace{5mm}\textrm{for all $i$}
\end{eqnarray}
and
\begin{eqnarray}\label{QSEP'_Requirement2}
\{v(\tilde{\alpha}_i\tilde{\alpha}_i^\dagger\otimes\tilde{\beta}_i\tilde{\beta}_i^\dagger)\}_i\text{
is affinely independent}
\end{eqnarray}
and
\begin{eqnarray}\label{QSEP'_Requirement3}
[\rho] \in
S(\conv\{v(\tilde{\alpha}_i\tilde{\alpha}_i^\dagger\otimes\tilde{\beta}_i\tilde{\beta}_i^\dagger)\}_i,\epsilon')?\end{eqnarray}
\end{definition}

\noindent Note that (\ref{QSEP'_Requirement1}) ensures that
$\tilde{\alpha}_i\tilde{\alpha}_i^\dagger\otimes\tilde{\beta}_i\tilde{\beta}_i^\dagger$
is $\epsilon'$-close to an actual state
$\hat{\alpha}_i\hat{\alpha}_i^\dagger\otimes\hat{\beta}_i\hat{\beta}_i^\dagger$,
where $\hat{\alpha}_i:=\tilde{\alpha}_i/||\tilde{\alpha}||$ and
$\hat{\beta}_i:=\tilde{\beta}_i/||\tilde{\beta}||$.  The check in
line (\ref{QSEP'_Requirement3}) is an easy modification of the
check described in the previous subsection. Let $p:=\lceil\log_2(
1/\delta_p)\rceil$.

Suppose that, for some $\sigma\in\sep$,
$\sigma\in\conv\{\alpha_i\alpha_i^\dagger\otimes\beta_i\beta_i^\dagger\}_{i=1}^{\dima^2\dimb^2-1}$
for normalised pure states $\alpha_i\in\mathbb{C}^\dima$ and
$\beta_i\in\mathbb{C}^\dimb$.  Let $\tilde{\alpha}_i$ and
$\tilde{\beta}_i$ be the $p$-bit truncations of $\alpha_i$ and
$\beta_i$, and let
$\gamma_i:=\alpha_i\alpha_i^\dagger\otimes\beta_i\beta_i^\dagger-
\tilde{\alpha}_i\tilde{\alpha}_i^\dagger\otimes\tilde{\beta}_i\tilde{\beta}_i^\dagger$.
The rectangular coordinates of the entries in $[\gamma_i]$ are no
bigger than $2^{-(p-6)}$.  It follows that
$\sqrt{\tr(\gamma_i^2)}$ is not larger than $MN2^{-(p-6.5)}$:
\begin{eqnarray}
||\alpha_i\alpha_i^\dagger\otimes\beta_i\beta_i^\dagger-
\tilde{\alpha}_i\tilde{\alpha}_i^\dagger\otimes\tilde{\beta}_i\tilde{\beta}_i^\dagger||\leq
MN2^{-(p-6.5)}.
\end{eqnarray}
Thus, setting $\epsilon':=MN2^{-(p-7)}$ and setting $p$ such that
$2\epsilon'<\delta$, it follows that QSEP' solves WMEM($\sep$)
with accuracy parameter $\delta$.  This gives
\begin{eqnarray}
p>\log_2(2MN/\delta)+7.
\end{eqnarray}

\vspace{3mm}

Therefore, to solve WMEM($\sep$), it suffices to loop through all
$(\dima^2\dimb^2)$-subsets of
$\lceil\log_2(2MN/\delta)+7\rceil$-bit unnormalised pure product
states, checking the three conditions in QSEP'\@.  Define
$\Omega_p$ as the number of $p$-bit unnormalised pure product
states resulting from the truncation (to $p$ bits) of all
normalised pure product states.  The complexity of this algorithm
is
\begin{eqnarray}
\nchoosek{\Omega_{\lceil\log_2(2MN/\delta)+7\rceil}}{\dima^2\dimb^2}\poly(\dima,\dimb,\log(1/\delta)).
\end{eqnarray}
Since the pure product states can be parametrised by
$2(\dima+\dimb)-4$ real parameters, we have the estimate
\begin{eqnarray}\label{eqn_EstimateOfOmega}
\Omega_p \gtrsim 2^{p(2(\dima+\dimb)-4)}.
\end{eqnarray}
Combined with the estimate $\nchoosek{n}{k}\sim n^k$, we get a
\emph{rough} asymptotic complexity estimate for the algorithm
 of
\begin{eqnarray}
\left(\frac{2^{6.5}\dima\dimb}{\delta}\right)^{2(\dima^3\dimb^2+\dima^2\dimb^3)-4\dima^2\dimb^2}\poly(\dima,\dimb,\log(1/\delta)).
\end{eqnarray}

In the interest of getting a rough \emph{lower} bound on the
complexity of this algorithm, I have underestimated $\Omega_p$.
The number $2^{p(2(\dima+\dimb)-4)}$ corresponds to the number of
different $p$-bit settings of the $2(\dima+\dimb)-4$ angles
(phases and amplitudes) that parametrise the normalised pure
product states.  The truncation-error analysis was done with
respect to rectangular coordinates, so this method of generating
the elements $\tilde{\alpha}_j\otimes \tilde{\beta}_j$ may miss
some elements that would have resulted from a $p$-bit truncation
of rectangular coordinates of normalised pure product states.  On
the other hand, if we use all $p$-bit settings of the $2(\dima +
\dimb)$ rectangular coordinates to generate elements
$\tilde{\alpha}_j\otimes \tilde{\beta}_j$, then many of the
elements generated will not satisfy $|1-
||\tilde{\alpha}_j||^2||\tilde{\beta}_j||^2| <\epsilon'$.  The
most efficient way to systematically generate the elements
$\tilde{\alpha}_j\otimes \tilde{\beta}_j$ is left as an open
problem:
\begin{problem}
What is the most efficient way to generate the $j$th element
$\tilde{\alpha}_j\otimes \tilde{\beta}_j$ of the set of $\Omega_p$
unnormalised pure product states resulting from the $p$-bit
truncation of all normalised pure product states?
\end{problem}

We take the algorithm of this section as the best exhaustive
search approach to solving the approximate quantum separability
problem. For example, it is better than searching all of $\sep$ in
order to calculate $E_{d^2_2}(\rho)$ of Section
\ref{subsec_EisertsEtalApproach}; and it is better than searching
all pure decompositions of $\rho$ in order to calculate
$E'_F(\rho)$ of Section \ref{subsec_EntanglementMeasures}.

\subsection{Bounded search for symmetric extensions}\label{sec_DohertyEtalKonigRenner}

In Section \ref{sec_DohertyEtalApproach}, we considered two tests
-- one that searches for symmetric extensions of $\rho$, and a
stronger one that searches for PPT symmetric extensions.  Now we
continue that exposition, showing that recent results can put an
upper bound on the number $k$ of copies of subsystem $\A$ when
solving an approximate formulation of the separability problem.
The bound only assumes symmetric extensions, \emph{not} PPT
symmetric extensions, so it is possible that a better bound may be
found for the stronger test.

If a symmetric state
$\varrho\in\densopsgen{(\mathbb{C}^d)^{\otimes n}}$ has a
symmetric extension to $\densopsgen{(\mathbb{C}^d)^{\otimes
(n+m)}}$ for all $m>0$, then it is called \emph{(infinitely)
exchangeable}.  The quantum de Finetti theorem\footnote{References
for material in this paragraph may be found in \cite{DPS04}.} says
that the infinitely exchangeable state $\varrho$ is separable.
Recalling the terminology of Section
\ref{sec_DohertyEtalApproach}, it is also possible to derive that,
for $\rho\in\densopsgen{\CMotimesCN}$, if there exists a symmetric
extension of $\rho$ to $k$ copies of subsystem $\A$ for all $k>0$,
then $\rho\in\sep$.  This is the result that proves that Doherty
et al.'s hierarchy of tests is complete:  if $\rho$ is entangled,
then the SDP at some level $k_0$ of the hierarchy will not be
feasible (i.e. will not find a symmetric extension of $\rho$ to
$k_0$ copies of subsystem $\A$). K{\"{o}}nig and Renner
\cite{qphKR04} derived quite general results about states $\rho$
that have symmetric extensions to $k$ copies of subsystem $\A$.
Their results give us our upper bound on $k$.

The upper bound follows directly from the main theorem in
\cite{qphKR04}.  The result is too technical to summarise
meaningfully without diverging from the aim of this thesis.  We
require the following corollary:
\begin{theorem}[Corollary of Theorem 6.1 in \cite{qphKR04}]
Suppose $\rho\in\densops$ and there exists a symmetric extension
of $\rho$ to $k\geq 2$ copies of subsystem $\A$.  Then
\begin{eqnarray}
\tr|\rho - \sigma|\leq \frac{4\dima^6}{\sqrt{k-1}},
\end{eqnarray}
for some $\sigma\in\sep$.
\end{theorem}
\noindent The proof of this theorem is similar to the proof of
Corollary 6.2 in \cite{qphKR04}.  Note that the result uses the
\emph{trace distance}, $\tr|X-Y|$, between two operators $X$ and
$Y$.  Let us assume we are solving the weak membership formulation
of the quantum separability problem with respect to the trace
distance, and with accuracy parameter $\delta$.  Then, setting
$\delta={4\dima^6}/{\sqrt{k-1}}$, we get the following upper bound
for $k$:

\begin{corollary} To solve WMEM($\sep$) (with respect to the trace distance)
with accuracy parameter $\delta$ by searching for symmetric
extensions (as described in Section
\ref{sec_DohertyEtalApproach}), it suffices to look for symmetric
extensions to
\begin{eqnarray}
\bar{k}:=\lceil 16\dima^{12}/\delta^2 +1 \rceil
\end{eqnarray}
copies of subsystem A.
\end{corollary}

To estimate the total complexity of the algorithm, note that
\begin{eqnarray}
d_{S_k}=
\frac{[(\dima-1)+k][(\dima-2)+k]\cdots[(1)+k]}{(\dima-1)!}>
k^{\dima -1}/(\dima -1)!.
\end{eqnarray}
Substituting $\bar{k}$ for $k$, we get
\begin{eqnarray}\label{eqn_MainDohertEtalKonigRennerComplexity}
d_{S_{\bar{k}}}> \left(\frac{16\dima^{11}}{\delta^2}\right)^{\dima
-1}.
\end{eqnarray}
Just to solve the first constraint in (\ref{prob_DohertyEtalSDP})
requires $\sqrt{n}$ (but usually far fewer) iterations of a
procedure that requires $O(m^2n^2)$ arithmetic operations, for
$m=(d_{S_{\bar{k}}}^2-\dima^2)\dimb^2$ and
$n=d_{S_{\bar{k}}}^2\dimb^2$.
\begin{problem}
Can the upper bound $\bar{k}$ be improved by taking into
consideration the PPT constraints in (\ref{prob_DohertyEtalSDP})?
\end{problem}
\noindent Despite this unattractive worst-case bound, the
hierarchy of tests has proved to be efficient in practice for
confirming that certain states are entangled (i.e. small $k$
suffices).

\subsection{Cross-norm criterion via linear programming}\label{sec_Rudolf&PerezGarcia}

Rudolph \cite{Rud00} derived a simple characterisation of
separable states in terms of a computationally complex operator
norm $||\cdot||_\gamma$.\footnote{The mathematical arguments
behind the results in this section are nontrivial in that they
involve notions from operator theory, which are tough-going for
the nonexpert (me).  Luckily, the results themselves can be stated
and understood, at least superficially, with relative ease.} For a
finite-dimensional vector space $V$, let $\mathcal{T}(V)$ be the
class of all linear operators on $V$. The norm is defined on
$\mathcal{T}(\mathbb{C}^\dima) \otimes
\mathcal{T}(\mathbb{C}^\dimb)$ as
\begin{eqnarray}
||t||_\gamma := \inf \lbrace \sum_{i=1}^{k}||u_i||_1 ||v_i
||_1:\hspace{1mm} t = \sum_{i=1}^{k}u_i\otimes v_i \rbrace,
\end{eqnarray}
where the infimum is taken over all decompositions of $t$ into
finite summations of elementary tensors, and
$||X||_1:=\tr(\sqrt{X^\dagger X})$.  Rudolph showed that
$||\rho||_\gamma \leq 1$ if and only if $||\rho||_\gamma = 1$, and
that a state $\rho$ is separable if and only if $||\rho||_\gamma =
1$.

P{\'{e}}rez-Garcia \cite{Per04} showed that approximately
computing this norm can be reduced to a linear program (which is a
special case of a semidefinite program): $\min\{ c^Tx :
\hspace{2mm}Ax=b, x\geq 0\}$,
where $A\in\mathbb{R}^{n\times m}$, $b\in\mathbb{R}^n$,
$c\in\mathbb{R}^m$, and $x$ is a vector of $m$ real variables;
here, $x\geq 0$ means that all entries in the vector are
nonnegative.  An LP can be solved in $O(m^{3}L')$ arithmetic
operations, where $L'$ is the length of the binary encoding of the
LP \cite{Ye97}. The linear program has on the order of
$\dima^2\dimb^2$ variables and
$\dima^{2\dima}\dimb^{2\dimb}(2k)^{2(\dima + \dimb)}$ constraints,
where $k$ is an integer that determines the relative
error\footnote{The \emph{relative error} of an approximation
$\tilde{x}$ of $x$ is defined as $|x-\tilde{x}|/x$.}
$(k/(k-1))^4-1$ on the computation of the norm.  Thus it may be
solved in
\begin{eqnarray}
O(\dima^{2\dima+2}\dimb^{2\dimb+2}(2k)^{2(\dima + \dimb)})
\end{eqnarray}
arithmetic operations.

Suppose $|| \rho ||_\gamma$ is found to be no greater than
$1+\eta$. Then, we would like to use $\eta$ to upper-bound the
distance, with respect to either trace or Euclidean norm, from
$\rho$ to $\sep$.  Unfortunately, we do not know how to do this.
This drawback, along with the fact that the error on the computed
norm is relative as opposed to absolute, does not allow this
algorithm to be easily compared to the other algorithms I
consider.  Still, there may be a way to overcome this problem, as
follows.

Following Rudolph \cite{qphRud02}, a norm closely related to
$||\cdot||_\gamma$ is
\begin{eqnarray}
||t||_\S := \inf \lbrace \sum_{i=1}^{k}||u_i||_1 ||v_i
||_1:\hspace{1mm} t = \sum_{i=1}^{k}u_i\otimes v_i \rbrace,
\end{eqnarray}
where the infimum is taken over all decompositions of $t$ into
finite summations of elementary \emph{Hermitian} tensors.  This
restriction on the decomposition implies that $||t||_\gamma\leq
||t||_\S$; thus, if $||\rho||_\S\leq 1$, then $\rho\in\sep$.
Conversely, if $\rho\in\sep$, then $\rho=\sum_i (p_i
\rho^\A_i)\otimes \rho^\B_i$; and this decomposition ensures
$||\rho||_\S\leq 1$.  Thus $||\rho||_\S\leq 1$ if and only if
$\rho\in\sep$.  The norm $||\cdot||_\S$ is related to an
entanglement measure called ``robustness''.

The \emph{robustness of entanglement} \cite{VT99} of
$\rho\in\densops$ is defined as
\begin{eqnarray}
R(\rho):=\inf \{a^- :\hspace{2mm} \rho = a^+\sigma^+
-a^-\sigma^-,\hspace{1mm} a^\pm\geq0,\sigma^\pm\in\sep\}.
\end{eqnarray}
In other words, the robustness is (a simple function of) the
minimal $p$, $0\leq p\leq 1$, such that
\begin{eqnarray}
\sigma^+ = p\sigma^- + (1-p)\rho
\end{eqnarray}
for separable states $\sigma^\pm$; the minimal $p$ is
$p_{R(\rho)}:=R(\rho)/(R(\rho)+1)$. Thus, $R(\rho)$ corresponds to
the minimal amount of separable ``noise'' ($\sigma^-$) that must
be added to $\rho$ in order to eliminate all the entanglement in
$\rho$.

Using properties of ``subcross norms'' (see references in
\cite{qphRud02}), Rudolph shows \cite{qphRud02} that for
$\rho\in\densops$
\begin{eqnarray}
R(\rho) \equiv \frac{1}{2}(||\rho||_\S -1);
\end{eqnarray}
the proof is based on the ideas of ``base norm'' used in
\cite{VW02}.

The point is that \emph{if} we could modify P{\'{e}}rez-Garcia's
algorithm so that it approximately computes $||\cdot||_\S$, then
we could relate the result to a standard norm, as follows. Suppose
the algorithm allows us to assert that $||\rho||_\S\leq 1+2\eta$.
Then $R(\rho)\leq \eta$.  Now, we have
\begin{eqnarray}
||\rho - \sigma^+|| &=& ||p_{R(\rho)}(\rho - \sigma^-)|| \\
&=& p_{R(\rho)}||\rho - \sigma^-|| \\
&=& \frac{R(\rho)}{1+R(\rho)}||\rho - \sigma^-||\\
&\leq& \frac{2\eta}{1+\eta} ,
\end{eqnarray}
where 2 is a an upper bound on the Euclidean diameter of the set
of (normalised) density operators (see Figure \ref{DensityOps} on
page \pageref{DensityOps}).\footnote{Actually, from the diagram,
we could get a slightly better bound than 2. But, since this
discussion is purely ``academic'', it does not matter.}

\begin{problem} Can the algorithm of P{\'{e}}rez-Garcia be
modified so that it approximately computes the norm
$||\cdot||_\S$?
\end{problem}

Continuing with our hypothetical run-time analysis, how would we
assert $||\rho||_\S\leq 1+2\eta$?  The actual algorithm returns an
approximation $x$ such that $||\rho||_\gamma\leq x\leq
(k/(k-1))^4||\rho||_\gamma$.  Let us assume that a modification of
the algorithm which computes $||\rho||_\S$ would do the same.  If
the modified algorithm returns a number that is less than 1, then
we know that $||\rho||_\S\leq 1$.  Otherwise, all that we need is
an upper bound $\Abs$ on the \emph{absolute} error of the
computation of $||\rho||_\S$, since, if $\Abs\leq\eta$, then we
can comfortably conclude that either $||\rho||_\S\leq 1+2\eta$, or
$||\rho||_\S>1$. Using the canonical basis $\mathcal{B}$ of
$\hermops$ described in Section \ref{sec_ConvexBodyProblems}, we
have $\M:=\max_{\rho\in\densops}||\rho||_\S\in
O(\poly(\dima,\dimb))$, which says the absolute error
$||\rho||_\S((k/(k-1))^4-1)$ is upper-bounded by $\Abs\in
O(((k/(k-1))^4-1)\poly(\dima,\dimb))$. The requirement $\Abs<\eta$
leads to a lower bound for $k$ of
\begin{eqnarray}
k > \frac{\M^{1/4}}{(\eta + \M)^{1/4}-\M^{1/4}}.
\end{eqnarray}

Rudolph \cite{qphRud02} has also shown that, for
$\rho\in\densops$,
\begin{eqnarray}\label{eqn_GammaNorm}
R(\rho)\geq ||\rho||_\gamma -1.
\end{eqnarray}
If equality holds in equation (\ref{eqn_GammaNorm}), then an
argument similar to the one above could be used. Rudolph notes
that equality holds for pure states and ``Werner'' and
``isotropic'' states (see \cite{VW02}).

\subsection{Fixed-point iterative method}\label{sec_Zapatrin}

Zapatrin \cite{qphZap05c} suggests an iterative method that solves
the separability problem.\footnote{Facts about iterative methods:
First, the basic Newton-Raphson method in one variable. Suppose
$\xi$ is a zero of a function $f:\mathbb{R}\rightarrow\mathbb{R}$
and that $f$ is twice differentiable in a neighbourhood $U(\xi)$
of $\xi$.  Then the Taylor expansion of $f$ about $x_0\in U(\xi)$
gives
\begin{eqnarray}
 0 =f(\xi) &=& f(x_0)+(\xi-x_0)f'(x_0)+ \cdots\\
&=& f(x_0)+(\tilde{\xi}-x_0)f'(x_0),
\end{eqnarray}
where $\tilde{\xi}= x_0 -f(x_0)/f'(x_0)$ is an approximation of
$\xi$. Repeating the process, with a truncated Taylor expansion of
$f$ about $\tilde{\xi}$, gives a different approximation
$\tilde{\tilde{\xi}}=\tilde{\xi}-f(\tilde{\xi})/f'(\tilde{\xi})$.
This suggests the iterative method $x_{i+1}=\Phi(x_i)$, for
$\Phi(x):=x-f(x)/f'(x)$.  If $f'(\xi)\neq 0$, the sequence
$(x_i)_i$ converges to $\xi$ if $x_0$ is sufficiently close to
$\xi$. More generally, if $\Phi(x):\mathbb{R}^n\rightarrow
\mathbb{R}^n$ is a contractive mapping on $B(x_0,r)$, then the
sequence $(x_0, \Phi(x_0),\Phi(\Phi(x_0)),\ldots)$ converges to
the unique fixed point in $B(x_0,r)$ (as long as $\Phi(x_0)\in
B(x_0,r)$) \cite{SB02}.} He defines the function
$\Phi:\hermops\rightarrow\hermops$:
\begin{eqnarray}
\Phi(X):= X + \lambda\left( \rho - \int\int
e^{\bra{\psi^\A}\otimes\bra{\psi^\B}X\ket{\psi^\A}\otimes\ket{\psi^\B}}\ketbra{\psi^\A}{\psi^\A}\otimes\ketbra{\psi^\B}{\psi^\B}d\mathbf{S}_\dima
d\mathbf{S}_\dimb\right),
\end{eqnarray}
where $\mathbf{S}_\dima$ and $\mathbf{S}_\dimb$ are the complex
origin-centred unit spheres (containing, respectively,
$\ket{\psi^\A}$ and $\ket{\psi^\B}$), and $\lambda$ is a constant
dependent on the derivative (with respect to $X$) of the quantity
in parentheses ($\lambda$ is chosen so that $\Phi$ is a
contraction mapping). In earlier work \cite{qphZap04, qphZap05a,
qphZap05b}, Zapatrin proves that any state $\sigma$ in the
interior $\sep^\circ$ of $\sep$ may be expressed
\begin{eqnarray}
 \sigma=\int\int
e^{\bra{\psi^\A}\otimes\bra{\psi^\B}X_\sigma\ket{\psi^\A}\otimes\ket{\psi^\B}}\ketbra{\psi^\A}{\psi^\A}\otimes\ketbra{\psi^\B}{\psi^\B}d\mathbf{S}_\dima
d\mathbf{S}_\dimb \in \sep,
\end{eqnarray}
for some Hermitian $X_\sigma$. Thus the function $\Phi$ has a
fixed point $X_\rho=\Phi(X_\rho)$ if and only if
$\rho\in\sep^\circ$. When $\rho\in\sep^\circ$, then a
neighbourhood (containing $0$) in the domain of $\Phi$ can be
found where iterating $X_{i+1}:=\Phi(X_i)$, starting at $X_0:=0$,
will produce a sequence $(X_i)_i$ that converges to $X_\rho$ when
$\rho\in\sep^\circ$, but diverges otherwise.

Each evaluation of $\Phi(X)$ requires $\dima^2\dimb^2/2
+\dima\dimb$ integrations of the form
\begin{eqnarray}
\int\int
e^{\bra{\psi^\A}\otimes\bra{\psi^\B}X\ket{\psi^\A}\otimes\ket{\psi^\B}}
\braket{\textbf{e}^\A_j}{\psi^\A}\braket{\textbf{e}^\B_{j'}}{\psi^\B}
\braket{\psi^\A}{\textbf{e}^\A_k}\braket{\psi^\B}{\textbf{e}^\B_{k'}}
d\mathbf{S}_\dima d\mathbf{S}_\dimb,
\end{eqnarray}
where $\{\textbf{e}^\A_j\}_j$ and $\{\textbf{e}^\B_k\}_k$ are the
standard bases for $\mathbb{C}^\dima$ and $\mathbb{C}^\dimb$.
However, the off-diagonal ($j\neq k$, $j'\neq k'$) integrals have
a complex integrand so are each really two real integrals; thus
the total number of real integrations is $\dima^2\dimb^2$.  Let
$\Xi_\delta$ represent the number of pure states at which the
integrand needs to be evaluated in order to perform each real
numerical integration, in order to solve the overall separability
problem with accuracy parameter $\delta$. Zapatrin shows that the
approximate number of iterations required is upper-bounded by
$2\dimb(\dimb+1)L(\log(1/\delta),\log(\dimb))$, where $L$ is a
bilinear function of its arguments.  The complexity of the entire
algorithm is roughly (ignoring $\log(\dimb)$ factors)
\begin{eqnarray}
\Xi_\delta \poly(\dima,\dimb,\log(1/\delta)).
\end{eqnarray}

In numerical integration, the final result of the integration
depends on the truncation-error at each point at which the
integrand is numerically evaluated. This is detrimental to the
complexity of Zapatrin's algorithm and I just make the reasonable
presumption that $\Xi_\delta$, whatever it is, is far greater than
$\Omega_{p}$ in the other algorithms analysed in this thesis (for
the same values of $\dima$, $\dimb$, and $\delta$). I do not
consider Monte-Carlo integration methods (i.e. methods based on
random sampling), because randomised algorithms for the
separability problem are outside our scope.

%
%

\chapter{Reduction to Entanglement Witness Search}\label{ChapterReductionToEWSearch}

In Section \ref{sec_SurveyOfAlgorithms}, we saw four proposed
algorithms for solving an approximate formulation of the quantum
separability problem, all of which have analytically bounded
running times. This chapter introduces a fifth, which is based on
the simple idea of searching for an entanglement witness for the
given state.  In the language of convex body problems set up in
Chapter \ref{ChapterSepAsDecisionProblem}, it solves the in-biased
weak separation problem for $K=\sep$:
\begin{definition}[In-biased weak separation problem ($\text{WSEP}_{\text{In}}$)]\label{def_WSEP}
Given a rational vector $p\in\mathbb{R}^n$ and rational
$\delta>0$, either
\begin{itemize}
\item assert $p\in S(K,\delta)$,
\hspace{2mm}\text{or}\label{eqn_WSEPSepAssertion}

\item find a rational vector $c\in\mathbb{R}^n$ with
$||c||_\infty= 1$ such that $c^Tx\leq c^Tp$ for every $x\in
K$\footnote{The $l_\infty$ norm appears here as a technicality, so
that $c$ need not be normalised by a possibly irrational
multiplier.  We will just use the Euclidean norm in what follows
and have $||c|| \approx 1$.}.\label{eqn_WSEPEntAssertion}
\end{itemize}
\end{definition}
\noindent Of the algorithms of the previous chapter, this fifth
algorithm is most closely related, in spirit, to Zapatrin's
algorithm of Section \ref{sec_Zapatrin}.  This is because both
algorithms reduce the quantum separability problem to
$\poly(\dima,\dimb,\log(1/\delta))$ iterations of a difficult
function evaluation: in Zapatrin's case, the difficulty is a
numerical integration; in the following algorithm, the difficulty
is the computation of a global maximum.

\section{Overview}

In Section \ref{sec_ReductionToOpt}, I explain that the quantum
separability problem can be reduced to the computation of
$b^*(A):= \max_{\sigma\in \sep} \{\tr(A\sigma)\}$ from Chapter
\ref{ChapterConvexity}.  Recall that exactly computing $b^*(A)$
corresponds precisely to the strong optimisation problem for
$K=\sep$ (see Definition \ref{def_SOPT} on page
\pageref{def_SOPT}). The main algorithm of this thesis (henceforth
referred to as ``the new algorithm'') is a new polynomial-time
reduction from $\text{WSEP}_{\text{In}}(K)$ to the (weak)
optimisation problem for $K$; the algorithm works for \emph{any}
convex set $K$ that satisfies certain conditions -- not just
$\sep$.   Section \ref{sec_DetectEntPartial} explains how such an
algorithm can be utilised in an experimental setting when faced
with the problem of deciding whether an unknown state, of which
many copies are available, is entangled; such an algorithm can be
applied to give a one-sided test for separability even when only
partial information about the state is available.

Recall that to solve $\text{WSEP}_{\text{In}}(\sep)$, in the case
where the given state $\rho$ is entangled, means to provide a
right entanglement witness that detects $\rho$.  The new algorithm
can be viewed as an exhaustive \emph{search} for an entanglement
witness for the given $\rho$; and if no entanglement witness is
found, then the algorithm concludes that $\rho$ is close to
separable.  In Section \ref{sec_SearchHeuristic}, I give the basic
idea behind the search method employed by the new algorithm. This
search method is a variant of a well-known method in convex
analysis, which I explain in Section
\ref{sec_ConnToOtherKnownMethods}.  Both search methods yield
oracle-polynomial-time reductions of the same asymptotic
complexity. I discuss the general form of such reductions in
Section \ref{sec_RemarksAboutCuttingPlaneAlgs}.  Indeed, the new
algorithm is not an improvement on known reductions of its kind.
The novelty of the work in this chapter, with regard to the
quantum information processing community, is the discovery that
the best known algorithm for the quantum separability problem (in
the case $\dima=\dimb$) is obtained by a reduction to the weak
optimisation problem over $\sep$: Section \ref{sec_NewSepAlg}
gives an upper bound on the complexity of the weak optimisation
problem over $\sep$ and Section \ref{sec_ComplexityComparison}
contains the comparison of the complexities of the new algorithm
and the algorithms of Sections \ref{sec_BasicAlgorithm} and
\ref{sec_DohertyEtalKonigRenner}. With regard to the convex
programming community, the new algorithm (whose details are
presented in Chapter \ref{ChapterAlgorithm}) is a variant of
well-known algorithms which, while perhaps not offering any
computational advantage, arguably holds intrinsic beauty because
it is based on a simple, intuitive heuristic (explained in Section
\ref{sec_SearchHeuristic}).

\section{Reduction to optimisation}\label{sec_ReductionToOpt}

Recall the function $b^*(A):= \max_{\sigma\in \sep}
\{\tr(A\sigma)\}$ from Chapter \ref{ChapterConvexity}. This
function leads naturally to an algorithm for quantum separability
as follows.  For $A\in\hermops$ such that $tr(A^2)=1$, define the
function $d_\rho (A)$ as
\begin{eqnarray}\label{eqn_dA}
d_\rho (A):= b^*(A) - tr(A\rho).
\end{eqnarray}
Geometrically, $d_\rho (A)$ is the signed distance from the state
$\rho$ to the hyperplane $\pi_{A,b^*(A)}$. It follows that $\rho$
is entangled if and only if there exists an $A$ such that $d_\rho
(A)<0$.  Any algorithm that determines whether the global minimum
(over the unit sphere $\{x\in\hermops: \tr(x^2)=1\}$)\footnote{The
minimum need only be over the $(\dima^2\dimb^2-2)$-dimensional
sphere $\{x\in\hermops: \tr(x^2)=1, \tr(x)=0\}$.  As well, as we
will see in Section \ref{sec_SearchHeuristic}, we can further
restrict to the hemisphere that has positive inner product with
$\rho$. Note that, based solely on the convexity of $\sep$,
$d_\rho(A)$ may have many local minimisers in this hemisphere.} of
$d_\rho (A)$ is negative thus solves the separability problem.
Any such algorithm would need a subroutine that approximately
computes $b^*(A)$ for any $A$. Since $b^*(A)$ is just the global
maximum of a linear functional $\tr(A\sigma)$ over all
$\sigma\in\sep$, we have reduced the approximate quantum
separability problem to the \emph{weak optimisation problem for
$K=\sep$}:
\begin{definition}[Weak optimisation problem (WOPT)]\label{def_WOPT}
Given a rational vector $c\in\mathbb{R}^n$ and rational
$\epsilon>0$, either
\begin{itemize}
\item find a rational vector $y\in\mathbb{R}^n$ such that $y\in
S(K,\epsilon)$ and  $c^Tx\leq c^Ty +\epsilon$ for every $x\in K$;
or

\item assert that $S(K,-\epsilon)$ is empty.\footnote{This will
never be the case for us, as $\sep$ is not
empty.}\label{eqn_WSEPEntAssertion}
\end{itemize}
\end{definition}

Theorem 4.4.7 from \cite{GLS88} says that $\text{WSEP}(\sep)
\leq_\T \text{WOPT}(\sep)$.  Thus, the NP-hardness of the quantum
separability problem is contained in the hardness of $b^*(A)$;
that is, WOPT($\sep$) is NP-hard.

The rest of this thesis develops an oracle-polynomial-time
algorithm for $\text{WSEP}_{\text{In}}(\sep)$ assuming an oracle
for WOPT($\sep$), which differs from those already in the
literature (as fully explained in Section
\ref{sec_ConnToOtherKnownMethods}).  In terms of attempting to
find a practical algorithm for $\text{WSEP}_\text{In}(\sep)$, the
skeptic notices that such an algorithm may not offer any advantage
over more direct or naive approaches to solving
$\text{WSEP}_{\text{In}}(\sep)$: instead of having to solve one
instance of an NP-hard problem, we now have to solve many!  We
will see at the end of this chapter that the theoretical
complexity of such an algorithm compares favourably with the
others. This is, in part, because the optimisation in $b^*(A)$
need only be carried out over the extreme points of $\sep$, which
are parametrised by only $2(\dima + \dimb) -4$ (free) variables;
the entire $\sep$ is parametrised by $\n-1$ (constrained)
variables. From a practical point of view, there are many
algorithms available for optimising functions -- far more than for
computing the separation problem. Options for computing
WOPT($\sep$) include the SDP-relaxation method of Lasserre, as in
Section \ref{subsec_EisertsEtalApproach}; Lipschitz optimisation
\cite{HP95}; and Hansen's global optimisation algorithm using
interval analysis \cite{Han92}.  I discuss the complexity of
computing WOPT($\sep$) in more detail in Section
\ref{sec_NewSepAlg}.


\section{Detecting Entanglement of an Unknown
State Using Partial Information}\label{sec_DetectEntPartial}

I now consider the task of trying to decide whether a completely
unknown physical state $\rho$, of which many copies are available,
is entangled.  For simplicity, we restrict to
$\rho\in\mathcal{H}_{2,2}$ but the discussion can be applied to a
bipartite system of any dimension, replacing Pauli operators with
canonical generators of SU(M) and SU(N) or any orthonormal
Hermitian product basis. For such $\rho$, this problem has already
been addressed in \cite{HE02}, where the so-called ``structural
physical approximation of an unphysical map'' \cite{qphHor01} was
used to implement the Peres-Horodecki positive partial transpose
(PPT) test \cite{Per96,HHH96}.  While the structural physical
approximation is experimentally viable in principle, it is very
difficult to do so. Thus, the easiest way to test for entanglement
at present is to perform ``state tomography'' in order to get good
estimates of 15 real parameters that define $\rho$, then
reconstruct the density matrix for $\rho$ and carry out the PPT
test on this matrix.

An experimentalist has many choices of which 15 parameters to
estimate: the expectations of any 15 linearly independent
observables qualify, as do the probability distributions of any 5
mutually unbiased (four-outcome) measurements \cite{Iva81,WF89}.
Whatever 15 parameters are chosen, we assume that the basic tool
of the experimentalist is the ability to perform local two-outcome
measurements on each qubit, e.g. measuring $\sigma_1$ on the first
qubit and $\sigma_2$ on the second.  Under this assumption, the
scenario where the two qubits of $\rho$ are far apart is easily
handled if classical communication is allowed between the two
labs.  We further assume, for simplicity, that the set of these
local two-outcome measurements is the set of Pauli operators
$\{\sigma_i\}_{i=0,1,2,3}$ (defined on page
\pageref{eqn_PauliOperators}). If $\sigma_i$ is measured on the
first qubit and $\sigma_j$ on the second, repeating this procedure
on many copies of $\rho$ gives good estimations of the three
expectations $\langle\sigma_i\otimes\sigma_0\rangle$,
$\langle\sigma_0\otimes\sigma_j\rangle$, and
$\langle\sigma_i\otimes\sigma_j\rangle$ (where the subscript
``$\rho$'' is omitted for readability). Let us call this procedure
\emph{measuring $\sigma_i\sigma_j$}.

Suppose the experimentalist sets out to solve our problem and
begins the data collection by measuring $\sigma_1\sigma_1$ and
then $\sigma_2\sigma_2$.  Even though only 6 of the 15 independent
parameters defining $\rho$ have been found, the example in Section
\ref{sec_NoisyBell} shows that $\rho$ is entangled if one of the
four inequalities (\ref{ineq_SuffEnt}) is true. It is
straightforward to show that if none of these inequalities is
true, then no entanglement witness in the span of
$\{\sigma_1\otimes\sigma_1,\sigma_2\otimes\sigma_2\}$ can detect
$\rho$ if it is entangled.\footnote{To show this, it suffices to
find four separable states whose projections onto
$\sp\{\sigma_1\otimes\sigma_1,\sigma_2\otimes\sigma_2\}$ are the
four vertices of the square with vertices $(\frac{1}{2},0)$,
$(0,\frac{1}{2})$, $(-\frac{1}{2},0)$, and $(0,-\frac{1}{2})$;
such states are $\frac{1}{4}I \pm
\frac{1}{2}\sigma_i\otimes\sigma_i$ for $i=1,2$. The result then
follows from convexity of $\mathcal{S}_{2,2}$.} However, there may
be an entanglement witness in the span of
\begin{eqnarray}\nonumber
\{\sigma_0\otimes\sigma_1, \sigma_0\otimes\sigma_2,
\sigma_1\otimes\sigma_1, \sigma_2\otimes\sigma_2,
\sigma_1\otimes\sigma_0, \sigma_1\otimes\sigma_0\}
\end{eqnarray}
that does detect $\rho$.\footnote{The idea of searching for an
entanglement witness in the span of operators whose expected
values are known was discovered independently and applied, in a
special case, to quantum cryptographic protocols in \cite{CLL04}.}

More generally, at any stage of the data-gathering process, if we
have the set of expectations
$\{\langle\sigma_i\otimes\sigma_j\rangle: (i,j)\in T\}$, then
$\rho$ is entangled if there is an entanglement witness in the
span of $\{\sigma_i\otimes\sigma_j: (i,j)\in T\}$ that detects
$\rho$ ($T\subset \{(k,l):k,l\in\{0,1,2,3\}\}\setminus (0,0)$). If
the experimentalist has access to a computer program that can
quickly discover such an entanglement witness (if it exists), then
the data-gathering process can be terminated early and no more
qubits have to be used to detect that $\rho$ is entangled. The new
algorithm is just such a program.  To see this, note that the
projection $\overline{\mathcal{S}_{2,2}}$ of $\mathcal{S}_{2,2}$
onto $\sp\{\sigma_i\otimes\sigma_j: (i,j)\in T\}$ is a
full-dimensional convex subset of $\mathbb{R}^{|T|}$, and the
projection $\overline{\rho}$ of $\rho$ onto
$\sp\{\sigma_i\otimes\sigma_j: (i,j)\in T\}$ is a point in
$\mathbb{R}^{|T|}$ such that
$\overline{\rho}\notin\overline{\mathcal{S}_{2,2}}$ if and only if
there is an entanglement witness in the span of
$\{\sigma_i\otimes\sigma_j: (i,j)\in T\}$ that detects $\rho$.
Since the new algorithm can be applied to any full-dimensional
convex set (satisfying certain conditions), we can apply it to
$\overline{\mathcal{S}_{2,2}}$.

We view the new algorithm as an extra tool that an experimentalist
can use to facilitate entanglement detection and minimise the
number of copies of $\rho$ that must be measured -- essentially,
trading classical resources for quantum resources. In Section
\ref{sec_Application}, I detail how the new algorithm is applied
to this experimental scenario.

\section{New method to solve separation with optimisation}\label{sec_SearchHeuristic}

Now we shed the quantum physical notation, in favour of the
simpler and more general convex analysis notation.  To reconcile
the two notations, recall the discussion at the beginning of
Section \ref{sec_ConvexBodyProblems} that relates the trace inner
product in $\hermops$ to the dot product in $\mathbb{R}^{\n-1}$
and explains that $\sep$ may be viewed as a convex subset of
$\mathbb{R}^{\n-1}$ that properly contains the origin (which
corresponds to the maximally mixed state $I_{\dima,\dimb}$).

So, assume we have a full-dimensional convex set $K\subset
\mathbb{R}^n$ that properly contains the origin.  The ultimate
goal is to develop a new algorithm for $\text{WSEP}_\text{In}$,
given an oracle for WOPT($K$).  Until Chapter
\ref{ChapterAlgorithm}, we ignore the weakness of the separation
and optimisation problems, as it obfuscates the main idea; that
is, we assume we are solving SSEP($K$) with an oracle for
SOPT($K$).

\begin{figure}[ht]
\centering \resizebox{250mm}{!}{\includegraphics{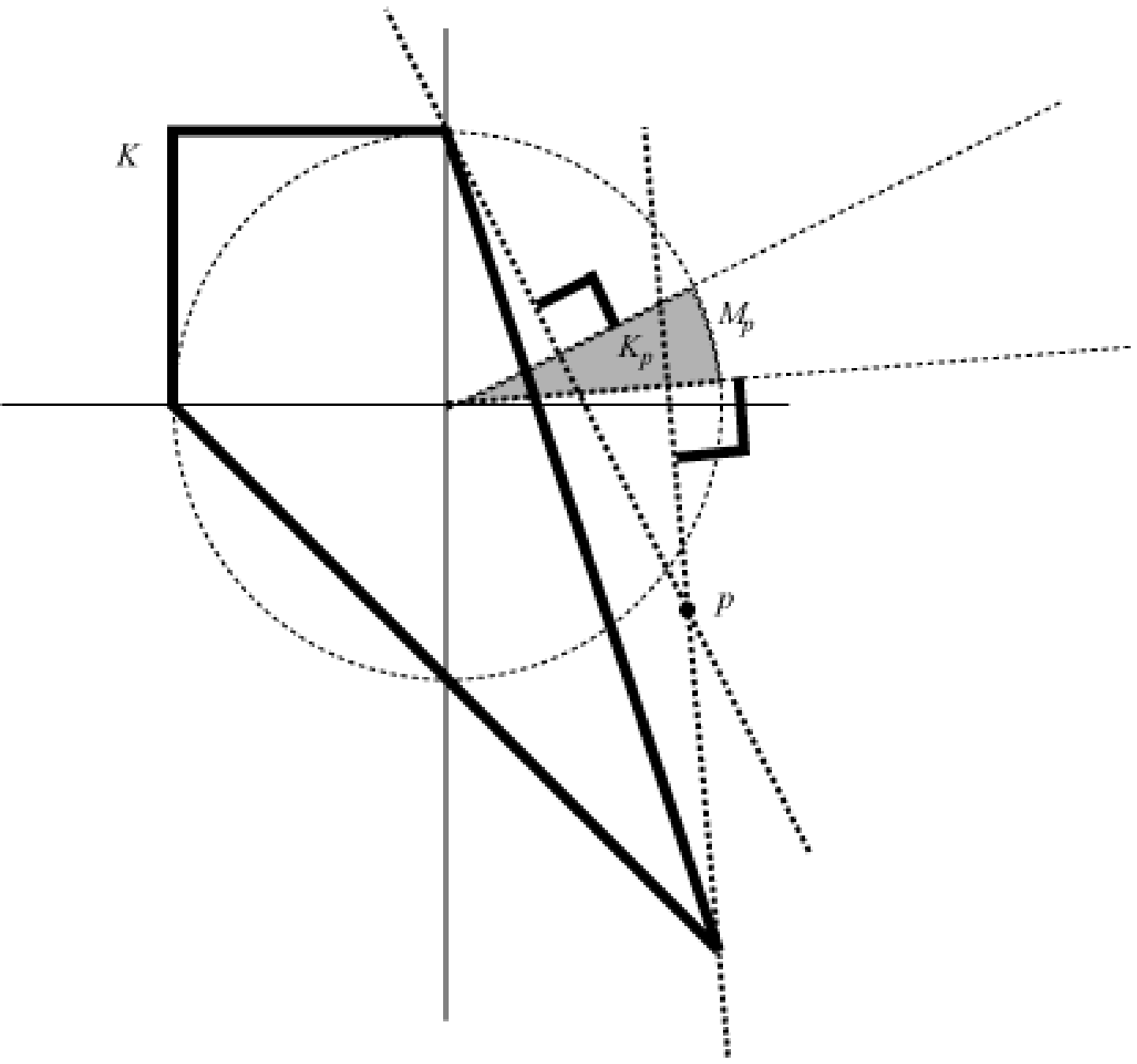}}
\caption[The sets $M_p$ and $K_p$ in $\mathbb{R}^2$]{The sets
$M_p$ and $K_p$ in $\mathbb{R}^2$.  Pictured in heavy outline is a
set $K$ in $\mathbb{R}^2$, where $K:=\conv\{(0,1), (-1,1), (-1,0),
(1,-2)\}$.  A point $p=(-7/8,-3/4)$ is shown as a heavy dot.  The
unit circle is drawn in a dashed line.  The set $M_p$ is the arc
of the unit circle that the shaded pie-slice subtends; the set
$K_p$ is the shaded pie-slice.  In two dimensions, the set $M_p$
($K_p$) is easy to construct.  This construction has been
illustrated: draw the two distinct lines through $p$ that are
tangent to $K$; the lines that determine the pie-slice are the two
straight lines that are perpendicular to the lines through $p$.
The idea behind this geometrical construction easily generalises
to $\mathbb{R}^3$.} \label{K_p}
\end{figure}

Suppose we have an oracle $\OSOPTK$ for the optimisation problem
over $K$ such that, given a nonzero input vector $c$, $\OSOPTK$
outputs a point $\OSOPTK(c)\equiv k_c\in K$ that maximises $c^Tx$
for all $x\in K$. An important step in developing the algorithm is
noting that, given $\OSOPTK$, the search for a separating
hyperplane reduces to the search for a region on the
$(n-1)$-dimensional surface of the \emph{unit hypersphere} $S_n$
(embedded in $\mathbf{R}^n$) centered at the origin. For $p\notin
K$, this region $M_p$ is simply $\{c\in S_n:\hspace{2mm} c^T
k_c<c^T p\}$ (see Figure \ref{K_p}).

The first observation is that, since $K$ properly contains the
origin, $M_p$ is contained in the hemisphere defined by $\{x: p^T
x\geq 0\}$:
\begin{fact}\label{Fact_InitialCut} For all $m\in M_p$, $m^Tp>0$.
\end{fact}
\begin{proof}
Let $m\in M_p$.  Then $m^Tp>m^Tk$ for all $k\in K$.  But the fact
that the 0-vector is properly contained in $K$ implies that there
exists $k\in K$ such that $m^Tk>0$.\end{proof}

\begin{figure}[ht]
\centering \resizebox{250mm}{!}{\includegraphics{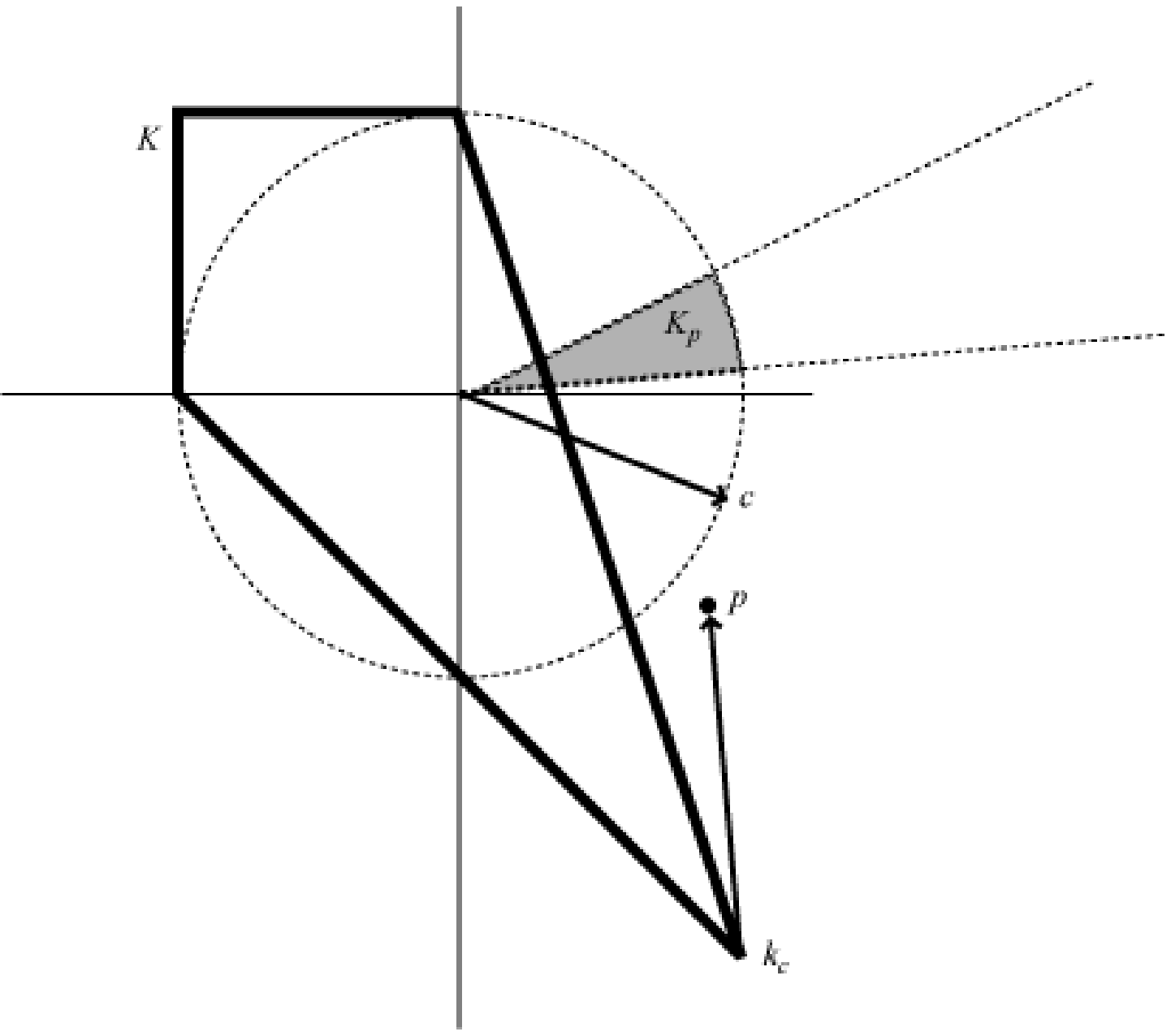}}
\caption[Illustration of intuitive heuristic behind Lemma
\ref{lem_Donny}]{Illustration of intuitive heuristic behind Lemma
\ref{lem_Donny}. Continuing from Figure \ref{K_p}, the unit vector
$c$ is a test vector that is close to $M_p$ but not in $M_p$.
Evidently, adding a component of $(p-k_c)$ to $c$ moves it closer
to $M_p$.} \label{DonnyLemma}
\end{figure}

The second observation, Lemma \ref{lem_Donny}, is based on the
following heuristic, which can be pictured in $\mathbb{R}^2$ and
$\mathbb{R}^3$. Suppose $c$, $||c||=1$, is not in $M_p$ (but is
reasonably close to $M_p$) and that the oracle returns $k_c$. What
is a natural way to modify the vector $c$, so that it gets closer
to $M_p$? Intuition dictates moving $c$ \emph{away from} $k_c$ and
\emph{towards} $p$, that is, add a small component of the vector
$(p-k_c)$ to $c$, in order to generate a new guess
$c'=c+\lambda(p-k_c)/||p-k_c||$, for some $\lambda>0$, which we
could then give to the oracle again (see Figure \ref{DonnyLemma}).
Incidentally, I have found that this heuristic actually works: the
following little program, in the context of the quantum
separability problem, always found entanglement witnesses for
entangled states in $\mathbb{H}_{2,2}$, even with very tiny
entanglement concurrence \cite{Woo98} (the value of $N'$ required
depends on the concurrence):

\vspace{2mm} \fbox{\begin{minipage}{13cm}
\begin{description} \item
$c:=p/||p||;\hspace{2mm}d:=1;\hspace{2mm}i:=0;$ \item
\textsc{while}\hspace{2mm}$(d>0$\hspace{2mm}\textsc{and}\hspace{2mm}$i<N')$\hspace{2mm}\textsc{do}\hspace{2mm}$\lbrace$
\item\hspace{5mm}$k_c:=\OSOPTK(c);$
\item\hspace{5mm}$d:=c^Tk_c-c^Tp;$
\item\hspace{5mm}\textsc{if}$\hspace{2mm}(d<0)$\hspace{2mm}\textsc{then}\hspace{2mm}$\lbrace$
\textsc{return} $c$ $\rbrace$
\item\hspace{10mm}\text{\textsc{else}\hspace{2mm}$\lbrace$ $c:=
c+d(p-k_c)/||p-k_c||$};\hspace{2mm}$c:=c/||c||$;\hspace{2mm}$i:=i+1\rbrace\rbrace;$
\item\text{\textsc{return} ``INCONCLUSIVE''}
\end{description}
\end{minipage}}
\vspace{2mm}

\noindent Notice the connection of the above program to the
function $d_\rho(A)$ of Section \ref{sec_ReductionToOpt}. This
program can be regarded as an \emph{extremely} simple heuristic
algorithm for the separation problem when given an optimisation
oracle and promised that $p\notin K$ (of course, it may give
inconclusive results; in practice, one should set $N'$ as large as
is practically feasible).

Interestingly, the above heuristic can be formalised as follows.
If $c$ is not in $M_p$ but is sufficiently close to $M_p$, then
$c$, $p$, and $k_c$ can be used to define a hemisphere which
contains $M_p$ and whose great circle cuts through $c$. More
precisely:
\begin{lemma}\label{lem_Donny} Suppose $m\in M_p$,
$c\notin M_p$, and let $\bar{a}:=(p-k_c)-\mathrm{Proj}_c(p-k_c)$.
If $m^Tc\geq 0$ then $m^T\bar{a}> 0$.
\end{lemma}
\begin{proof}
Note that $m^T\bar{a} = m^T(p-k_c)-[c^T(p-k_c)](m^Tc)$.  The
hypotheses of the lemma immediately imply that $m^T(p-k_c)>0$ and
$c^T(p-k_c)\leq 0$.  Thus, if $m^Tc\geq 0$, then $m^T\bar{a}> 0$.
\end{proof}
\noindent The lemma gives a method for reducing the search space
after each query to $\OSOPTK$ by giving a \emph{cutting plane},
$\{x:\bar{a}^Tx=0\}$, that slices off a portion of the search
space. The idea is that at each iteration a vector $c\in S_n$ is
chosen that is approximately in the centre of the remaining search
space. Then $c$ is given to the oracle which returns $k_c$. If
$c^T p>c^T k_c$, then a separating hyperplane for $p$ has been
found and the algorithm terminates. Otherwise, as long as $m^T
c\geq 0$ \emph{for all} $m\in M_p$, the lemma says that the
current search space may be sliced through its centre $c$ and the
origin, and one half discarded. Because the search space is being
approximately halved at each step, the algorithm quickly either
finds a separating hyperplane for $p$ or concludes that $p\in K$.

The above search problem can easily be reduced to an instance of
the convex \emph{feasibility problem}:

\smallskip\noindent\textbf{Feasibility Problem:} Given a convex set
$K'\subset\mathbf{R}^n$, either \vspace{-1mm}\begin{enumerate}
\item find a point $k'\in K'$, or \item assert that $K'$ is empty.
\end{enumerate}
\noindent In this case, the convex set $K'$ is the set $K_p$ which
is defined as
\begin{eqnarray}\label{def_K_p}
K_p:=\left[\mathrm{ConvexHull}\left( M_p\cup \{\bar{0}\}\right)
\right] \setminus \{\bar{0}\},
\end{eqnarray}
where $\bar{0}\in \mathbf{R}^n$ denotes the origin.  The set
$K_p$, if not empty, can be viewed as a cone-like object,
emanating from the origin and cut off by the unit hypersphere (see
Figure \ref{K_p}). Several well-known oracle-polynomial-time
algorithms exist for the feasibility problem for $K'$ in the case
where there is a separation oracle for $K'$ that, given a test
point $y\in\mathbf{R}^n$, returns either a hyperplane that
separates $y$ from $K'$ or asserts that $y\in K'$. The oracle
$\OSOPTK$, along with Lemma \ref{lem_Donny}, essentially gives a
separation oracle for $K_p$, as long as the test vectors $c$ given
to $\OSOPTK$ satisfy $m^T c\geq 0$ for all $m\in M_p$. Because of
this last requirement, none of the existing algorithms can be
applied directly. However, the analytic-center algorithm due to
Atkinson and Vaidya \cite{AV95} beautifully lends itself to a
modification that allows the requirement $m^T c\geq 0$ for all
$m\in M_p$ to be satisfied.  I will say more about such algorithms
in Section \ref{sec_RemarksAboutCuttingPlaneAlgs}.

Finding a vector in $M_p$ and finding a nonzero point in $K_p$ are
equivalent for our purpose.  From now on, we regard the ``search
space'' as the full-dimensional origin-centred hyperball $B_n$ in
$\mathbf{R}^n$; however, to make the analysis more transparent, we
will always normalise each test point before giving it to the
oracle.

How could we ensure that all our test vectors $c$ satisfy $m^T
c\geq 0$ for all $m\in M_p$?  Recall Fact \ref{Fact_InitialCut},
which says that the set $K_p$ is contained in the halfspace
$\lbrace x:p^T x\geq 0\rbrace$. Let $a_1:=p/||p||$. Thus, straight
away, the search space is reduced to the hemisphere
$B_n\cap\lbrace x:a_1^T x\geq 0\rbrace$. The first test vector to
give to the oracle $\OSOPTK$ is $p/||p||$, which clearly has
nonnegative dot-product with all points in $K_p$ and hence all
$m\in M_p$.  By way of induction, assume that, at some later stage
in the algorithm, the current search space has been reduced to
$P:=B_n\bigcap \cap_{i=1}^{h}\lbrace x:a_i^T x\geq b_i \rbrace$ by
the generation of cutting planes $\lbrace x: a_i^T x=b_i \rbrace$,
where the $a_i$, for $i=2,3,\ldots, h$, are the normalised
$\bar{a}$ from $h-1$ invocations of Lemma \ref{lem_Donny}.  Let
$\omega$ be the ``centre'' of $P$, and suppose that this
``centre'' is a \emph{positive linear combination of the normal
vectors} $a_i$, that is,
\begin{eqnarray}\label{eqn_CentreIsConicCombinationOfNormals}
\omega = \sum_{i=1}^h \lambda_i a_i,\hspace{2mm}\text{where
$\lambda_i\geq 0$ for all $i=1,2,\ldots, h$.}
\end{eqnarray}
Then, by inductive hypothesis, this implies that $m^T \omega\geq
0$ for all $m\in M_p$. Thus, $c:=\omega/||\omega||$ is a suitable
vector to give to the oracle $\OSOPTK$ and use in Lemma
\ref{lem_Donny}. Therefore, it suffices to find a definition of
``centre $\omega$ of $P$'' that satisfies
(\ref{eqn_CentreIsConicCombinationOfNormals}), in order that all
our test vectors $c$ satisfy $m^T c\geq 0$ for all $m\in M_p$.

Reducing the separation problem for $K$ to the convex feasibility
problem for some $K'$, while using the optimisation oracle for $K$
as a separation oracle for $K'$, is not a new concept in convex
analysis.  But the precise way that Lemma \ref{lem_Donny}
generates each new cutting plane, incorporating the intuitive
correction heuristic, does not appear in the literature.  This is
likely because there is a well-known, standard way to carry out
such a reduction, which I cover in the next section.

\section{Connection to standard method}\label{sec_ConnToOtherKnownMethods}

The standard way to perform the reduction of the last section may
be found in the synthesis of Lemma 4.4.2 and Theorem 4.2.2 in
\cite{GLS88}.

\begin{definition}[Polar of $K$] The \emph{polar} $K^\star$ of a full-dimensional convex set
$K\subset \mathbb{R}^n$ that contains the origin is defined
as\footnote{In some textbooks, e.g. \cite{NW88}, $K^\star$ is
called the ``1-polar''.}
\begin{eqnarray}
K^\star := \{c\in\mathbb{R}^n: c^Tx\leq 1 \hspace{2mm}\forall x
\in K\}.
\end{eqnarray}
\end{definition} \noindent If $c\in K^\star$, then the plane $\pi_{c,1}\equiv\{x:
c^Tx=1\}$ separates $p\in\mathbb{R}^n$ from $K$ when $c^Tp >1$.
Thus, the separation problem for $p$ is equivalent to the
feasibility problem for $Q_p$, defined as\footnote{Note that $Q$
is guaranteed not to be empty when $p\notin K$.  For, then, there
certainly exists some plane $\pi_{c',b'}$ separating $p$ from $K$.
But since $K$ contains the origin, $b'$ may be taken to be
positive. Thus $\pi_{c'/b', 1}$ separates $p$ from $K$.}
\begin{eqnarray}
Q_p := K^\star \cap \{c: p^Tc \geq 1\}.
\end{eqnarray}
As mentioned in the previous section (and elaborated on in the
next section), to solve the feasibility problem for any $K'$, it
suffices to have a separation routine for $K'$.  Because we can
easily build a separation routine $\OSSEPQp$ for $Q_p$ out of
$\OSSEPKstar$, it suffices to have a separation routine
$\OSSEPKstar$ for $K^\star$ in order to solve the feasibility
problem for $Q_p$.\footnote{I slightly abuse the oracular
``$\mathcal{O}$'' notation, introduced in Section
\ref{sec_ReviewNPCness}, by using it for both truly oracular
(black-boxed) routines and for other (possibly not completely
black-boxed) routines.}  Building $\OSSEPQp$ out of
$\OSSEPKstar$ is done as follows:\\

\indent\indent\indent\indent\indent\indent\indent Routine $\OSSEPQp(y)$:\\
\indent\indent\indent\indent\indent\indent\indent \textsc{case:} $p^Ty < 1$\\
\indent\indent\indent\indent\indent\indent\indent\indent \textsc{return} $-p$\\
\indent\indent\indent\indent\indent\indent\indent \textsc{else:} $p^Ty \geq 1$\\
\indent\indent\indent\indent\indent\indent\indent\indent \textsc{call} $\OSSEPKstar(y)$\\
\indent\indent\indent\indent\indent\indent\indent\indent \textsc{case:} $\OSSEPKstar(y)$ returns separating vector $q$\\
\indent\indent\indent\indent\indent\indent\indent\indent\indent \textsc{return} $q$\\
\indent\indent\indent\indent\indent\indent\indent\indent \textsc{else:} $\OSSEPKstar(y)$ asserts $y\in K^\star$\\
\indent\indent\indent\indent\indent\indent\indent\indent\indent \textsc{return} ``$y\in Q$''\\

\noindent It remains to show that the optimisation routine
$\OSOPTK$ for $K$ gives a separation routine $\OSSEPKstar$ for
$K^\star$. Suppose $y$ is given to $\OSOPTK$, which returns $k\in
K$ such that $y^Tx \leq y^Tk =:b$ for all $x\in K$. If $b\leq 1$,
then $\OSSEPKstar$ may assert $y\in K^\star$. Otherwise,
$\OSSEPKstar$ may return $k$, because $\pi_{k,1}$ (and hence
$\pi_{k,b}$) separates $y$ from $K^\star$: since $k^Ty=b > 1$, it
suffices to note that $k^Tc=c^Tk\leq 1$ for all $c\in K^\star$ by
the definition of $K^\star$ and the fact that $k\in K$.

\begin{figure}[ht]
\centering \resizebox{250mm}{!}{\includegraphics{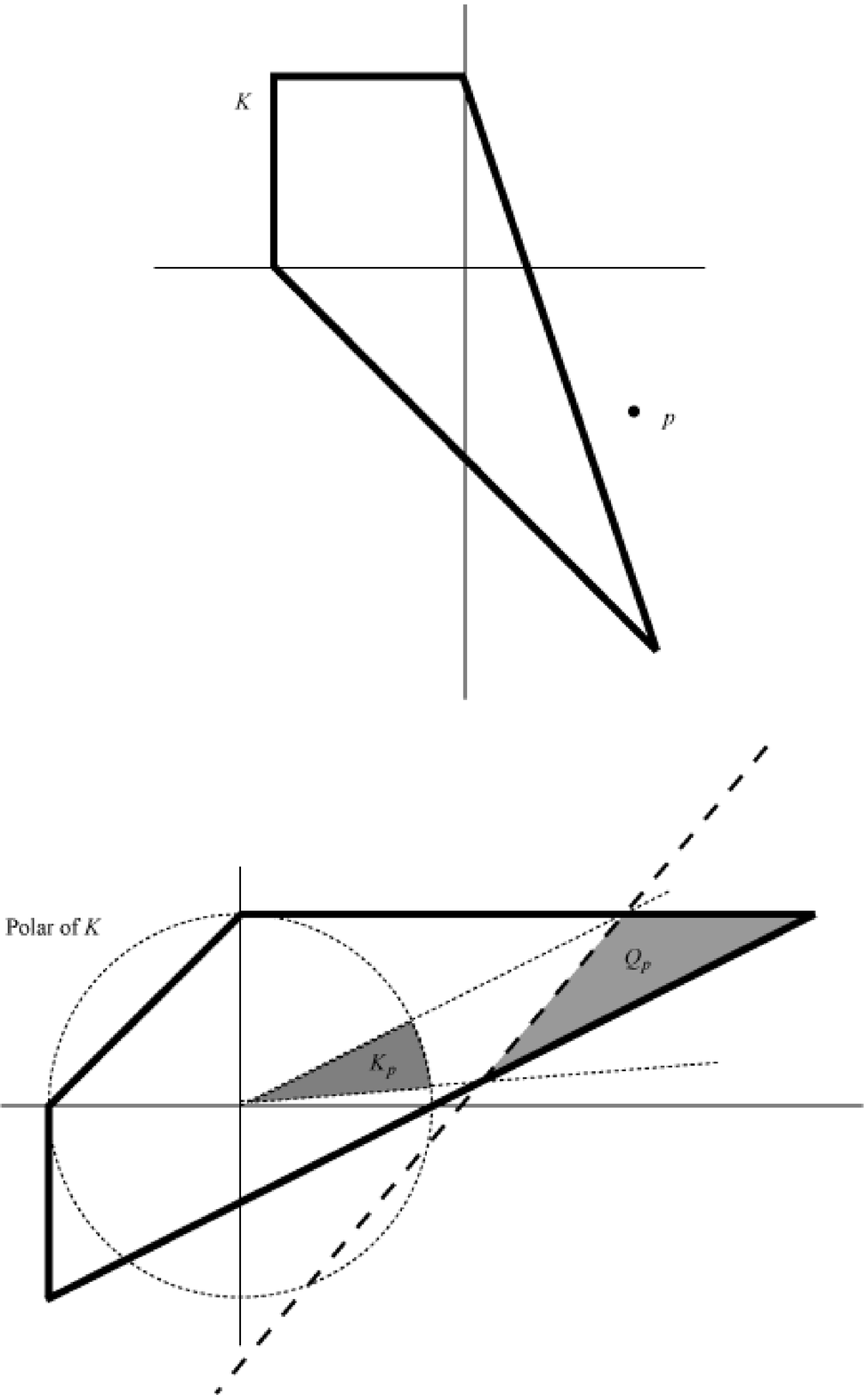}}
\caption[Relationship between convex separation methods]{The upper
picture is a set $K$ in $\mathbb{R}^2$, where  $K:=\conv\{(0,1),
(-1,1), (-1,0), (1,-2)\}$.  A point $p=(-7/8,-3/4)$ is shown.  The
polar $K^\star$ of $K$ is shown in heavy outline in the lower
picture; $K^\star = \conv \{(0,1),(-1,0),(-1,-1),(3,1)\}$.  The
set $Q_p$ is the shaded polytope, bounded by the long-dashed plane
$\{c: p^Tc =1\}$. The set $K_p$ is the shaded pie-slice and is the
radial projection of $Q_p$ onto the origin-centred unit ball
(whose boundary is shown as a short-dashed circle).  The
particular $K$ and $K^\star$ are taken from \cite{NW88}.}
\label{Connection}
\end{figure}

Figure \ref{Connection} shows the relationship between the method
of Section \ref{sec_SearchHeuristic} and the above method, by
illustrating that the set $K_p$ (defined in (\ref{def_K_p})) is
just the radial projection of $Q_p$ onto $B_n$.  Thus,
unsurprisingly, both methods test the feasibility of virtually the
same thing.  The novelty of the method of Section
\ref{sec_SearchHeuristic} lies in the \emph{way} the cutting
planes are generated.

\section{Cutting-plane algorithms for convex feasibility for
$K'$}\label{sec_RemarksAboutCuttingPlaneAlgs}

Some remarks about convex feasibility cutting-plane algorithms for
$K'\subset\mathbb{R}^n$, relative to a separation oracle
$\OSSEPKprime$, are in order. All such algorithms have the same
basic structure:

\begin{enumerate}
\item Define a (possibly very large) regular bounded convex set
$P_0$ which is guaranteed to contain $K'$, such that, for some
reasonable definition of ``centre'', the centre $\omega_0$ of
$P_0$ is easily computed.  The set $P_0$ is called an \emph{outer
approximation} to $K'$.  Common choices for $P_0$ are the
origin-centred hyperbox, $\{x\in\mathbb{R}^n: -2^L \leq x_i \leq
2^L, \hspace{2mm}1\leq i\leq n \}$ and the origin-centred
hyperball, $\{x: x^Tx \leq 2^L\}$ (where $2^L$ is a trivially
large bound).

\item Give the centre $\omega$ of the current outer approximation
$P$ to $\OSSEPKprime$.

\item If $\OSSEPKprime$ asserts ``$\omega\in K'$'', then HALT.

\item Otherwise, say $\OSSEPKprime$ returns the hyperplane
$\pi_{c,b}$ such that $K'\subset \{x: c^Tx \leq b\}$.  Update
(shrink) the outer approximation $P:=P\cap \{x: c^Tx \leq b'\}$
for some $b'\geq b$. Possibly perform other computations to
further update $P$. Check stopping conditions; if they are met,
then HALT. Otherwise, go to step (ii).
\end{enumerate}
\noindent  The difficulty with such algorithms is knowing when to
halt in step (iv).  Generally, the stopping conditions are related
to the size of the current outer approximation.  Because it is
always an approximate (weak) feasibility problem that is solved,
the associated accuracy parameter $\delta$ can be exploited to get
a ``lower bound'' $V$ on the ``size'' of $K'$, with the
understanding that if $K'$ is smaller than this bound, then the
algorithm can correctly assert that $S(K',-\delta)$ is empty. Thus
the algorithm stops in step (iv) when the current outer
approximation is smaller than $V$.

The cutting-plane algorithm is called \emph{(oracle-)
polynomial-time} if it runs in time \\
$O(\poly(n,\log(1/\delta)))$ with unit cost for the oracle.  It is
called \emph{(oracle-) fully polynomial} if it runs in time
$O(\poly(n,1/\delta))$.  This thesis is concerned primarily with
polynomial-time cutting-plane algorithms.

Using the standard cut-generation rule, there are a number of
polynomial-time convex feasibility algorithms that can be applied
(see \cite{AV95} for a discussion of all of them).  The three most
important are the \emph{ellipsoid method}, the \emph{volumetric
centre method}, and the \emph{analytic centre method}.  The
ellipsoid method has $P_0=\{x:x^Tx\leq 2^L\}$ and is the only one
which requires ``further update'' of the outer approximation $P$
in step (iv) after a cut has been made -- a new minimal-volume
ellipse is drawn around $P:=P\cap \{x: c^Tx \leq b'\}$.  The
ellipsoid method, unfortunately, suffers badly from gigantic
precision requirements, making it unusable in practice.  The
volumetric centre and analytic centre algorithms are more
efficient than the ellipsoid algorithm and are very similar to
each other in complexity and precision requirements, with the
analytic centre algorithm having some supposed practical
advantages.\footnote{To date, no one has implemented a
polynomial-time cutting plane algorithm. For an implementation of
a fully polynomial algorithm, see
http://ecolu-info.unige.ch/logilab.}

The cutting plane $\{x: c^Tx = b'\}$ requires further definition:
\begin{equation}
\text{If } \left\{ \begin{array}{rcl}  b'&<& c^T\omega\\
b'&=& c^T\omega
\\  b'&>& c^T\omega\end{array}\right\} \text{ then the above is a } \left\{ \begin{array}{c} \text{\emph{deep-cut}} \\
\text{\emph{central-cut}}
\\ \text{\emph{shallow-cut}} \end{array}\right\}\text{ algorithm.}
\end{equation}
\noindent Intuitively, deep-cut algorithms should be fastest.
Ironically, though, except for the case of ellipsoidal algorithms
(which are practically inefficient), the algorithms that are
provably polynomial-time are central- or even shallow-cut
algorithms. For instance, even though $\OSSEPKstar$, built on
$\OSOPTK$, gives deep cuts $\pi_{k,1}$, it is not known how to
utilise the deep cuts to get a polynomial-time algorithm using
analytic or volumetric centers.  Note that the new cut-generation
method in Section \ref{sec_SearchHeuristic} is capable only of
giving central cuts; but this does not, \emph{a priori}, put it at
any disadvantage (relative to the standard cut-generation method)
with regard to polynomial-time analytic or volumetric centre
algorithms. We will see in Chapter \ref{ChapterAlgorithm} that
this new cut-generation rule indeed yields a polynomial-time
algorithm.

\section{A new quantum separability algorithm}\label{sec_NewSepAlg}

The algorithm in Chapter \ref{ChapterAlgorithm}, which is based on
analytic centres, gives a new method for solving the quantum
separability problem by solving $\text{WSEP}_\text{In}(\sep)$.  As
we will see, the number of arithmetic operations required by the
algorithm is
\begin{eqnarray}
O((T+
\dima^6\dimb^6\log(1/\delta))\dima^2\dimb^2\log^2(\dima^2\dimb^2/\delta)),
\end{eqnarray}
where $T$ is the cost of one call to the WOPT($\sep$) routine.

Now consider the complexity of computing an instance
$(A,\epsilon)$ of WOPT($\sep$). The only way to get an upper bound
on this complexity is to assume the most naive way to carry out
this computation, which is to one-by-one calculate $\tr(A\sigma)$
for each of the pure separable states $\sigma$ to a sufficiently
high precision, and then return the $\sigma$ that produced the
largest value of $\tr(A\sigma)$.

I use the same framework and notation of Section
\ref{subsubsec_ReformulationQSEP}.  Suppose
$\sigma=\alpha\alpha^\dagger\otimes\beta\beta^\dagger$ maximises
$\tr(A\sigma)$, and, as before, let $\tilde{\alpha}$ and
$\tilde{\beta}$ be the $p'$-bit truncations of $\alpha$ and
$\beta$. Let
$\gamma:=\alpha\alpha^\dagger\otimes\beta\beta^\dagger-
\tilde{\alpha}\tilde{\alpha}^\dagger\otimes\tilde{\beta}\tilde{\beta}^\dagger$.
The real coordinates of the entries of $[\gamma]$ have absolute
value no greater than $2^{-(p'-6)}$.  Since we give to the
WOPT($\sep$) routine an $A$ such that $||A||_2=1$, we have
$||A||_1\leq \sqrt{\dima\dimb}||A||_2=\sqrt{\dima\dimb}$ which,
since $A$ is normal, is equivalent to $\sum_{ij}|A_{ij}|\leq
\sqrt{\dima\dimb}$ \cite{HJ85}, where $A_{ij}$ are the entries of
$[A]$.  This gives a bound of $|A_{ij}|\leq \sqrt{\dima\dimb}$. It
follows that
\begin{eqnarray}
\tr(A(\alpha\alpha^\dagger\otimes\beta\beta^\dagger))-\tr(A(\tilde{\alpha}\tilde{\alpha}^\dagger\otimes\tilde{\beta}\tilde{\beta}^\dagger))&=&\tr(A\gamma)\\
&\leq& \dima^{2.5}\dimb^{2.5}2^{-(p'-7)}.
\end{eqnarray}
We set $p'$ such that
$\dima^{2.5}\dimb^{2.5}2^{-(p'-7)}<\epsilon$, which gives
\begin{eqnarray}
p'>\log_2\left( \frac{\dima^{2.5}\dimb^{2.5}}{\epsilon}\right)+7.
\end{eqnarray}
This gives\footnote{When looping through all the elements
$\tilde{\alpha}$ and $\tilde{\beta}$ in practice, we would skip
all $\tilde{\alpha}$ and $\tilde{\beta}$ whose norms are greater
than 1, so as not to report an inflated global maximum.}
\begin{eqnarray}
T &\backsim& \Omega_{p'}
\poly(\dima,\dimb,1/\delta)\\\label{DominantFactorNewAlgorithm}
&\lesssim&
\left(\frac{2^7\dima^{2.5}\dimb^{2.5}}{\epsilon}\right)^{2(\dima+\dimb)}\poly(\dima,\dimb,1/\delta).
\end{eqnarray}

In practice, however, it need not be so bad.  We can formulate the
optimisation problem as the (constrained or unconstrained)
maximisation of a real function $f(\sigma):=\tr(A\sigma)$ of real
variables parametrising $\sigma$, and then apply continuous
optimization methods to $f$.   Denote by $f^*$ the global maximum
of $f$.  As the global optimisation algorithm proceeds, it may
give progressively better lower and upper bounds on
$f^*$.\footnote{Upper bounds on $f^*$ are given by Hansen's
interval-analysis global optimisation algorithm \cite{Han92,
HW04}. This algorithm calculates bounds on the derivative of $f$
(over a bounded domain) in order to compute upper bounds on
$f^*$.} Call these bounds $\underline{f}$ and $\overline{f}$,
respectively.  A key advantage of the algorithm is that, during
any computation of $\mathcal{O}(A)$, the search for $f^*$ may be
halted early when either (\emph{i}) $\tr(A\rho) \leq
\underline{f}$, in which case Lemma \ref{lem_Donny} can be invoked
to generate a new cutting plane, or (\emph{ii}) $\overline{f}<
\tr(A\rho)$, in which case the algorithm has found an entanglement
witness for $\rho$. Note that lower bounds $\underline{f}$ can be
generated very quickly using \emph{local} optimisation routines
seeded at random points in the domain of $f$.  Thus, the
algorithm's run time may be significantly shorter than the
worst-case complexity of WOPT($\sep$) predicts.

\section{Complexity comparison of algorithms}\label{sec_ComplexityComparison}

All of the algorithms considered solve the weak membership problem
for $\sep$ with accuracy parameter $\delta$.  How does the new
separability algorithm of the previous section compare to the
others?

Recall the reasonable presumption that the numerical integration
in Zapatrin's algorithm (Section \ref{sec_Zapatrin}) is far more
computationally intensive than the global minimisation of the new
algorithm.  Recall also that P{\'{e}}rez-Garcia's algorithm
(Section \ref{sec_Rudolf&PerezGarcia}) is not clearly related to
the weak membership problem, barring new results about the
$||\cdot||_\gamma$-norm and robustness of entanglement.

The following table summarises the dominating factors (that are at
least factorial in $\dima$ or $\dimb$)\footnote{Recall Stirling's
approximation: $n^n \approx n!e^n/\sqrt{2\pi n}$.} in the
run-times of the new algorithm and the algorithms of Sections
\ref{sec_BasicAlgorithm} and \ref{sec_DohertyEtalKonigRenner}:

\vspace{5mm}

\begin{tabular}[h]{|l|c|}\hline
Search for separable decomposition (Section
\ref{sec_BasicAlgorithm}) & $(MN/\delta)^{O(M^3N^2 + M^2N^3
)}$\\\hline Bounded search for symmetric extensions (Section
\ref{sec_DohertyEtalKonigRenner}) & $(M/\delta)^{O(M)}$\\\hline
Search for entanglement witness (Section \ref{sec_NewSepAlg}) &
$(MN/\delta)^{O(M+N)}$
\\\hline
\end{tabular}

\vspace{5mm}

\noindent Of the three algorithms in the table, the search for
separable decompositions is, as expected, the most complex.

A few remarks are in order regarding the new algorithm and the
bounded search for symmetric extensions.  Right away, we can see
that if $\dima$ is a constant, then the bounded search for
symmetric extensions has a much lower complexity.  Note, however,
that if $\dima=\dimb$, then the two complexities, as summarised in
the table, become the same. As a related side point, note that
Gurvits \cite{Gur03} has actually shown WMEM$(\sep)_{\dima,\dimb}$
to be NP-hard when $\dima \leq \dimb \leq \dima(\dima-1)/2$; it is
an open problem as to whether, say,
WMEM$(\mathcal{S}_{2,\dimb})_{\dimb}$ is NP-hard.  So, if we want
to be absolutely sure we are solving a hard problem, we can
restrict to the case where $\dima=\dimb$. In this case, it is easy
to check that the detailed complexity estimates given previously
indicate that the new algorithm has a better complexity, even when
we take into account that the bounded search for symmetric
extensions uses the trace norm as opposed to the Euclidean norm.
Recall that the bounded search for symmetric extensions has
complexity on the order of $d^4_{S_k}$, where we can invoke the
lower bound $d_{S_{\bar{k}}}>
\left({16\dima^{11}}/{\delta^2}\right)^{\dima -1}$ from equation
(\ref{eqn_MainDohertEtalKonigRennerComplexity}) to get $d^4_{S_k}
> 2^{16\dima -4}\dima^{44\dima}/\delta^{8\dima-8}$.  But the
algorithm gets a complexity reduction for solving the weak
membership problem with respect to the trace distance instead of
the Euclidean distance. This reduction corresponds to substituting
$\dima\delta$ for $\delta$ in the above lower bound, which gives
the best known lower bound on the complexity of the bounded search
for symmetric extensions of
\begin{eqnarray}\label{LowerboundOnBoundedSearchForSymmExts}
2^{16\dima-4}\dima^{36\dima
+8}\left(\frac{1}{\delta}\right)^{8\dima -8}.
\end{eqnarray}
The dominant factor in the run-time estimate of the new algorithm,
which appears in (\ref{DominantFactorNewAlgorithm}), is
$\left({2^7\dima^{2.5}\dimb^{2.5}}/{\epsilon}\right)^{2(\dima+\dimb)}$.
In Chapter \ref{ChapterAlgorithm}, we will see that
$\epsilon:=\delta/5$.  Making this substitution and setting
$\dimb:=\dima$ gives an upper bound (ignoring polynomial factors)
on the run time of the new algorithm of
\begin{eqnarray}\label{UpperboundOnNewAlgorithm}
2^{40\dima}\dima^{20\dima}\left(\frac{1}{\delta}\right)^{4\dima}.
\end{eqnarray}
The factors in (\ref{LowerboundOnBoundedSearchForSymmExts}) and
(\ref{UpperboundOnNewAlgorithm}) that are at least factorial in
$\dima$ are, respectively, $\dima^{36\dima +8}$ and
$\dima^{20\dima}$, the former being larger.  As well, the
dependence on $\delta$ in
(\ref{LowerboundOnBoundedSearchForSymmExts}) is worse than that in
(\ref{UpperboundOnNewAlgorithm}).  Therefore, the new algorithm
has the smaller run-time estimate when $\dima=\dimb$.


\chapter{New polynomial-time reduction from WSEP to WOPT}\label{ChapterAlgorithm}

As promised, I now show that the cut-generation rule of Section
\ref{sec_SearchHeuristic}, which is based on an intuitive
heuristic, yields an oracle-polynomial-time algorithm for the
in-biased weak separation problem for a convex set
$K\subset\mathbb{R}^n$ relative to an oracle for the weak
optimization problem for $K$; we only assume that $K$ contains a
ball of finite radius centered at a known point $c_0$ and is
contained in a ball of finite radius $R$.  The algorithm uses
$O(\poly(n,\log(R/\delta)))$ calls to the weak optimisation
oracle, where $\delta$ is the accuracy parameter that appears in
Definition \ref{def_WSEP}.  For the remainder of this thesis,
$\mathcal{O}$ will denote the oracle for the weak optimisation
problem for $K$. One simplifying assumption that we will carry
through this chapter, without loss of generality, is that $c_0$ is
the origin. This new algorithm is based on the analytic centre
cutting-plane algorithm of Atkinson and Vaidya \cite{AV95}.

Continuing the discussion in the previous chapter, Section
\ref{sec_MainIdea} gives the main idea behind the new algorithm.
Section \ref{sec_TheAlgorithm} presents the algorithm in terms of
parameters that will be given in section \ref{sec_Proof}, which
contains the proof of correctness of the algorithm. Section
\ref{sec_ComplexityAndDiscussion} discusses complexity and relates
the algorithm to the standard cut-generation method of Section
\ref{sec_ConnToOtherKnownMethods}. Section \ref{sec_Application}
gives the algorithm's parameters for the specific case of the
quantum separability problem.

\section{The Main Idea of the Algorithm}\label{sec_MainIdea}

The general idea of the algorithm is as follows. Let $P$ be the
current outer approximation $P:=B_n\bigcap \cap_{i=1}^{h}\lbrace
x:a_i^T x\geq b_i \rbrace$, as described in the second-last
paragraph of Section \ref{sec_SearchHeuristic}. Recall that we
need a definition of ``centre $\omega$ of $P$'' that satisfies
(\ref{eqn_CentreIsConicCombinationOfNormals}). Define the
\emph{analytic centre} $\omega$ of $P$ as the unique minimiser of
the real convex function
\begin{eqnarray}\label{def_F}
F(x):= -\sum_{i=1}^{h}\log(a_i^T x-b_i) - \log(1-x^T x).
\end{eqnarray}
The relation $\nabla F(\omega)=0$ gives
\begin{eqnarray}\label{eqn_defw}
\omega=\frac{1-\omega^T\omega}{2}\sum_{i=1}^{h}\frac{a_i}{a_i^T
\omega - b_i},
\end{eqnarray}
which shows that $\omega$, defined as the analytic centre of $P$,
indeed satisfies (\ref{eqn_CentreIsConicCombinationOfNormals}).

The algorithm stops when the current outer approximation becomes
either too small (volume-wise) or too thin to contain $K_p$. For
this, a lower bound $r>0$ on the radius of the largest ball
contained in $K_p$ is needed. By exploiting the accuracy parameter
$\delta$ of the weak separability problem, such an $r$ exists and
is derived in section \ref{subsec_DerivationOf_r}.

The actual algorithm is not as straightforward.  For instance,
each time a new cutting plane is added, it is shifted by some
amount ($b_i<0$) so as to keep the analytic centre of the old $P$
in the new $P$. As well, cutting planes are occasionally discarded
so that $h$ does not exceed some prespecified number. This
shifting and discarding of hyperplanes is done exactly as in
\cite{AV95}. 
To facilitate comparison, we use notation that corresponds to the
notation used in \cite{AV95}.

\section{The Algorithm}\label{sec_TheAlgorithm}

Following \cite{AV95}, the algorithm utilises three types of
quantities ($\sigma_i(z)$, $\kappa(a_i,b_i)$, and $\mu_i(z)$),
whose significance we now briefly explain. Suppose that
$P=B_n\bigcap \cap_{i=1}^{h}\lbrace x:a_i^T x\geq b_i \rbrace$ is
the current search space at some stage during the algorithm; that
is, suppose a total of $h$ cutting planes have been generated.
Denote the hyperplane $\{x:a^T_ix-b_i=0\}$ by the ordered pair
$(a_i,b_i)$. Recall that for any positive definite matrix $A$, one
can define the ellipsoid $E(A,z,r)$ as
\begin{eqnarray}
E(A,z,r):=\{x\in\mathbf{R}^n: (x-z)^TA(x-z)\leq r^2\}.
\end{eqnarray}
When $A=\hess
F(z)$, we refer to $E(A,z,r)$ as the \emph{Hessian ellipsoid}.

We mentioned that one of the stopping conditions is that the
volume of $P$ gets too small to contain $K_p$.  Later we will see
that the volume of $P$ can be related to the determinant of
$\hessfw$, where $\w$ is the analytic center of $P$. Define the
quantities
\begin{eqnarray}
\sigma_i(x):=\frac{a_i^T(\hessfx)^{-1}a_i}{(a_i^Tx-b_i)^2},\hspace{2mm}
1\leq i\leq h
\end{eqnarray}
for $x\in P$.  The denominator is the square of the distance from
$x$ to the hyperplane $(a_i,b_i)$.  The numerator is the square of
the radius of the Hessian ellipsoid $E(\hessfx,x,1)$ in the
direction of $a_i$. In Lemma \ref{AV-lem1}, we will see that
$E(\hessfx,x,1)\subset P$.  The smaller the quantity
$\sigma_i(x)$, the further away the hyperplane $(a_i,b_i)$ is from
the ellipsoid $E(\hessfx,x,1)$.  If $z$ is an approximate analytic
center of $P$, then a sufficiently small value of $\sigma_i(z)$
will indicate that $(a_i,b_i)$ has a small effect on
$\det(\hessfz)$ and so it can be discarded because it does not
sufficiently affect the volume of $P$.

Computing $\sigma_i(z)$ values is relatively computationally
expensive, so there is a simple test that can trigger a check of
$\sigma_i(z)$.  When the hyperplane $(a_i,b_i)$ is first
introduced, the quantity $\kappa(a_i,b_i)$ is set to $a_i^Tz-b_i$,
which is the distance from $(a_i,b_i)$ to the approximate analytic
center $z$ of $P$.  If, at some later step, we find that the
distance from the current approximate analytic center $z$ to
$(a_i,b_i)$ has doubled, then the quantity $\sigma_i(z)$ is
computed and tested.  We denote the ratio of the current distance
to the original distance by
$\mu_i(z):=(a_i^Tz-b_i)/\kappa(a_i,b_i)$.  If $\sigma_i(z)$ is not
sufficiently small, then $\kappa(a_i,b_i)$ is reset to the current
distance.

To compute approximate analytic centers, we use the Newton method.
A useful function that measures the quality of the approximation
is
\begin{eqnarray}\label{def_Lambda}
\lambda(x):=\sqrt{\grad F(x)^T\hessfxinv\grad F(x)}.
\end{eqnarray}
As well, define the function $q_\lambda:=1-(1-3\lambda)^{1/3}$ for
$\lambda\in\mathbf{R}$, and the function
$\Psi(x):=(\lambda(x))^2$.

The subscripts `d' and `a' in the algorithm mean `after a
hyperplane is \emph{discarded}' or `after a hyperplane is
\emph{added}', respectively.

The algorithm is presented in terms of undefined constants (all
variables with the subscript ``$0$'', plus $\nu$) and parameters
($r$,$u$,$\tilde{\delta}$).  For a list of the definitions of the
parameters and suitable values of the constants, the reader may
consult subsection \ref{subseq_SelectingConstants}.

The stopping conditions in the following algorithm are required
for the proof of polynomial-time convergence, but they are not the
best conditions to use in practice. In subsection
\ref{subsec_DynamicStopConds}, we give tighter stopping conditions
that depend more heavily on $z$ and $\hessfz$.

The algorithm for the in-biased weak separation problem for $K$,
relative to an oracle for the weak optimization problem for $K$,
is as follows:

\noindent\textsc{begin}\\
\noindent\textsc{initialise}$\lbrace$\\
\indent $a_1:=p/||p||$ \\
\indent $P:=B_n\cap\{x:a_1^Tx\geq 0\}$\\
\indent $z:=a_1/\sqrt{3}$\\
\indent $\kappa(a_1,b_1):=1/\sqrt{3}\rbrace$\\
\noindent\textsc{do}$\lbrace$\\
\indent\textsc{if} $\max_i\mu_i(z)>2$ \textsc{then}\\
\indent\indent \textit{Case} 1:\\
\indent\indent\textsc{if} there is an index $j$ such that
$\mu_j(z)>2$ \textsc{and} $\sigma_j(z)<\sigma_0$ \textsc{then}\\
\indent\indent\indent \textit{Subcase} 1.1:\\
\indent\indent\indent Discard $(a_j,b_j)$ from the set of
hyperplanes defining $P$, yielding a new \\
\indent\indent\indent region $P_\d$; $P_\mathrm{new}:=P_\d$.\\
\indent\indent\indent Starting at $x_0:=z$, iterate Newton steps
$x_i$
until both \\
\indent\indent\indent $\lambda(x_i)<\rho_0$ and $q_{\lambda(x_i)}
< \frac{\tilde{\delta}}{1+\tilde{\delta}}\frac{||x_i||}{\sqrt{2}}$
to get a new approximation
$z_\d:=x_i$ \\
\indent\indent\indent to the new analytic center $\w_\d$ of
$P_\d$; $z_\mathrm{new}:=z_\mathrm{d}$.\\
\indent\indent\textsc{else}\\
\indent\indent\indent\textit{Subcase} 1.2:\\
\indent\indent\indent Let $(a_j,b_j)$ be any hyperplane such that
$\mu_j(z)>2$.\\
\indent\indent\indent Reset $\kappa(a_j,b_j):= a_j^Tz-b_j$.\\
\indent\indent \textsc{endif}\\
\indent\textsc{else}\\
\indent\indent \textit{Case} 2:\\
\indent\indent Call weak optimization oracle on $c:=z/||z||$ with $\epsilon:=\delta/5$.\\
\indent\indent \textsc{if} oracle outputs $k_c\in K$ such that $c^Tp\geq c^Tk_c+\delta/5$ \textsc{then}\\
\indent\indent\indent \textsc{return} $c$.\\
\indent\indent\textsc{endif}\\
\indent\indent $a:=(p-k_c)-c^T(p-k_c)c$; $a:=a/||a||$.\\
\indent\indent  Compute $\beta<0$ such that
$\gamma^2:=(a^T[\hessfz]^{-1}a)/(a^Tz-\beta)^2=\gamma_0^2$.\\
\indent\indent Add $(a,\beta)$ to the set of hyperplanes
defining $P$, that is, \\
\indent\indent set $P_\a:=P\cap\lbrace x:a^Tx\geq\beta\rbrace$; $P_\mathrm{new}:=P_\a$.\\
\indent\indent Starting at $x_0:=z$, iterate Newton steps $x_i$
until both \\
\indent\indent $\lambda(x_i)<\rho_0$ and $q_{\lambda(x_i)} <
\frac{\tilde{\delta}}{1+\tilde{\delta}}\frac{||x_i||}{\sqrt{2}}$
to get a new approximation
$z_\a:=x_i$ \\
\indent\indent to the new analytic center $\w_\a$ of
$P_\a$; $z_\mathrm{new}:=z_\mathrm{a}$.\\
\indent\indent Set $\kappa(a,\beta):= a^Tz_\a - \beta$.\\
\indent \textsc{endif}\\
\indent $P:=P_\mathrm{new}$; $z:=z_\mathrm{new}$.$\rbrace$\\
\noindent\textsc{until}$\lbrace$\\
\indent \textit{Stopping Condition 1:} $h\geq \nu n u(n,\delta)$, \textsc{or}\\
\indent \textit{Stopping Condition 2:} $2r >  \frac{[\min_i\{a_i^Tz-b_i\}]}{1-\zeta_0}(3h+4)$$\rbrace$\\
\textsc{enddo}\\
\textsc{return} ``$p\in S(K,\delta)$''\\
\textsc{end}\\

\section{Proof of Correctness of the Algorithm}\label{sec_Proof}

To prove that the algorithm is correct, we need to deal with the
fact that the algorithm is run on a computer with fixed precision.
If the volume and width of $K_p$ are to be lower-bounded, then
clearly we need to exploit the weakness of the separability
problem; that is, we only need to find a separating hyperplane for
$p$ when $p$ is outside of $S(K,\delta)$.  This would give a lower
bound on the volume and width of $K_p$ in terms of $n$, $R$, and
$\delta$. We present the convergence proofs next, assuming that we
have a lower bound $r$ on the maximum radius of a ball contained
in $K_p$:
\begin{eqnarray}
r<\sup_{x}\lbrace r'\in\mathbf{R^+}: B(x,r')\subset
K_p\rbrace\label{def_rLowerBdInnerRadKp},
\end{eqnarray}
where $B(x,r):=\{y\in\mathbf{R}^n: ||y-x||\leq r\}$ and
$\mathbf{R}^+$ denotes the positive real numbers.  In subsection
\ref{subseq_ProducingGoodCutPlanes}, we will derive a suitable
$r=r(n,R,\delta)$.  The volume of a hypersphere of radius $r$ in
$\mathbf{R}^n$ is lower-bounded by $(r/n)^n$ \cite{GLS88}. Thus,
inequality (\ref{def_rLowerBdInnerRadKp}) gives
\begin{eqnarray}
\textrm{volume($K_p$)}\geq
\left(\frac{r}{n}\right)^n\label{ineq_LBdVolKp}.
\end{eqnarray}

We note here expressions for the gradient $\grad F(x)$ and Hessian
$\hess F(x)$ of the function $F(x)$ as defined in (\ref{def_F}):
\begin{eqnarray}
\grad F(x) &=& -\sum_{i=1}^{h}\frac{a_i}{a_i^Tx-b_i} +
\frac{2x}{1-x^Tx}\nonumber\\
\hess F(x) &=&
\sum_{i=1}^{h}\frac{a_ia_i^T}{(a_i^Tx-b_i)^2}+\frac{4xx^T}{(1-x^Tx)^2}+\frac{2I}{1-x^Tx}\nonumber,
\end{eqnarray}
where $I$ denotes the identity operator.

The full proof will be given in stages.  In subsection
\ref{subsec_Convergence}, we will present the results required to
prove that the algorithm works with the assumptions that the
cutting planes generated do not cut into the set $K_p$ and that
sufficiently good approximations of the analytic centers are at
hand. The proofs (mostly appearing in the Appendix) will be left
in terms of parameters including various constants and the inner
radius $r$. In subsection \ref{subseq_ProducingGoodCutPlanes}, we
show that such correct cutting planes can be generated. In
subsection \ref{subsec_DerivationOf_r}, we derive a suitable value
for $r$. In subsection \ref{subsec_Newton}, we describe the Newton
method used to calculate approximate analytic centers and show
that the number of required Newton iterations is small. In
subsection \ref{subseq_SelectingConstants}, we give concrete
values for all constants.

Before diving into the tough stuff, I show that the initialisation
of the analytic centre $z:=a_1/\sqrt{3}$ is correct.  I actually
prove something slightly more general, which will come up in the
discussion in Section \ref{sec_ComplexityAndDiscussion}.
\begin{fact}\label{Fact_InitialCentre}
For $||a_1||=1$, the analytic centre $\omega$ of $\{x: x^Tx\leq
R^\star\}\cap\lbrace x:a_1^T x-s\geq 0 \rbrace$, for $s\geq0$, is
\begin{eqnarray}
\omega=\frac{s +\sqrt{s^2+3R^\star}}{3}a_1.
\end{eqnarray}
\end{fact}

\begin{proof}
The equation $\grad F(\omega)=0$ (for the barrier of radius
$R^\star$) gives
\begin{eqnarray}
\frac{2\omega}{R^\star-\omega^T\omega} = \frac{a_1}{a_1^T\omega -
s}.
\end{eqnarray}
This implies that $\omega = \lambda a_1$ for some $\lambda>0$.
Making this substitution and solving for $\lambda$ gives
$3\lambda^2-2s\lambda -R^\star=0$, which gives the required
result.
\end{proof}

\subsection{Convergence}\label{subsec_Convergence}

The new algorithm for the feasibility problem for $K_p$ differs
from the one in \cite{AV95} in two essential ways:
\begin{enumerate}

\item I do not assume that we have an unrestricted, unweakened
separation oracle for $K_p$.  Rather, we assume that we have a
weakened separation oracle (built from the weak optimization
oracle for $K$ and Lemma \ref{lem_Donny}) which is restricted in
that it can only handle queries $c$ satisfying $m^Tc\geq 0$ for
all $m\in K_p$.

\item To accommodate the above restriction, I use the
$\bar{0}$-centered unit hyperball $B_n$ containing $K_p$ as the
initial search space instead of a $\bar{0}$-centered hyperbox
$\lbrace x\in\mathbf{R}^n: -2^L\leq x_i\leq 2^L, 1\leq i\leq n
\rbrace$.

\end{enumerate}

The second item above means that the current search space $P$ is
never a polytope.  Consequently, most of the lemmas of \cite{AV95}
that are properties of the function $F(x)$ cannot be used without
modification. Luckily, though, the function $F(x)$ is a
self-concordant functional \cite{Ren01} which has all the
analogous properties necessary to make the proofs of \cite{AV95}
work for our algorithm.  I present these fundamental lemmas below;
the corresponding label number in \cite{AV95} will appear in
parentheses after our label number.  In the following, assume
$P=B_n\bigcap \cap_{i=1}^{h}\lbrace x:a_i^T x\geq b_i \rbrace$ and
$F(x):= -\sum_{i=1}^{h}\log(a_i^T x-b_i) - \log(1-x^T x)$ for
$h\geq 0$, so that the interior of $P$ is the domain of $F$.  As
always, $\w$ denotes the analytic center (unique minimiser) of $P$
($F(x)$).

\begin{lemma}[Line (2) in \cite{AV95}]\label{AV-KKT}
Let $A$ be positive definite.  For any fixed vector $w$ in
$\mathbf{R}^n$,
\begin{eqnarray}
\max_{x\in E(A,z,r)}w^T(x-z)=r\sqrt{w^T
A^{-1}w}.\nonumber\end{eqnarray}
\end{lemma}
\begin{proof} See \cite{GLS88}, for example.
\end{proof}

\begin{lemma}[Lemma 1 in \cite{AV95}]\label{AV-lem1}
For every $z\in P$, $E(\hessfz,z,1)\subset P$.
\end{lemma}
\begin{proof} Follows from definition of self-concordance; see
\cite{Ren01} or \cite{NN94}.
\end{proof}

\begin{lemma}[Lemma 3 in \cite{AV95}]\label{AV-lem3}
If $\alpha<1$ and $y\in E(\hessfz,z,\alpha)$, then
\begin{eqnarray}
(1-\alpha)^2\xi^T\hessfz\xi \leq \xi^T\hess F(y)\xi \leq
(1-\alpha)^{-2}\xi^T\hessfz\xi
\end{eqnarray}
for all $\xi\in\mathbf{R}^n$.
\end{lemma}
\begin{proof} Follows from definition of self-concordance; see
\cite{Ren01} or \cite{NN94}.
\end{proof}

\begin{lemma}[Corollary 4 in \cite{AV95}]\label{AV-cor4}
Suppose $A$ and $B$ are positive definite $n\times n$ matrices
such that $\xi^T A\xi\geq \theta\xi^T B\xi$ for some $\theta>0$
and for all $\xi\in\mathbf{R}^n$.  Then $\xi^T
A^{-1}\xi\leq\theta^{-1}\xi^T B^{-1}\xi$ for all
$\xi\in\mathbf{R}^n$.
\end{lemma}
\begin{proof}
See proof of Lemma 2 in \cite{AV92}.
\end{proof}

Recall the second-degree Taylor expansion of $F(y)$ about
$z\in\mathbf{R}^n$:
\begin{eqnarray}
F(y)-F(z)=\grad F(z)^T(y-z)+\frac{1}{2}(y-z)^T\hessfz (y-z)+
\mathrm{Error}.
\end{eqnarray}

\begin{lemma}[Lemma 5 in \cite{AV95}]\label{AV-lem5}
If $y\in E(\hessfz,z,\alpha)$ where $\alpha<1$, then the error in
using the second-degree Taylor polynomial constructed about $z$ to
approximate $F(y)$ satisfies
$|\mathrm{Error}|\leq\frac{\alpha^3}{3(1-\alpha)}$.
\end{lemma}
\begin{proof}
See proof of Theorem 2.2.2 in \cite{Ren01}.
\end{proof}

\begin{lemma}[Lemma 6 in \cite{AV95}]\label{AV-lem6}
If $\lambda(z)<\frac{1}{3}$, then $F(z)-F(\w)\leq
\frac{1}{2}q^2_{\lambda(z)}\frac{1+q_{\lambda(z)}}{1-q_{\lambda(z)}}$.
\end{lemma}
\begin{proof}
See proof of Theorem 2.2.2 (\textit{iii}) (line 2.2.15) in
\cite{NN94}.
\end{proof}

\begin{lemma}[Lemma 7 in \cite{AV95}]\label{AV-lem7}
Let $\alpha:=\sqrt{(\w-z)^T\hessfz (\w-z)}$.  If
$\lambda(z)<\frac{1}{3}$, then $\alpha\leq q_{\lambda(z)}$.
\end{lemma}
\begin{proof}
See proof of Theorem 2.2.2 (\textit{iii}) (line 2.2.17) in
\cite{NN94}.
\end{proof}

%

The next lemma gives a Hessian ellipsoid centered at the analytic
center $\w$ which contains the current search space $P$.  The
volume of the ellipsoid gives an upper bound on the volume of $P$
which is useful for knowing when $P$ is too small to contain
$K_p$.

\begin{lemma}[Lemma 9 in \cite{AV95}]\label{AV-lem9}
If $h>31$ then
$P\subset E(\hessfw,\w,\sqrt{14}h)$. 
\end{lemma}
\begin{proof}
Since $\w$ is the unique minimiser of $F(x)$, we have
\begin{eqnarray}
\bar{0}^T=(\grad
F(\w))^T=\sum_{i=1}^{h}\frac{-a^T_i}{a_i^T\w-b_i}+\frac{2\w^T}{1-\w^T\w}\Leftrightarrow
\frac{2\w^T}{1-\w^T\w}=\sum_{i=1}^{h}\frac{a_i^T}{a_i^T-b_i}.\nonumber
\end{eqnarray}
Therefore,
\begin{eqnarray}
h &=&
\sum_{i=1}^{h}\frac{a_i^T\w-b_i}{a_i^T\w-b_i}=\left(\sum_{i=1}^{h}\frac{a_i^T}{a_i^T\w-b_i}\right)(\w)-\sum_{i=1}^h
\frac{b_i}{a_i^T\w-b_i}\nonumber\\
&=& \frac{2\w^T\w}{1-\w^T\w} +\sum_{i=1}^h
\frac{-b_i}{a_i^T\w-b_i}\label{eqn_BoundOnSizeOfw}\\
&=& \frac{2\w^T\w}{1-\w^T\w} +\sum_{i=1}^h
\frac{-a^T_i\w+a^T_i\w -b_i}{a_i^T\w-b_i} +\underbrace{\sum_{i=1}^{h}\frac{a_i^Tx}{a_i^T\w-b_i}-\frac{2\w^Tx}{1-\w^T\w}}_{=\bar{0}^Tx=0}\nonumber\\
&=&
\sum_{i=1}^{h}\frac{a^T_i(x-\w)+a^T_i\w-b_i}{a^T_i\w-b_i}-\frac{2\w^T(x-\w)}{1-\w^T\w}\label{eqn_hInTermsOfx}\\
h^2 &=& \left(\sum_{i=1}^{h}\frac{a^T_i(x-\w)+a^T_i\w-b_i}{a^T_i\w-b_i}\right)^2\nonumber\\
&&
+\left(\frac{2\w^T(x-\w)}{1-\w^T\w}\right)^2-4\left(\sum_{i=1}^{h}\frac{a^T_i(x-\w)+a^T_i\w-b_i}{a^T_i\w-b_i}\right)\frac{\w^T(x-\w)}{1-\w^T\w}.\nonumber
\end{eqnarray}
Now, for $x\in P$, we have that $a^T_ix-b_i\geq 0$ and so
\begin{eqnarray}
\left(\sum_{i=1}^{h}\frac{a^T_i(x-\w)+a^T_i\w-b_i}{a^T_i\w-b_i}\right)^2\nonumber
&\geq&
\sum_{i=1}^{h}\left(\frac{a^T_i(x-\w)+a^T_i\w-b_i}{a^T_i\w-b_i}\right)^2\nonumber.
\end{eqnarray}
Therefore,
\begin{eqnarray}
h^2 &\geq&\sum_{i=1}^{h}\frac{(a^T_i(x-\w)+(a^T_i\w-b_i))^2}{(a^T_i\w-b_i)^2}\nonumber\\
&&
+\left(\frac{2\w^T(x-\w)}{1-\w^T\w}\right)^2-4\left(\sum_{i=1}^{h}\frac{a^T_i(x-\w)+a^T_i\w-b_i}{a^T_i\w-b_i}\right)\frac{\w^T(x-\w)}{1-\w^T\w}\nonumber\\
&=&\sum_{i=1}^{h}\frac{(a^T_i(x-\w))^2}{(a^T_i\w-b_i)^2}+2\sum_{i=1}^{h}\frac{a^T_i(x-\w)}{a^T_i\w-b_i}+h\nonumber\\
&&
+\left(\frac{2\w^T(x-\w)}{1-\w^T\w}\right)^2-4\left(\sum_{i=1}^{h}\frac{a^T_i(x-\w)+a^T_i\w-b_i}{a^T_i\w-b_i}\right)\frac{\w^T(x-\w)}{1-\w^T\w}\nonumber\\
&=&\overbrace{\sum_{i=1}^{h}\frac{(a^T_i(x-\w))^2}{(a^T_i\w-b_i)^2}+\left(\frac{2\w^T(x-\w)}{1-\w^T\w}\right)^2}^{=(x-\w)^T\hessfw(x-\w)-\frac{2||x-\w||^2}{1-\w^T\w}}+\underbrace{\left(\frac{2||x-\w||^2}{1-\w^T\w}-\frac{2||x-\w||^2}{1-\w^T\w}\right)}_{=0}\nonumber\\
&&
+2\sum_{i=1}^{h}\frac{a^T_i(x-\w)}{a^T_i\w-b_i}+h-4\left(\sum_{i=1}^{h}\frac{a^T_i(x-\w)+a^T_i\w-b_i}{a^T_i\w-b_i}\right)\frac{\w^T(x-\w)}{1-\w^T\w}\nonumber\\
&=& (x-\w)^T\hessfw(x-\w)+h-\frac{2||x-\w||^2}{1-\w^T\w}\nonumber\\
&&
+2\sum_{i=1}^{h}\frac{a^T_i(x-\w)}{a^T_i\w-b_i}-4\left(\sum_{i=1}^{h}\frac{a^T_i(x-\w)+a^T_i\w-b_i}{a^T_i\w-b_i}\right)\frac{\w^T(x-\w)}{1-\w^T\w}\nonumber\\
&=& (x-\w)^T\hessfw(x-\w)+h-\frac{2(x-\w)^T(x-\w)}{1-\w^T\w}\nonumber\\
&&
+\frac{4\w^T(x-\w)}{1-\w^T\w}-4\left(\frac{2\w^T(x-\w)}{1-\w^T\w}+h\right)\frac{\w^T(x-\w)}{1-\w^T\w}\nonumber\\
&=& (x-\w)^T\hessfw(x-\w)+h-\frac{2x^T(x-\w)}{1-\w^T\w}+\frac{2\w^T(x-\w)}{1-\w^T\w}\nonumber\\
&&
+\frac{4\w^T(x-\w)}{1-\w^T\w}-4\left(\frac{2\w^T(x-\w)}{1-\w^T\w}+h\right)\frac{\w^T(x-\w)}{1-\w^T\w}\nonumber\\
&=& (x-\w)^T\hessfw(x-\w)+h-\frac{2x^T(x-\w)}{1-\w^T\w}\nonumber\\
&&
-(2h-3)\frac{2\w^T(x-\w)}{1-\w^T\w}-8\left(\frac{\w^T(x-\w)}{1-\w^T\w}\right)^2\nonumber.
\end{eqnarray}
Let $s:=\frac{1}{1-\w^T\w}$ and $t:=\w^T(x-\w)$. Thus,
$x^T\w=\w^Tx=t+\w^T\w$, $\w^T\w=\frac{s-1}{s}$, and $|t|<2$ since
$x,\w\in B_n$.  All this gives
\begin{eqnarray}
h^2&\geq& (x-\w)^T\hessfw(x-\w)+h-2sx^Tx+2sx^T\w
-(2h-3)2st-8s^2t^2\nonumber\\
&\geq&(x-\w)^T\hessfw(x-\w)+h-2s+2sx^T\w
-(2h-3)2st-8s^2t^2\hspace{2mm}[\mathrm{since}\hspace{2mm} x\in B_n]\nonumber\\
&=&(x-\w)^T\hessfw(x-\w)+h-2s+2s(t+\w^T\w) -(2h-3)2st-8s^2t^2\nonumber\\
&=&(x-\w)^T\hessfw(x-\w)+h-2s+2s\frac{s-1}{s} -(2h-4)2st-8s^2t^2\nonumber\\
&=&(x-\w)^T\hessfw(x-\w)+h -(h-2)4st-8s^2t^2-2\nonumber\\
&\geq&(x-\w)^T\hessfw(x-\w)+h
-(h-2)8s-32s^2-2\label{AV-lem9_LineForDynamicStopCond1}.
\end{eqnarray}
Because in the
algorithm $b_i<0$ for all $i$, equation (\ref{eqn_BoundOnSizeOfw})
gives
\begin{eqnarray}
s=\frac{1}{1-\w^T\w}\leq \frac{h+2}{2}\label{ineq_UpperBdw}.
\end{eqnarray}
Plugging in this bound gives
\begin{eqnarray}
(x-\w)^T\hessfw(x-\w)&\leq& 13h^2+31h+18.
\end{eqnarray}
The right side of the above inequality is less than
$14h^2$ if $h>31$.
\end{proof}

The next lemma is required for the stopping condition based on
$P$'s becoming too thin to contain $K_p$.  Define the width of $P$
in the direction of $a_i$ as $\mathrm{width}(a_i):=\max_{x,y\in
P}a_i^T(x-y)$.

\begin{lemma}[Lemma 10 in \cite{AV95}]\label{AV-lem10}
For every $i$, $\mathrm{width}(a_i)\leq (a_i^T\w-b_i)(3h+4)$.  As
well, for every $i$, $\mathrm{width}(a_i)\leq
(a_i^T\w-b_i)(h+4/(1-||\w||^2))$.
\end{lemma}
\begin{proof}
From equation \ref{eqn_hInTermsOfx}, it follows that
\begin{eqnarray}
h=\sum_{i=1}^{h}\frac{a_i^Tx-b_i}{a_i^T\w-b_i}-\frac{2\w^T(x-\w)}{1-w^T\w}\nonumber
\end{eqnarray}
for all $x\in P$.  Since for every index $j$ there exists some
$x^j$ in $P$ satisfying $\mathrm{width}(a_j)\leq a_j^Tx^j-b_j$, we
have
\begin{eqnarray}
\frac{\mathrm{width}(a_j)}{a_j^T\w-b_j}&\leq&
\frac{a_j^Tx^j-b_j}{a_j^T\w-b_j}\nonumber\\
&\leq& \sum_{i=1}^{h}\frac{a_i^Tx^j-b_i}{a_i^T\w-b_i}\nonumber\\
&=& h+\frac{2\w^T(x^j-\w)}{1-\w^T\w}\nonumber\\
&\leq& h+\frac{4}{1-\w^T\w}\nonumber,
\end{eqnarray}
where the last inequality follows from $x^j,\w\in B_n$.  This
proves the second statement of the lemma. Employing the bound
$\frac{1}{1-\w^T\w}\leq \frac{h+2}{2}$, as in the proof of Lemma
\ref{AV-lem9}, proves the first statement.
\end{proof}


Now we state the main results needed to derive the stopping
conditions of the algorithm.  At each iteration, we assume that we
have an approximate analytic center $z$ that satisfies
$\lambda(z)=\sqrt{\Psi(z)}\leq \p \leq \p_0 <\frac{1}{3}$. In
section \ref{subsec_Newton}, we will explain how to achieve this
approximation using Newton iterates.  Lemma \ref{AV-lem7} gives
\begin{eqnarray}
\w\in E(\hessfz,z,q_\p).\label{AV-eqn65}
\end{eqnarray}
In what follows, we will set $\zeta:=q_\p$ and
$\zeta_0:=q_{\p_0}$. We also assume the approximation satisfies
$\zeta\leq\zeta_0<1$.  We regard $\rho$ and $\zeta$ as varying
parameters with respective tight upper bounds $\rho_0$ and
$\zeta_0$, which are constants, to be selected after the analysis
is complete. As such, our $\p$ and $\zeta$ correspond to those in
\cite{AV95}.

The structure of the argument is exactly as in \cite{AV95}
\emph{mutatis mutandis}. Hence, the proofs are in the appendix;
they are included for completeness and to provide justification
for the constants we use in the algorithm, since our constants
differ from those in \cite{AV95}.


\subsubsection{Derivation of Stopping Condition 1: Volume
Argument}\label{subsubsec_StopCond1Volume}

\begin{lemma}[Lemma 17 in \cite{AV95}]\label{AV-lem17}
Let $z$ be an approximation to $\w$ such that $\w\in
E(\hessfz,z,\zeta)$.  Suppose the hyperplane $(a,\beta)$ is added
in Case 2 with
$\gamma_0^2=\gamma^2=\frac{a^T(\hessfz)^{-1}a}{(a^Tz-\beta)^2}$.
Then,
\begin{eqnarray}
&\mathrm{(a)}&\frac{|a^T(z-\w)|}{a^Tz-\beta}\leq \zeta\gamma,\nonumber\\
&\mathrm{(b)}&\frac{|a^T(z-\w)|}{a^T\w-\beta}\leq
\zeta\gamma/(1-\zeta\gamma),\nonumber\\
&\mathrm{(c)}&\Psi_\a(\w)\leq\tilde{\gamma}^2:=\gamma^2\left(\frac{1}{1-\zeta\gamma}\right)^2\left(\frac{1}{1-\zeta}\right)^2.\nonumber\\
\end{eqnarray}
\end{lemma}

With $\zeta$ suitably small enough that
$\tilde{\gamma}<\frac{1}{3}$, we have by Lemma \ref{AV-lem7} that
\begin{eqnarray}
\w_\a \in E(\hessfaw,\w,q_{\tilde{\gamma}}).\label{AV-eqn81}
\end{eqnarray}

\begin{lemma}[Lemma 18 in \cite{AV95}]\label{AV-lem18}
Suppose a hyperplane is added in Case 2, and the analytic center
moves from $\w$ to $\w_\a$.  Let $\gamma=\sqrt{a^T\hessfzinv
a}/(a^Tz-\beta)^2$.  If $\tilde{\gamma}<\frac{1}{3}$, then
\begin{eqnarray}
\frac{a^T\hessfzinv
a}{(a^Tz_\a-\beta)^2}\geq\gamma^2\left(\frac{1-\zeta}{1+\gamma
q_{\tilde{\gamma}}/(1-\zeta) + \zeta\gamma}\right)^2\nonumber.
\end{eqnarray}
\end{lemma}

\begin{theorem}[Approximation version of Theorem 13 in
\cite{AV95}]\label{AV-thm13approx}Suppose that $\max_{1 \leq i
\leq h}\mu_i(z)\leq 2$ at the beginning of an iteration, i.e. Case
2 is about to occur.  If the current search space $P$ is
determined by $h$ hyperplanes (in addition to the unit
hypersphere), then
\begin{eqnarray}
\mathrm{det}(\hessfz)>2^{-n}(1+C_2)^h=2^{(\log_{2}(1+C_2))h-n},
\end{eqnarray}
for some positive constant $C_2$ which depends on the parameters
$\sigma_0$ and $\gamma_0$ of the algorithm and the ``minimal
goodness'' $\zeta_0$ of the approximation to the analytic centers.
This can be improved to
\begin{eqnarray}
\mathrm{det}(\hessfz)>2^{-n}(2.5)(1+C_2)^{h-1}.
\end{eqnarray}
\end{theorem}

\begin{lemma}[Lemma 19 in \cite{AV95}]\label{AV-lem19}
For the approximate analytic center $z$ with $\w\in
E(\hessfz,z,\zeta)$, we have
\begin{eqnarray}
P\subset E(\hessfz,z,\vartheta),\nonumber
\end{eqnarray}
where
\begin{eqnarray}
\vartheta &:=& \sqrt{2\left(\frac{14h^2}{(1-\zeta)^2}
+\zeta^2\right)} ,\hspace{5mm}\textrm{if $h>31$.}\nonumber
\end{eqnarray}
\end{lemma}

From here on we will assume that $h>31$, that is, that the minimum
number of total hyperplanes will be $31$.  We will also assume
that $\zeta<1/16$, in which case, $\vartheta$ in the above lemma
satisfies $\vartheta\leq 6h$.

\begin{theorem}[Approximation version of Theorem 14 in \cite{AV95}]\label{AV-thm14approx} There exists a constant
$\nu$, independent of $h$, $n$, $R$, and $\delta$, and there
exists a function
$u(n,\delta)\in\Theta(\mathrm{poly}(n,\log(\frac{R}{\delta})))$
such that if $h=\nu n u(n,\delta)$, then the volume of $K_p$ is
sufficiently small so as to assert that $p\in S(K,\delta)$.
\end{theorem}

This completes the derivation of Stopping Condition 1.


\subsubsection{Derivation of Stopping Condition 2: Width
Argument}\label{subsubsec_StopCond2Width}

\begin{lemma}[Lemma 16 in \cite{AV95}]\label{AV-lem16}
Let $\zeta<1$.  If $\w\in E(\hessfz,z,\zeta)$, then for all $i$,
$1\leq i\leq h$,
\begin{eqnarray}
\sigma_i(\w) \leq \frac{\sigma_i(z)}{(1-\zeta)^4}.
\end{eqnarray}
\end{lemma}

Define
\begin{eqnarray}\label{AV-eqn14}
N(x):= -\sum_{i=1}^{h}\ln
\left(\frac{a_i^Tx-b_i}{\kappa(a_i,b_i)}\right)-\ln(1-x^Tx)=F(x)+\sum_{i=1}^h\ln(\kappa(a_i,b_i)).
\end{eqnarray}
Note that $N(x)-N(y)=F(x)-F(y)$ in any given iteration.

\begin{theorem}[Approximation version of Theorem 11 in
\cite{AV95}]\label{AV-thm11approx} There exists a positive
constant $\theta$, independent of $h$, $n$, $R$, and $\delta$,
such that after $\iota$ iterations of the algorithm,
$N(\w)\geq\theta \iota$.  The constant $\theta$ will depend on the
parameters of the algorithm.
\end{theorem}

\begin{theorem}[Approximation version of Theorem 15 in
\cite{AV95}]\label{AV-thm15approx} If the algorithm does not first
find a separating hyperplane or halt by Stopping Condition 1,
then, within $O(nu\log(nuR/\delta))$ iterations, Stopping
Condition 2 must be met. If Stopping Condition 2 is met, then the
set $K_p$ is negligibly small and the algorithm may return ``$p\in
S(K,\delta)$''.
\end{theorem}

This completes the derivation of Stopping Condition 2.

\subsection{Producing Good Cutting Planes}\label{subseq_ProducingGoodCutPlanes}

Suppose that $M_p$ is large enough that the algorithm must return
an element of $M_p$.  Up until this point, we have assumed that
the cutting planes generated by the algorithm do not accidentally
slice off any portion of $K_p$, that is, that $m^Ta_i>0$ for all
$i=1,\ldots,h$ and for all $m\in M_p$. With finite-precision
computations, this condition is not sufficient. In order to combat
the effects of round-off, we would ideally require something
stronger: for all $m\in M_p$,
\begin{eqnarray}
m^Ta_i>\tilde{\delta},\hspace{5mm}\textrm{for all $i=1,\ldots
,h$,}\label{ineq_Require_mTai_geq_delta'}
\end{eqnarray}
for some $\tilde{\delta}>0$.  As it stands, this requirement is
tricky to achieve.  However, if we merely insist that
(\ref{ineq_Require_mTai_geq_delta'}) holds for all $m$ in the
smaller set
\begin{eqnarray}
M_p':=\{c\in S_n:\hspace{2mm} c^T k+\delta'<c^T p
\hspace{2mm}\forall k\in S(K,\epsilon)\},
\end{eqnarray}
for some $\delta'>0$, then we can ensure that the cutting planes
do not accidentally slice off any portion of
$K_p':=\left[\mathrm{ConvexHull}\left( M_p'\cup \{\bar{0}\}\right)
\right] \setminus \bar{0}$.  The size of $K_p'$ is still large
enough to give the asymptotic behaviour we desire from our
algorithm.  

\begin{lemma}\label{lem_DeltaCushionOnNormalisedCenter}
Let $P$ be the current search space, defined by $h$ cutting planes
$\{x:a_i^Tx=b_i\}$, where $||a_i||=1$, for $i=1,\ldots,h$. Assume
that $z$ is an approximate analytic center of $P$ satisfying
$\lambda(z)\leq \rho$ such that $\zeta:=q_{\rho}<1$. Assume
further that
\begin{eqnarray}
\zeta <
\frac{\tilde{\delta}}{1+\tilde{\delta}}\frac{||z||}{\sqrt{2}}.\nonumber
\end{eqnarray}
Let $c:=z/||z||$. If $m^Ta_i\geq\tilde{\delta}$ for all
$i=1,\ldots,h$, then $m^Tc>\frac{\tilde{\delta}}{2}$.
\end{lemma}
\begin{proof}
Equation (\ref{eqn_defw}) says that $c':=\w/||\w||$ can be written
as $\sum_{i=1}^{h}\eta_i a_i$ with $\eta_i\geq 0$ for all $i$.
Thus,
\begin{eqnarray}\nonumber
\frac{m^T\w}{||\w||}=m^Tc' = \sum_{i=1}^{h}\eta_i (m^Ta_i)\geq
\tilde{\delta}\sum_{i=1}^{h}\eta_i > \tilde{\delta},
\end{eqnarray}
because
\begin{eqnarray}\nonumber
1=c'^Tc' = \sum_{i=1}^{h}\eta_i (a_i^Tc')\leq \sum_{i=1}^{h}\eta_i
|a_i^Tc'| < \sum_{i=1}^{h}\eta_i.
\end{eqnarray}
Since $\w\in E(\hessfz,z,\zeta)$, we have
$2||z-\w||^2/(1-||z||^2)<\zeta^2$ which implies
\begin{eqnarray}
||z-\w||<\zeta/\sqrt{2}.\label{ineq_BoundOnDistBetween_w_and_z}
\end{eqnarray}
Consider the two cases:\\
\noindent Case A: $m^Tz\geq m^T\w$\\
\indent In this case, we have
\begin{eqnarray}
m^Tc\geq\frac{m^T\w}{||z||}>\tilde{\delta}\frac{||\w||}{||z||}\nonumber.
\end{eqnarray}
\noindent Case B: $m^Tz< m^T\w$\\
\indent In this case, $0\leq m^T\w-m^Tz=m^T(\w-z)\leq
||\w-z||<\zeta/\sqrt{2}$ gives
\begin{eqnarray}
m^Tc\geq\frac{m^T\w-\zeta/\sqrt{2}}{||z||}>\frac{\tilde{\delta}||\w||-\zeta/\sqrt{2}}{||z||}\nonumber.
\end{eqnarray}
Now consider two other cases:\\
\noindent Case I: $||\w||\geq ||z||$\\
\indent In this case, we have
\begin{eqnarray}
\frac{||\w||}{||z||}\geq 1\nonumber.
\end{eqnarray}
\noindent Case II: $||\w||< ||z||$\\
\indent In this case, (\ref{ineq_BoundOnDistBetween_w_and_z})
gives
\begin{eqnarray}
||\w||>||z||-\zeta/\sqrt{2}\nonumber.
\end{eqnarray}
Examining all four combinations of the above cases:\\
Case AI:
\begin{eqnarray}
m^Tc>\tilde{\delta}||\w||/||z||\geq \tilde{\delta}\nonumber
\end{eqnarray}
Case AII:
\begin{eqnarray}
m^Tc>\frac{\tilde{\delta}(||z||-\zeta/\sqrt{2})}{||z||}=\tilde{\delta}-\frac{\tilde{\delta}\zeta/\sqrt{2}}{||z||}\nonumber
\end{eqnarray}
so that as long as $\zeta<||z||/\sqrt{2}$, we have
$m^Tc>\tilde{\delta}/2$;\\
Case BI:
\begin{eqnarray}
m^Tc>
\frac{\tilde{\delta}||\w||}{||z||}-\frac{\zeta/\sqrt{2}}{||z||}\geq
\tilde{\delta}-\frac{\zeta/\sqrt{2}}{||z||} \nonumber
\end{eqnarray}
so that as long as $\zeta<||z||\tilde{\delta}/\sqrt{2}$, we have
$m^Tc>\tilde{\delta}/2$;\\
Case BII:
\begin{eqnarray}
m^Tc>
\frac{\tilde{\delta}(||z||-\zeta/\sqrt{2})}{||z||}-\frac{\zeta/\sqrt{2}}{||z||}=\tilde{\delta}-(1+\tilde{\delta})\frac{\zeta/\sqrt{2}}{||z||}\nonumber
\end{eqnarray}
so that as long as
$\zeta<\frac{\tilde{\delta}}{1+\tilde{\delta}}||z||/\sqrt{2}$, we
have
$m^Tc>\tilde{\delta}/2$.\\
The last case imposes the smallest upper bound on $\zeta$, which
is the upper bound in the statement of the lemma.
\end{proof}

Assume that the hypotheses of Lemma
\ref{lem_DeltaCushionOnNormalisedCenter} hold for all $m\in M_p'$
so that $m^Tc>\tilde{\delta}/2$ for all $m\in M_p'$ and some
$\tilde{\delta}>0$. Suppose the test point $c$ is given to the
weak optimization oracle which returns $k_c$. Then
\begin{eqnarray}\label{implic_AcceptanceCrit}
c^Tk_c +\epsilon \leq c^Tp \hspace{5mm}\Rightarrow\hspace{5mm}
c^Tx\leq c^Tp \hspace{2mm}\forall x\in K,
\end{eqnarray}
so that the left-hand side of (\ref{implic_AcceptanceCrit}) is a
valid acceptance criterion (appearing in the algorithm) if we are
solving the \emph{in-biased} weak separation problem.  For a
worst-case analysis, we assume that $p$ has distance $\delta$ from
the boundary of $K$. It is convenient to divide this distance into
three parts such that $\delta'+\epsilon < \delta$ (see Figure
\ref{DeriveTheta} on page \pageref{DeriveTheta}). The rejection
criterion for a test vector $c$ is simply the logical negation of
the left-hand side of (\ref{implic_AcceptanceCrit}):
\begin{eqnarray}
-\epsilon < -c^T(p -k_c)\label{ineq_RejectCrit}.
\end{eqnarray}
Thus, we have a revised version of Lemma \ref{lem_Donny}:
\begin{lemma}
Suppose that $m\in M_p'$ and that $c$ satisfies the rejection
criterion \emph{(\ref{ineq_RejectCrit})}.  Let
$\bar{a}:=(p-k_c)-\mathrm{Proj}_c(p-k_c)$. If $m^Tc\geq 0$ then
$m^T\bar{a}> \delta'-\epsilon$.
\end{lemma}
\begin{proof}
Case $-c^T(p-k_c)\geq 0$:\\
\begin{eqnarray}
m^T\bar{a} = m^T(p-k_c) +[ -c^T(p-k_c)](m^Tc) > \delta' + 0 =
\delta'
\end{eqnarray}
Else $-c^T(p-k_c)<0$:\\
\begin{eqnarray}
m^T\bar{a} = m^T(p-k_c) +[ -c^T(p-k_c)](m^Tc) > \delta' -\epsilon
|m^Tc|=  \delta' -\epsilon
\end{eqnarray}
\end{proof}
\noindent  Therefore, we set $\epsilon := \delta'/2$ so that
$m^T\bar{a}>\delta'/2$ in the conclusion of the lemma.  Since we
can assume that $p\in B(0,R)$, and since $k_c\in B(0,R)$, we have
$||p-k_c||\leq 2R$.  Thus, $||\bar{a}||\leq 2R$.  Letting $a$ be
the normal vector to the new cutting plane, we have
\begin{eqnarray}\nonumber
m^Ta=\frac{m^T\bar{a}}{||\bar{a}||} >
\frac{\delta'/2}{||\bar{a}||}\geq\frac{\delta'}{4R}.
\end{eqnarray}
If we set $\tilde{\delta}:=\delta'/{4R}$, then, as long as the
machine precision is sufficiently high so that the error in $m^Tc$
(due to round-off error of $c$) is less than $\tilde{\delta}/2$,
the cutting planes do not accidentally slice off any bit of
$K_p'$. We have assumed that the first normalised analytic center
$c_1:=p/||p||$ used in the algorithm satisfies
$m^Tc_1\geq\tilde{\delta}$ for all $m\in M_p'$.  Note we actually
have that $m^Tc_1\geq{\delta'}$ for all $m\in M_p'$, because
$m^Tp\geq 0$ for all $m\in M_p$ (Fact \ref{Fact_InitialCut}).
Therefore, it makes sense to set $\delta':=2\delta/5$, and thus
$\epsilon:=\delta'/2=\delta/5$.

\subsection{Derivation of $r$}\label{subsec_DerivationOf_r}

Now we derive the radius $r$ as a function of $R$ and $\delta$. In
light of the previous subsection, $r$ is redefined as a lower
bound on the maximum radius of a ball that fits inside $K_p'$.

First, we derive a lower bound $\theta$ on the one-dimensional
angle that defines the maximum-size hypercircular-based cone
(emanating from the origin) that fits inside $K_p'$. The bound
will assume only that $K$ is convex, centered at the origin
$\bar{0}$, and contained in $B(\bar{0},R)$.

\begin{figure}[ht]
 \centering
 \resizebox{100mm}{!}{\includegraphics{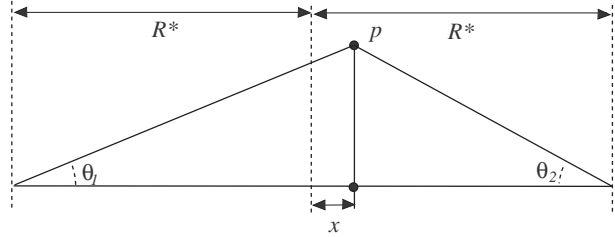}}
 \caption[The solid angle $\theta_1+\theta_2$ of the hypercircular-based cone]{The solid angle $\theta_1+\theta_2$ of the hypercircular-based cone as a function of displacement $x$ from center of $p$.}
 \label{WorstcaseDisplacement}
\end{figure}

To get this lower bound, we need to derive a worst-case scenario
for $p$ and $K$ that makes $K_p$ as small as possible.  Suppose
$p$ has minimal distance $\delta$ from the boundary of $K$.  Thus,
the ball $B(p,\delta)$ intersects $K$ only at one point $k^*\in
K$.  Consider the hyperplane $H:=\{x: (p-k^*)^Tx =(p-k^*)^Tk^*
\}$; it is tangent to $B(p,\delta)$ at $k^*$.  No point $k$ in $K$
is on the same side of $H$ as $p$ (that is, satisfies $(p-k^*)^Tk>
(p-k^*)^Tk^*$), else the line from $k$ to $k^*$ would contain
points in $K$ that intersect $B(p,\delta)$ and hence contradict
the minimality of the distance from $p$ to $k^*$.  If we let
\begin{eqnarray}
K^* := B(\bar{0},R)\cap \{x: (p-k^*)^Tx
\leq(p-k^*)^Tk^*\},\nonumber
\end{eqnarray}
then we have shown that $K\subset K^*$.  Let $M_p^*$ be $\{c\in
S_n:\hspace{2mm} c^T k<c^T p\hspace{2mm}\forall k\in K^*\}$. It
follows that $M_p^*\subset M_p$.  Finally, we show that if $p$ is
centered next to the set $C^*:=H\cap B(\bar{0},R)$, the set
$M_p^*$ is as small as possible.  Note that $C^*$ is a hyperdisc
of radius $R^*$, where $R^*\leq R$.  \fig{WorstcaseDisplacement}
defines the angles $\theta_1$ and $\theta_2$ as a function of the
displacement $x$ of $p$ from the center of $C^*$, for $x\in
[0,R^*]$.  For a lower bound on $M_p^*$, we want to minimise the
sum $\theta_1+\theta_2$. Since $\partial\theta_1/\partial
x<\partial\theta_2/\partial x$, this sum is minimised at $x=0$,
that is, when $p$ is centered next to $C^*$.  As well, the value
of $R^*$ that minimises the sum is $R^*=R$.  Define
$M_p^{*\prime}$ with respect to $K^*$ just as $M_p'$ was defined
with respect to $K$.  Since
\begin{eqnarray}
M_p^{*\prime} \subset M_p^* \subset M_p,\nonumber
\end{eqnarray}
calculating a lower bound on the size of $M_p^{*\prime}$ is
sufficient. Below, instead of working with $K^*$ explicitly, we
assume the worst case where $K$ is $K^*$ with $R^*=R$ and $p$
centered next to $C^*$.

\begin{figure}[ht]
 \centering
 \resizebox{100mm}{!}{\includegraphics{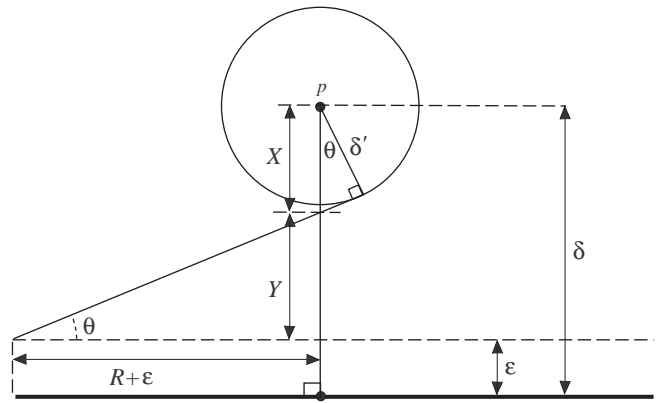}}
 \caption{Derivation of solid angle $2\theta$ of hypercircular-based cone in terms of $R$ and $\delta$.}
 \label{DeriveTheta}
\end{figure}

The angle $\theta$ can be seen in \fig{DeriveTheta}. We have
\begin{eqnarray}
\tan\theta &=& \frac{Y}{R +\epsilon}\nonumber\\
\cos\theta &=& \frac{\delta'}{X }\nonumber\\
2\delta'= X+Y
&=&(R+\epsilon)\tan\theta+\frac{\delta'}{\cos\theta}\nonumber
\end{eqnarray}
so that
\begin{eqnarray}
\tan\theta = \frac{\delta'}{R+\epsilon}(2-1/\cos\theta)
\label{eqn_DefTanTheta}
\end{eqnarray}
or
\begin{eqnarray}
\sin\theta = \frac{\delta'}{R+\epsilon}\left(2\cos\theta
-1\right). \label{eqn_DefSinTheta}
\end{eqnarray}

\begin{figure}[ht]
 \centering
 \resizebox{100mm}{!}{\includegraphics{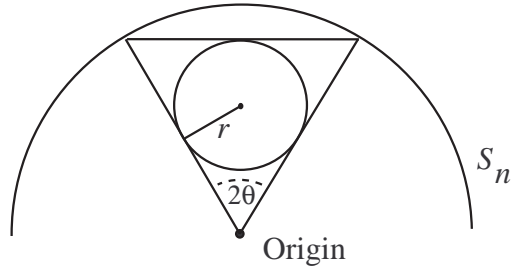}}
 \caption{Derivation of radius $r$ as a function of $\theta$.}
 \label{UnitSphere}
\end{figure}

Now, we derive $r$ as a lower bound on the maximum radius of a
ball that fits inside the hypercircular-based cone defined by
$\theta$.  From \fig{UnitSphere}, we have
\begin{eqnarray}
r=
\sin\theta\tan\left(\frac{\pi}{4}-\frac{\theta}{2}\right)\nonumber.
\end{eqnarray}
Since $\tan( \upsilon-\psi)=(\tan\upsilon -
\tan\psi)/(1+\tan\upsilon\tan\psi)$,
\begin{eqnarray}
r=
\sin\theta\frac{1-\tan(\theta/2)}{1+\tan(\theta/2)}\label{eqn_Def_r}.
\end{eqnarray}

As $\delta\rightarrow 0$ (and hence $\delta',\epsilon\rightarrow
0$), equation (\ref{eqn_DefTanTheta}) tends to $\tan\theta =
\delta'/R$ and equation (\ref{eqn_Def_r}) tends to $r=2\delta/5R$.
For convenience of exposition, we use the approximation $r\approx
2\delta/5R$.  In practice, equation (\ref{eqn_Def_r}) (in
conjunction with a numerical solution for $\theta$) may be used in
the derivation of Stopping Conditions 1 and 2.

\subsection{Newton Iterates}\label{subsec_Newton}

The next theorem says that, with respect to $F$, the new (actual)
analytic center and the old (approximate) analytic center are
never too far apart, so that the Newton procedure for finding the
new approximate analytic center terminates quickly (see the
Appendix for a proof).

\begin{theorem}[Theorem 20 in \cite{AV95}]\label{AV-thm20}
There exists some constant $C_\d$ such that any time a hyperplane
is discarded in Subcase 1.1, $F_\d(z)-F_\d(\w_\d)\leq C_\d$.
Likewise, there exists some constant $C_\a$ such that any time a
hyperplane is added in Case 2, $F_\a(z)-F_\a(\w_\a)\leq C_\a$.
\end{theorem}

In Subcase 1.1 or Case 2, to calculate new approximations
$z_\mathrm{new}$ to the new analytic center $\w_\mathrm{new}$, we
perform damped Newton iterations, as defined in \cite{NN94},
starting at the old approximate analytic center $z$.  Denote the
sequence of ensuing Newton iterates by $\{x_i:i=0,1,\ldots\}$. The
starting point is $x_0 :=z$.  Define
$\lambda_*:=2-\sqrt{3}=0.2679...$.  For $i\geq 0$, define the
Newton iterates as:
\begin{eqnarray}
x_{i+1} := x_i - \varsigma_i(\hess F(x_i))^{-1}\grad
F(x_i),\nonumber
\end{eqnarray}
where
\begin{eqnarray}
\varsigma_i := \left\{ \begin{array}{ll}
                (1+\lambda(x_i))^{-1} & \mbox{ if $\lambda(x_i)\geq\lambda_*$ }, \nonumber\\
                1 & \mbox{ if $\lambda(x_i)<\lambda_*$}.\nonumber
                \end{array}
                \right.
\end{eqnarray}
Theorem 2.2.3 in \cite{NN94} shows that, in the first stage of the
Newton process ($\lambda(x_i) \geq \lambda_*$), the difference
$F(x_i)-F(x_{i+1})$ is at least $\lambda_*$ and, in the second
stage of the Newton process ($\lambda(x_i) < \lambda_*$),
$\lambda(x_{i+1})<\lambda(x_{i})/2$. Thus, Theorem \ref{AV-thm20}
says that, within $O(1)$ iterations, the value of $\lambda(x_{i})$
will start decreasing quadratically.  The total number of Newton
iterations required is no more than
\begin{eqnarray}
\lceil{C_\d/\lambda_*}\rceil+\lceil{\log_2
(\lambda_*/\rho_0)}\rceil, \nonumber
\end{eqnarray}
in Subcase 1.1, and
\begin{eqnarray}
\lceil C_\a/\lambda_*\rceil +\lceil{\log_2
(\lambda_*/\rho_0)}\rceil, \nonumber
\end{eqnarray}
in Case 2.

\subsection{Selecting the Constants}\label{subseq_SelectingConstants}

Finally, we summarise the values of all the parameters of the
algorithm and give values of the constants that work in general
and for some special cases.

The parameters have been defined as follows:
\begin{eqnarray}\nonumber
r&:=&2\delta/5R \nonumber\\
u&:=& 2\log_2(n)+\log_2(1/r)\nonumber\\
\tilde{\delta}&:=& \delta'/2R = \delta/5R. \nonumber
\end{eqnarray}

For the constants, we have to summarise the strongest conditions
that the convergence analysis placed on them:

\begin{eqnarray}
\lambda(z)&<&1/3\nonumber\\
\tilde{\gamma}&<&1/3\nonumber\\
\zeta &<& 0.02 \textrm{ [see proof of Theorem \ref{AV-thm11approx}, Case 2]}\nonumber\\
C_1 &>& 0\nonumber\\
C_2&>&0\nonumber\\
C_3 &<& 1/3  \textrm{ [to invoke Lemma \ref{AV-lem6}]} \nonumber\\
C_4&<&0.615 \nonumber\\
C_5 &<& 1 \textrm{ [to invoke Lemma \ref{AV-lem5}]}\nonumber\\
C_6 &>& 0 \nonumber \\
3+(\log_2(12)+1/2)/2 &<& \frac{1}{2}\left(\nu \log_2(1+C_2)
-{\log_2(\nu)}\right). \nonumber
\end{eqnarray}

The following list of values can be shown to satisfy the above
constraints:
\begin{eqnarray}
\rho_0&:=&0.001 \nonumber\\
\zeta_0&:=& q_{\rho_0}=0.00101\nonumber\\
\gamma_0&:=& 0.25 \nonumber\\
\sigma_0&:=&0.08 \nonumber\\
\nu &:=& 1078.\nonumber
\end{eqnarray}

%

The potentially smallest upper bound imposed on $\zeta$ is
\begin{eqnarray}\label{ineq_DynamicBoundOnZeta}
\zeta <
\frac{\tilde{\delta}}{1+\tilde{\delta}}\frac{||z||}{\sqrt{2}},
\end{eqnarray}
in Lemma \ref{lem_DeltaCushionOnNormalisedCenter}.  We now show
that this upper bound is never so small as to require an
unreasonable number of Newton iterates, by deriving a lower bound
on $||z||$ based on Stopping Condition 2. While Stopping Condition
2 is not satisfied, we have
\begin{eqnarray}\nonumber
2r < (a_j^Tz-b_j)(3\nu n u +4)/(1-\zeta_0)\hspace{5mm}\forall j,
\end{eqnarray}
thus, in particular, for $j=1$,
\begin{eqnarray}\nonumber
||z||\geq a_1^Tz>\frac{2r(1-\zeta_0)}{3\nu n u +4}.
\end{eqnarray}
Thus the lowest upper bound ever imposed on $\zeta$ will be
\begin{eqnarray}\label{ineq_WorstBoundOnZeta}
\zeta < \frac{\tilde{\delta}}{1+\tilde{\delta}}
\frac{\sqrt{2}r(1-\zeta_0)}{3\nu n u +4}.
\end{eqnarray}
Let $t$ be the righthand side of the above inequality; note that
$t$ is lower-bounded by a polynomial in $\frac{1}{n}$ and
$\frac{\delta}{R}$. This gives a tight, worst-case upper bound on
$\rho$ of $t-t^2+t^3/3$
which is still a polynomial in $\frac{1}{n}$ and
$\frac{\delta}{R}$.  Thus, in the worst case, the required number
of Newton iterates is $O(\textrm{polylog}(n,\frac{R}{\delta}))$.

\subsection{Tighter Stopping Conditions}\label{subsec_DynamicStopConds}

The upper bound $(h+2)/2$ on $(1-||\omega||^2)^{-1}$ in
(\ref{ineq_UpperBdw}) is not tight because it throws away the
entire summation in (\ref{eqn_BoundOnSizeOfw}).  Line
(\ref{AV-eqn65}) gives
\begin{eqnarray}\nonumber
||\w||\leq ||z||+\zeta_0 \sqrt{\lambda_{\mathrm{max}}
(\hessfzinv)},
\end{eqnarray}
where $\lambda_{\mathrm{max}}(\hessfzinv)$ is the largest
eigenvalue of $\hessfzinv$; which gives
\begin{eqnarray}\nonumber
(1-||\omega||^2)^{-1} \leq \varpi(z) := \left( 1- \left(
||z||+\zeta_0 \sqrt{\lambda_{\mathrm{max}} (\hessfzinv)}
\right)^2 \right)^{-1}.
\end{eqnarray}
\noindent Recalling Lemma \ref{AV-lem10}, Stopping Condition 2 can
be immediately tightened to
\begin{eqnarray}\nonumber
2r >  \frac{[\min_i\{a_i^Tz-b_i\}]}{1-\zeta_0}(h+4\varpi(z)).
\end{eqnarray}

To tighten Stopping Condition 1, we go back to line
(\ref{AV-lem9_LineForDynamicStopCond1}), which gives
\begin{eqnarray}\nonumber
(x-\w)^T\hessfw(x-\w)\leq
h^2+h(8\varpi(z)-1)-16\varpi(z)+32\varpi(z)^2+2.
\end{eqnarray}
In conjunction with the proof of Lemma \ref{AV-lem19}, we get
\begin{eqnarray}\nonumber
P \subset E(\hessfz,z,\vartheta')
\end{eqnarray}
where
\begin{eqnarray}\nonumber
\vartheta' :=
\sqrt{2\left(\frac{h^2+h(8\varpi(z)-1)-16\varpi(z)+32\varpi(z)^2+2}{(1-\zeta_0)^2}+\zeta_0^2\right)}.
\end{eqnarray}
\noindent Using this and Theorem \ref{AV-thm13approx}, line
(\ref{AV-eqn49approx}) becomes
\begin{eqnarray}
\frac{(2\vartheta')^n}{2^{[\log_2((2.5)(1+C_2)^{h-1})]/2-n/2}}
&<&\left(\frac{r}{n}\right)^n \nonumber\\
\frac{2\vartheta'}{2^{[\log_2((2.5)(1+C_2)^{h-1})]/2n-1/2}}
&<&\frac{r}{n} \nonumber\\
\log_2({2\vartheta'})   - \left[
{\log_2((2.5)(1+C_2)^{h-1})/2n-1/2} \right] &<&\log_2({r}/{n}),
\nonumber
\end{eqnarray}
to give the stopping condition
\begin{eqnarray}
h> \frac{1}{\log_2(1+C_2)}\left[ 2n\log_2({2n\vartheta'}/{r})+n
\right] +\log_2(4/5).\nonumber
\end{eqnarray}

Employing these dynamic stopping conditions ensures that the
number of calls to the WOPT oracle is minimised.  When WOPT is
NP-hard, as in the quantum separability problem, this is important
in practice.

\section{Complexity and Discussion}\label{sec_ComplexityAndDiscussion}

As in \cite{AV95}, we only ever have to compute $\sigma_0^{-1}n+1$
of the $\sigma_i(z)$ values, regardless of the number of
hyperplanes $h$.

\begin{theorem}[Theorem 21 in \cite{AV95}]\label{AV-thm21}
In Case 1, if there is at least one hyperplane satisfying the
conditions of Subcase 1.1, then we must discover such a hyperplane
in at most $\sigma_0^{-1}n+1$ evaluations of the $\sigma_j(z)$
values.
\end{theorem}
\begin{proof} See \cite{AV95}.
\end{proof}

The total arithmetic complexity of the algorithm is
$O((T+n^3\log(R/\delta))n\log^2(nR/\delta))$, where $T$ is the
cost of one call to WOPT. See \cite{AV95} for a detailed
discussion of the arithmetic complexity of the algorithm,
including the complexity of calculating the inverse Hessians.

Note that -- in the worst case -- the algorithm requires more
machine precision than the algorithm in \cite{AV95} due to
(\ref{ineq_WorstBoundOnZeta}). However, I conjecture that, in the
vast majority of instances, the magnitude $||z||$ of the
approximate analytic center remains larger than a constant; hence
the algorithm, which incorporates the dynamic bound
(\ref{ineq_DynamicBoundOnZeta}), does not require excessive
precision. Some evidence for this conjecture is based on the
following result:
\begin{fact} If $b_i=0$ for all $i$, that is, if all cuts are
central (through the origin, in our case), then, when a cut is
added, the new analytic centre $\omega_\a$ is always bigger than
the old analytic centre $\omega$; that is, $||\omega_\a||\geq
||\omega||$.
\end{fact}
\begin{proof}
From left-multiplying $\grad F(\omega)=0$ by $\omega^T$, we get
\begin{eqnarray}
\frac{||\omega||^2}{1-||\omega||^2} = \frac{1}{2}\sum_{i=1}^h
\frac{a_i^T\omega}{a_i^T\omega -b_i}= \frac{1}{2}\sum_{i=1}^h
\frac{a_i^T\omega}{a_i^T\omega} = h/2.
\end{eqnarray}
Similarly, we get
\begin{eqnarray}
\frac{||\omega_\a||^2}{1-||\omega_a||^2} = (h+1)/2.
\end{eqnarray}
Since the quantity ${||\omega||^2}/{(1-||\omega||^2)}$ increases
as $||\omega||$ increases (and similarly for $||\omega_\a||$), we
have $||\omega_\a||\geq ||\omega||$.
\end{proof}
\noindent Thus, if all the cutting planes go through the origin
(all $b_j=0$), then the analytic center of $P$ grows in magnitude
with each additional cutting plane. Since the shifts $b_j$ of the
cutting planes tend to zero as the algorithm proceeds (because the
eigenvalues of $\hessfzinv$ tend to zero), the behaviour of the
analytic center tends to the case of all cutting planes going
through the origin. If the requirement for shallow cutting in
\cite{AV95} could be removed somehow, then our algorithm would be
free of this worst case.  It is an open problem whether there
exists a polynomial-time, analytic centre algorithm for the convex
feasibility problem that does not require shallow cutting.

With respect to the number of calls to WOPT, how might the
algorithm compare with the unmodified analytic centre algorithm of
\cite{AV95} applied to $Q_p$ and (a weakened) $\text{SEP}_{Q_p}$
(as outlined in Section
\ref{sec_ConnToOtherKnownMethods})?\footnote{Actually, finding any
point $c$ such that $t c\in Q_p$, for $t>0$, suffices.} The new
cut-generation rule elegantly combines the routine
$\text{SEP}_{K^\star}$ and the constraint $\{c: p^Tc \geq 1\}$.


\begin{problem}  Analyse, and compare more carefully, the two cut-generation rules.
\end{problem}

I have shown that the Atkinson-Vaidya algorithm works with an
initial bounding sphere, in place of a hyperbox.  This result
means that we can use this modified algorithm with
$\text{SEP}_{K^\star}$ and initial outer approximation equal to
\begin{eqnarray}
P_0 = B(\bar{0},R^\star) \cap \{c: p^Tc \geq 1\},
\end{eqnarray}
where $R^\star$ is the radius of the smallest origin-centred ball
that contains the polar $K^\star$.  Note that such a radius is
available when the original set $K$ is known to contain a ball of
radius $r_0$; in which case, $R^\star = 1/r_0$.  In the quantum
separability problem, we have such a radius, given by the maximum
separable ball centred at the maximally mixed state (see Section
\ref{sec_OneSidedTestsAndRestrictions}). Fact
\ref{Fact_InitialCentre} gives the analytic centre of $P_0$ (note
that if a hyperbox was used instead, the analytic centre may not
be easily computable because of a lack of symmetry e.g. the centre
is not necessarily a scalar multiple of $p$). This likely makes
the standard method more efficient. Using this method with deep
cuts may, in practice, yield the fastest fully polynomial
algorithm.

\section{Application to Quantum Separability Problem}\label{sec_Application}

We can handle two scenarios -- one experimental, as described in
Section \ref{sec_DetectEntPartial}, and the other theoretical. In
the theoretical scenario, we assume that we know the density
matrix for the given state $\rho\in\densops$, but we do not know
whether $\rho$ is separable; knowing the density matrix
corresponds to having gathered all $\n-1$ independent expected
values of the physical state $\rho$ in the experimental scenario.
Since the algorithm finds an entanglement witness when $\rho$ is
entangled, it could also be applied when $\rho$ is known to be
entangled but an entanglement witness for $\rho$ is desired
(though one may want to apply the entanglement witness
optimization procedure \cite{LKCH00} to the result of the
algorithm, as our algorithm does not output optimal entanglement
witnesses).

We characterise all potential entanglement witnesses for $\rho$ by
\begin{eqnarray}
\mathcal{W}\equiv \{A\in\mathbb{H}_{M,N}: \tr(A)=0,\tr(A^2)\leq
1\}.
\end{eqnarray}
For entangled $\rho$, define $\mathcal{W}_\rho$ to be the subset
of $\mathcal{W}$ consisting of (right) entanglement witnesses that
detect $\rho$; if $\rho$ is separable, then define
$\mathcal{W}_\rho$ to be empty.

Let $\mathcal{B}=\{X_i:i=0,1,\ldots,\dima^2\dimb^2-1\}$ be a basis
as described in the beginning of Section
\ref{sec_ConvexBodyProblems}.  Let $j$ be the number of nontrivial
expected values of $\rho$ that are known, $2\leq j\leq
\dima^2\dimb^2-1$; that is, (without loss) assume we know the
expected values of the elements of $\mathcal{B}'=\{X_1,
X_2,\ldots, X_j\}$.  The algorithm either finds an entanglement
witness in $\sp(\mathcal{B}')$ for $\rho$, or it concludes that no
such witness exists.

\begin{figure}[ht]
\centering
\resizebox{150mm}{!}{\includegraphics{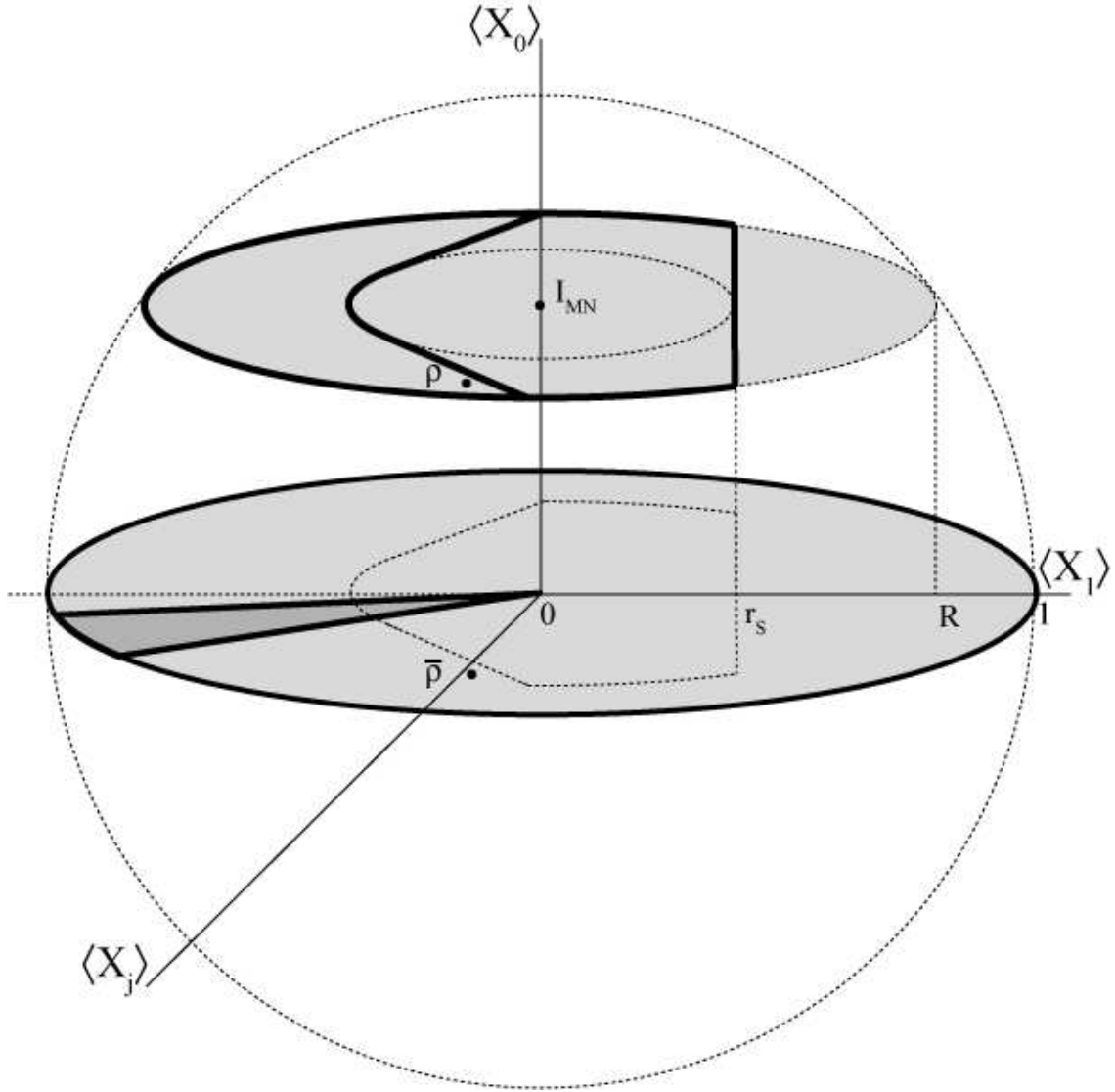}}
\caption[Schematic diagram of density operators]{Schematic diagram
(not to scale) of density operators and $\overline{\mathcal{W}}$
in $\mathbb{R}^{n}$, $n=M^2N^2$. The $\langle X_1\rangle-\langle
X_j\rangle$ plane is a two-dimensional representation of the space
$\sp(\{X_1,X_2,...,X_j\})$. The large dashed circle represents the
origin-centered unit hypersphere in $\mathbb{R}^n$. The upper
shaded ellipse represents the $(n-1)$-dimensional hyperball of
radius $R$ centered at the maximally mixed state $I_{MN}$ which is
the intersection of the hyperplane $\pi_{I,1}$ and the
origin-centered unit hyperball in $\mathbb{R}^n$. The density
operators are the heavy-outlined region in this
$(n-1)$-dimensional hyperball; the inner heavy outlined shape
represents the separable states $S_{M,N}$.  The boundary of the
maximal separable hyperball of radius $r_S$ centered at $I_{MN}$
is shown as a dashed ellipse. An entangled state $\rho$ is shown,
and its projection $\overline{\rho}$ is also shown.  The shaded
elliptical disk (in heavy outline) in the $\langle
X_1\rangle-\langle X_j\rangle$ plane is $\overline{\mathcal{W}}$
(i.e. is a representation of the origin-centered
$(n-1)$-dimensional unit hyperball); the darker shaded wedge is
$\overline{\mathcal{W}_\rho}$. The boundary of
$\overline{S}_{M,N}$ is shown as a dashed line.}
\label{DensityOps}
\end{figure}

For any $Y\in\mathbb{H}_{M,N}$ with $Y=\sum_{i=0}^{n-1} y_iX_i$,
let $\overline{Y}$ be the $j$-dimensional vector of the real
numbers $y_i$ for $i=1,2,\ldots,j$.  Conversely, for any
$\overline{Y}\in\mathbb{R}^j$ with elements $y_1,\ldots,y_j$, let
$Y=\sum_{i=1}^{j} y_iX_i$.  Define the hyperplanes
$\pi_{\overline{Y},b}=\{x\in\mathbb{R}^j: \overline{Y}^T x=b\}$
and halfspaces $H_{\overline{Y},b}=\{x\in\mathbb{R}^j:
\overline{Y}^T x\leq b\}$ similarly to before. Define
\begin{eqnarray}
\overline{S}_{M,N} = \{\overline{\sigma}: \sigma\in \sep\}.
\end{eqnarray}
\noindent Note $\overline{S}_{M,N}$ is a convex set in
$\mathbb{R}^j$ containing the origin. Define
\begin{eqnarray}
\overline{\mathcal{W}}&\equiv& \{\overline{A}\in\mathbb{R}^j: A\in\mathcal{W}\}\\
\overline{\mathcal{W}}_\rho&\equiv&
\{\overline{A}\in\overline{\mathcal{W}}: A\in\mathcal{W}_\rho\}.
\end{eqnarray}
\noindent Figure \ref{DensityOps} shows a schematic of the sets
$\sep$, $\overline{S}_{M,N}$, $\overline{\mathcal{W}}$, and
$\overline{\mathcal{W}}_\rho$.

The algorithm solves the following problem:\\
\emph{ {\bf Entanglement Witness Problem.} Given the expected
values of elements of $\mathcal{B}'$ for
$\rho\in\mathcal{D}_{M,N}$ and a precision parameter $\delta>0$,
either assert
\begin{tabbing}
\hspace*{0.4cm}\= ENTANGLED \=:\ \=\kill
\>   ``$\overline{\rho}\in\overline{S}_{M,N}$'': \>\hspace{9mm}there exists a separable state $\sigma$\\
\> \> \hspace{8mm}such that $||\bar{\rho}-\bar{\sigma}|| < \delta$; or return \\
\> $A\in\mathcal{W}$: \> \hspace{8mm}an operator such that \\
\> \> \hspace{8mm}$b^*(A) < tr(A\rho)$.
\end{tabbing}}
\noindent Note when $j=\dima^2\dimb^2-1$, this problem solves the
separability problem.

%

To reconcile the elements of the algorithm with the physics, note
the following main correspondences:
\begin{eqnarray}
n &\sim& j \nonumber\\
K &\sim& \overline{S}_{M,N} \nonumber\\
\nonumber K_p &\sim& \overline{\mathcal{W}}_\rho (=
\mathcal{W}_\rho \text{ if $j=M^2N^2-1$}).
\end{eqnarray}

To use this algorithm for the separability problem, it remains to
show that there exists an appropriate centre $c_0$ and outer
radius $R$ of $S_{M,N}$.  The maximally mixed quantum state
$I_{MN}=I/MN$ is properly contained in $S_{M,N}$
\cite{BCJLPS99,GB02}, thus $c_0$ is the 0-vector in
$\mathbb{R}^{j}$ corresponding to $I_{M,N}$:
\begin{eqnarray}
c_0 = (\tr(X_iI_MN))_{i=1,...,j} = \begin{bmatrix} 0 \\ 0 \\
\vdots \\ 0 \end{bmatrix} \in \mathbb{R}^j.
\end{eqnarray}
\noindent We can calculate an upper bound $R$ on the radius of the
smallest $c_0$-centered hypersphere that contains
$\overline{S}_{M,N}$ by referring to \fig{DensityOps}.  Because
the space is Euclidean, we have $R=\sqrt{1-1/MN}$.


For the quantum separability problem, we can derive a slightly
better lower bound $r$ than we could for the generic case in
Section \ref{subsec_DerivationOf_r}.  This is easily done by
making use of the radius $r_S=1/\sqrt{MN(MN-1)}$ of the largest
separable hyperball centered at $I_{M,N}$ contained in $S_{M,N}$,
which is derived in \cite{GB02}.  Using $r_S$, we no longer need
to resort to using $R^*=R$; rather, we can use
$R^*=\sqrt{R^2-r^2_S}$, as is easily seen from \fig{DensityOps}.
The result is
\begin{eqnarray}
r= \sin\theta\frac{1-\tan(\theta/2)}{1+\tan(\theta/2)},
\end{eqnarray}
where
\begin{eqnarray}
\sin\theta =
\frac{2\delta/5}{\sqrt{R^2-r^2_S}+\delta/5}\left(2\cos\theta
-1\right). \label{eqn_DefSinTheta}
\end{eqnarray}

\section{Closing remarks}

This chapter has given a new oracle-polynomial-time algorithm for
WSEP($K$) relative to an oracle for WOPT($K$), for any convex
$K\subset \mathbb{R}^n$ that properly contains a known point $c_0$
and is contained in a ball of finite known radius $R$.  The
novelty of the algorithm lies in the way in which the oracle is
used to generate cutting planes for the feasibility problem for
part of the polar of $K$.  This new cut-generation method is based
on an intuitive search heuristic (see Section
\ref{sec_SearchHeuristic}).

The new algorithm is based heavily on the cutting-plane algorithm
of Atkinson and Vaidya \cite{AV95}, which uses shallow cutting.  I
mentioned that it is an open problem whether there exists a
polynomial-time, analytic centre algorithm for the convex
feasibility problem that does not require shallow cutting.
Actually, it has been suggested by John E. Mitchell \cite{Mit05}
that central cutting can be used in the Atkinson-Vaidya algorithm,
if certain techniques from \cite{MT92} are employed to compute the
new analytic centre.  However, from correspondence with Mitchell
and Yinyu Ye, it is unclear whether modifying the Atkinson-Vaidya
algorithm in this way retains the polynomial-time convergence:
while it is clear that a new analytic centre can be efficiently
computed when central cuts are used, it is not clear that all the
other delicate machinery in the convergence argument emerges
unscathed.
\begin{problem}
Can the Atkinson-Vaidya algorithm be modified so as not to require
shallow cutting, while still being polynomial-time?
\end{problem}

Combined with the material in Chapter
\ref{ChapterReductionToEWSearch}, I have shown that the new
algorithm has a real potential with regard to practical
application to solving the quantum separability problem, both in
theoretical and experimental contexts -- the general strategy of
reducing the problem to the optimisation problem is the best known
(see Section \ref{sec_ComplexityComparison}).  In the case where
$\dima=\dimb=2$, Ben Travaglione and I developed a working
implementation of the algorithm, using Hansen's interval-analysis
global optimisation routine \cite{Han92, HW04}.  This
implementation runs fast enough that it could be used in practice.
However, for higher dimensions (which are of greater interest and
where precision and round-off considerations become extremely
important), we were not satisfied that the implementation is
optimal.  Currently, Donny Cheung and I are working on a more
robust implementation, which will hopefully run reasonably quickly
in the case $\dima=\dimb=3$; that is, the run time of the program
on a separable state is hopefully on the order of hours or days,
as opposed to years.

\begin{appendix}
\chapter{Convergence proofs}



\mylemma{AV-lem17}{Lemma 17 in \cite{AV95}}{Let $z$ be an
approximation to $\w$ such that $\w\in E(\hessfz,z,\zeta)$.
Suppose the hyperplane $(a,\beta)$ is added in Case 2 with
$\gamma_0^2=\gamma^2=\frac{a^T(\hessfz)^{-1}a}{(a^Tz-\beta)^2}$.
Then,
\begin{eqnarray}
&\mathrm{(a)}&\frac{|a^T(z-\w)|}{a^Tz-\beta}\leq \zeta\gamma,\nonumber\\
&\mathrm{(b)}&\frac{|a^T(z-\w)|}{a^T\w-\beta}\leq
\zeta\gamma/(1-\zeta\gamma),\nonumber\\
&\mathrm{(c)}&\Psi_\a(\w)\leq\tilde{\gamma}^2:=\gamma^2\left(\frac{1}{1-\zeta\gamma}\right)^2\left(\frac{1}{1-\zeta}\right)^2.\nonumber
\end{eqnarray}
}
\begin{proof}
\noindent (a) Since $\w\in E(\hessfz,z,\zeta)$, we know that
\begin{eqnarray}\label{AV-eqn78}
|a^T(z-\w)|\leq \zeta\sqrt{a^T\hessfzinv a }.
\end{eqnarray}
Therefore,
\begin{eqnarray}
\frac{|a^T(z-\w)|}{a^Tz-\beta}\leq\frac{\zeta\sqrt{a^T\hessfzinv
a}}{a^Tz-\beta} = \zeta\gamma.
\end{eqnarray}

\noindent (b) By (\ref{AV-eqn78}), we get
\begin{eqnarray}
\frac{|a^T(z-\w)|}{a^T\w-\beta} &\leq&
\frac{\zeta\sqrt{a^T\hessfzinv a}}{a^Tz -
\beta+a^T(\w-z)}\nonumber\\
&\leq& \frac{\zeta\sqrt{a^T\hessfzinv a}}{a^Tz - \beta
-\zeta\sqrt{a^T\hessfzinv a}}
=\frac{\zeta\gamma}{1-\zeta\gamma}.\label{AV-eqn79}
\end{eqnarray}

\noindent (c)We have \begin{eqnarray}\gamma^2 &=& \frac{a^T\hessfzinv a}{(a^Tz-\beta)^2} \nonumber\\
&\geq& (1-\zeta)^2\frac{a^T\hessfwinv a}{(a^Tz-\beta)^2}
\hspace{2mm}[\textrm{by Lemma \ref{AV-cor4}}]\nonumber\\
&=&(1-\zeta)^2\frac{a^T\hessfwinv a/(a^T\w-\beta)^2}{\left(1+
a^T(z-\w)/(a^T\w-\beta)\right)^2}\nonumber\\
&\geq& (1-\zeta)^2\frac{a^T\hessfwinv
a/(a^T\w-\beta)^2}{(1+\zeta\gamma/(1-\zeta\gamma))^2}\hspace{2mm}[\textrm{from
part (b)}]\nonumber\\
&=& \frac{(1-\zeta)^2}{\left( 1+\zeta\gamma/(1-\zeta\gamma)
\right)^2} \grad F_\a (\w)^T\hessfwinv\grad F_\a(\w) \nonumber\\
&\geq& \frac{(1-\zeta)^2}{\left( 1+\zeta\gamma/(1-\zeta\gamma)
\right)^2}  \underbrace{\grad F_\a (\w)^T\hessfawinv\grad
F_\a(\w)}_{\Psi_\a(\w)}\nonumber,
\end{eqnarray}
where the second-last line follows from $\grad F_\a (\w)=\grad
F(\w)-a/(a^T\w-\beta)=-a/(a^T\w-\beta)$, and the last line follows
by noting that $\hessfax=\hessfx+aa^T/(a^Tx-\beta)^2$ and applying
Lemma \ref{AV-cor4}.  Thus,
\begin{eqnarray}
\gamma^2 &\geq&
\frac{(1-\zeta)^2}{(1+\zeta\gamma/(1-\zeta\gamma))^2}\Psi_\a
(\w)\label{AV-eqn80}\\
\Psi_\a(\w)&\leq& \gamma^2
\frac{1}{(1-\zeta\gamma)^2(1-\zeta)^2}=\tilde{\gamma}^2.\nonumber
\end{eqnarray}
\qed
\end{proof}

\mylemma{AV-lem18}{Lemma 18 in \cite{AV95}}{Suppose a hyperplane
is added in Case 2, and the analytic center moves from $\w$ to
$\w_\a$. Let $\gamma=\sqrt{a^T\hessfzinv a}/(a^Tz-\beta)^2$.  If
$\tilde{\gamma}<\frac{1}{3}$, then
\begin{eqnarray}
\frac{a^T\hessfzinv
a}{(a^Tz_\a-\beta)^2}\geq\gamma^2\left(\frac{1-\zeta}{1+\gamma
q_{\tilde{\gamma}}/(1-\zeta) + \zeta\gamma}\right)^2\nonumber.
\end{eqnarray}}

\begin{proof}
We have
\begin{eqnarray}
a^Tz_\a-\beta &=& a^Tz-\beta+a^T(z_\a - \w_\a + \w_\a - \w +\w
-z)\nonumber\\
&\leq& a^Tz-\beta +
|a^T(z_\a-\w_\a)|+|a^T(\w_\a-\w)|+|a^T(\w-z)|\nonumber\\
&\leq& a^Tz-\beta +
\underbrace{\zeta\sqrt{a^T\hessfazainv a}}_{\textrm{by (\ref{AV-eqn65}) and Lemma \ref{AV-KKT}} }+\underbrace{q_{\tilde{\gamma}}\sqrt{a^T\hessfawinv a}}_{\textrm{by (\ref{AV-eqn81})}}\nonumber\\&&+\underbrace{\zeta\sqrt{a^T\hessfzinv a}}_{\textrm{by (\ref{AV-eqn65}) and Lemma \ref{AV-KKT}}}\nonumber\\
&\leq& a^Tz-\beta+\zeta\sqrt{a^T\hessfazainv
a}+\underbrace{q_{\tilde{\gamma}}\sqrt{a^T\hessfwinv
a}}_{\textrm{by Lemma
\ref{AV-cor4}}}\nonumber\\&&+\zeta\sqrt{a^T\hessfzinv
a}\nonumber\\
&\leq& a^Tz-\beta+\underbrace{\zeta(a^Tz_\a-\beta)}_{\textrm{by
Lemma \ref{AV-lem1}}}+q_{\tilde{\gamma}}\sqrt{a^T\hessfwinv
a}\nonumber\\&&+\zeta\sqrt{a^T\hessfzinv a}\nonumber.
\end{eqnarray}
And so,
\begin{eqnarray}
\frac{a^Tz_\a-\beta}{a^Tz-\beta}
&\leq&\frac{1}{1-\zeta}\left(1+q_{\tilde{\gamma}}\frac{\sqrt{a^T\hessfwinv
a}}{a^Tz-\beta}+\frac{\zeta\sqrt{a^T\hessfzinv
a}}{a^Tz-\beta}\right)\nonumber\\
&\leq&\frac{1}{1-\zeta}\left(1+\underbrace{\frac{q_{\tilde{\gamma}}}{1-\zeta}\frac{\sqrt{a^T\hessfzinv
a}}{a^Tz-\beta}}_{\textrm{by Lemmas \ref{AV-lem3} and
\ref{AV-cor4}}}+\frac{\zeta\sqrt{a^T\hessfzinv
a}}{a^Tz-\beta}\right)\nonumber\\
&=&\frac{1}{1-\zeta}\left(1+\frac{q_{\tilde{\gamma}}}{1-\zeta}\gamma+\zeta\gamma\right)=\frac{1+\gamma
q_{\tilde{\gamma}}/(1-\zeta)+\zeta\gamma}{1-\zeta}.\label{AV-eqn84}
\end{eqnarray}
It now follows that
\begin{eqnarray}
\frac{a^T\hessfzinv a}{(a^Tz_\a-\beta)^2} &=& \frac{a^T\hessfzinv
a}{(a^Tz-\beta)^2}\frac{(a^Tz-\beta)^2}{(a^Tz_\a-\beta)^2}\nonumber\\
&\geq& \gamma^2\left(\frac{1-\zeta}{1+\gamma
q_{\tilde{\gamma}}/(1-\zeta)+\zeta\gamma}\right)^2
\end{eqnarray}
as stated in the lemma.\qed
\end{proof}

\mytheorem{AV-thm13approx}{Approximation version of Theorem 13 in
\cite{AV95}}{Suppose that $\max_{1 \leq i \leq h}\mu_i(z)\leq 2$
at the beginning of an iteration, i.e. Case 2 is about to occur.
If the current search space $P$ is determined by $h$ hyperplanes
(in addition to the unit hypersphere), then
\begin{eqnarray}
\mathrm{det}(\hessfz)>2^{-n}(1+C_2)^h=2^{(\log_{2}(1+C_2))h-n},
\end{eqnarray}
for some positive constant $C_2$ which depends on the parameters
$\sigma_0$ and $\gamma_0$ of the algorithm and the ``minimal
goodness'' $\zeta_0$ of the approximation to the analytic centers.
 This can be improved to
\begin{eqnarray}
\mathrm{det}(\hessfz)>2^{-n}(2.5)(1+C_2)^{h-1}.
\end{eqnarray}
}
\begin{proof}
Let $\{(a_1,b_1),\ldots,(a_h,b_h)\}$ be the set of hyperplanes
describing $P$.  For each $i$, let $s_i$ be the number of the most
recent iteration in which $\kappa(a_i,b_i$) was changed.  Without
loss of generality, we assume that $s_1<s_2<\ldots <s_h$.  Let
$F_0(x)$ be our self-concordant barrier function over the
hyperball alone:
\begin{eqnarray}
\hess F_0(x)=\frac{4xx^T}{(1-x^Tx)^2} + \frac{2I}{(1-x^Tx)}.
\end{eqnarray}
Construct a set of auxiliary matrices as follows:
\begin{eqnarray}
M_0:=2I,\hspace{5mm}M_i:=M_{i-1} +
\frac{a_ia_i^T}{(\kappa(a_i,b_i))^2},\hspace{2mm}i=1,\ldots h.
\end{eqnarray}
The matrix $M_0$ is $\hess F_0(\bar{0})$.  The $M_i$'s add in the
terms corresponding to $(a_i,b_i)$ with the current settings of
$\kappa(a_i,b_i)$.

Let $z(s_k)$ represent the approximate analytic center at the
beginning of iteration $s_k$.  At the beginning of iteration
$s_k$, the $\kappa$ values corresponding to constraints in the set
$\{(a_1,b_1),\ldots,(a_{k-1},b_{k-1})\}$ have already experienced
their final change up to the time of the statement of the theorem.
Because $\kappa(a_k,b_k)$ changes in the iteration $s_k$,
iteration $s_k$ must be an occurrence of either Subcase 1.2 or
Case 2.  If it is an occurrence of Case 2, then we easily see that
\begin{eqnarray}
a_i^Tz(s_k)-b_i\leq
2\kappa(a_i,b_i),\hspace{2mm}i=1,\ldots,k-1,\label{AV-eqn36approx}
\end{eqnarray}
for otherwise Case 2 would not occur at all.  Inequality
(\ref{AV-eqn36approx}) also holds true, however, if iteration
$s_k$ is an occurrence of Subcase 1.2, for the following reason:
Suppose that, for some $\hat{\iota}$ in $\{1,\ldots k-1\}$,
$a^T_{\hat{\iota}}z(s_k)-b_{\hat{\iota}}>2\kappa(a_{\hat{\iota}},b_{\hat{\iota}})$.
Notice that Subcase 1.2 does not affect the approximate analytic
center at all, and no plane is in line to be discarded, else
Subcase 1.1 would occur instead of Subcase 1.2.  As a result,
instances of Subcase 1.2 continue to occur until
$\kappa(a_{\hat{\iota}},b_{\hat{\iota}})$ becomes reset.  This
contradicts the assumption that, at iteration $s_k$, the
$\hat{\iota}$th $\kappa$-value has experienced its last change.
So, regardless of whether iteration $s_k$ is an instance of
Subcase 1.2 or Case 2, inequality (\ref{AV-eqn36approx}) holds,
and, therefore,
\begin{eqnarray}
\frac{1}{(a^T_iz(s_k)-b_i)^2}\geq
\frac{1}{4(\kappa(a_i,b_i))^2},\hspace{2mm}i=1,\ldots,k-1.\label{Av-eqn37approx}
\end{eqnarray}
Since the Hessian of the barrier function $F$ takes the form
\begin{eqnarray}
\hess F(z(s_k)) &=&
\frac{4z(s_k)z(s_k)^T}{(1-z^T(s_k)z(s_k))^2}+\frac{2I}{1-z^T(s_k)z(s_k)}+\sum_{i=1}^{k-1}\frac{a_ia_i^T}{(a_i^Tz(s_k)-b_i)^2}\nonumber\\
&&+(\textrm{additional positive semi-definite terms}),
\end{eqnarray}
it follows that at the beginning of iteration $s_k$ we have
\begin{eqnarray}
\xi^T\hess F(z(s_k))\xi &\geq& \xi^TM_0\xi +
\sum_{i=1}^{k-1}\frac{(a_i^T\xi)^2}{4(\kappa(a_i,b_i))^2}\nonumber\\
&\geq&
\frac{\xi^TM_0\xi}{4}+\sum_{i=1}^{k-1}\frac{(a_i^T\xi)^2}{4(\kappa(a_i,b_i))^2}\nonumber\\
&=& \frac{1}{4}\xi^TM_{k-1}\xi,\hspace{2mm}\textrm{for every
$\xi\in\mathbf{R}^n$.}\label{AV-eqn38approx}
\end{eqnarray}
(The two terms corresponding to the hypersphere are minimised when
$z=\bar{0}$, which is the setting of $z$ in our definition of
$M_0$.)  Thus, by Lemma \ref{AV-cor4},
\begin{eqnarray}
\xi^T(\hess F(z(s_k)))^{-1}\xi \leq
4\xi^TM_{k-1}^{-1}\xi,\hspace{2mm}\textrm{for every
$\xi\in\mathbf{R}^n$}.\nonumber
\end{eqnarray}

Iteration $s_k$ is the last time $\kappa(a_k,b_k)$ was changed,
and this change occurred in Subcase 1.2 or Case 2.  If the change
occurred in Subcase 1.2, then, at the time of the change,
\begin{eqnarray}
\sigma_0 &\leq&  \frac{a^T_k(\hess
F(z(s_k)))^{-1}a_k}{(a_k^Tz(s_k)-b_k)^2}\hspace{5mm}\textrm{[by
definition of $\sigma_k$]}\nonumber\\
&\leq& 4\frac{a^T_k(M_{k-1})^{-1}a_k}{(a_k^Tz(s_k)-b_k)^2} =
4\frac{a^T_k(M_{k-1})^{-1}a_k}{(\kappa(a_k,b_k))^2}\nonumber
\end{eqnarray}
($\kappa(a_k,b_k)$ is the newly reset value) and we conclude that
\begin{eqnarray}
\frac{a_k^T(M_{k-1})^{-1}a_k}{(\kappa(a_k,b_k))^2}\geq\frac{\sigma_0}{4}.\label{AV-eqn39approx}
\end{eqnarray}

If the change occurred in Case 2, then the argument is harder.  We
employ the notation $(a,\beta)$ to refer to the hyperplane added
in Case 2, just as we did in the algorithm itself.  By Lemma
\ref{AV-lem18}, if $\tilde{\gamma}<\frac{1}{3}$, then
\begin{eqnarray}
\frac{a^T\hessfzskinv
a}{(a^Tz_\a-\beta)^2}\geq\gamma^2\left(\frac{1-\zeta}{1+\gamma
q_{\tilde{\gamma}}/(1-\zeta) +
\zeta\gamma}\right)^2\label{AV-eqn86}.
\end{eqnarray}
Thus, since we set $\beta$ to make $\gamma^2 = \gamma_0^2$,
\begin{eqnarray}
\frac{a^T\hessfzskinv a}{(a^Tz_\a-\beta)^2}\geq
C_1:=\gamma_0^2\left(\frac{1-\zeta_0}{1+\gamma_0
q_{\tilde{\gamma_0}}/(1-\zeta_0) +
\zeta\gamma_0}\right)^2\label{AV-eqn41approx_defC1},
\end{eqnarray}
where $\tilde{\gamma_0}:=
\gamma_0^2\left({1}/(1-\zeta_0\gamma_0)\right)^2\left({1}/(1-\zeta_0)\right)^2$,
consequently,
\begin{eqnarray}
C_1 \leq \frac{a^T(\hess F(z(s_k)))^{-1}a}{(a^Tz_\a -\beta)^2}
\leq 4\frac{a^T(M_{k-1})^{-1}a}{(a^Tz_\a
-\beta)^2}=4\frac{a^T(M_{k-1})^{-1}a}{(\kappa(a,\beta))^2},\label{AV-eqn42approx}
\end{eqnarray}
and it follows that
\begin{eqnarray}
\frac{a^T(M_{k-1})^{-1}a}{(\kappa(a,\beta))^2} \geq
\frac{C_1}{4}\label{AV-eqn43approx}.
\end{eqnarray}

Regardless of whether the change of $\kappa(a_k,b_k)$ occurs in
Subcase 1.2 or in Case 2, from (\ref{AV-eqn39approx}) and
(\ref{AV-eqn43approx}) we can assert that
\begin{eqnarray}
\frac{a_k^T (M_{k-1})^{-1}a_k}{(\kappa(a_k,b_k))^2} \geq
C_2:=\frac{1}{4}\min\{\sigma_0,C_1\}\label{AV-eqn44approx_defC2},\hspace{5mm}\textrm{for
$1\leq k\leq h$.}
\end{eqnarray}
In fact, since we add the first cutting plane $(a_1,0)$
``manually'', we know that, for $k=1$,
\begin{eqnarray}
\frac{a_1^T (M_{0})^{-1}a_1}{(\kappa(a_1,b_1))^2} =
\frac{1/2}{1/{3}} = 3/2= 1.5,\label{eqn_ManualBound}
\end{eqnarray}
where $1.5$ may be larger than the largest $C_2$ we can achieve.

Since each $M_i$, $i=1,\ldots,h$ is symmetric positive definite,
we have, for $i\geq 1$,
\begin{eqnarray}
\det(M_i) &=& \det(M_{i-1} +
a_ia_i^T/(\kappa(a_i,b_i))^2)\nonumber\\
&=&\det\left ((M_{i-1})^{1/2} \left( I + \frac{(M_{i-1})^{-1/2}a_ia_i^T (M_{i-1})^{-1/2} }{(\kappa(a_i,b_i))^2} \right)  (M_{i-1})^{1/2} \right)\nonumber\\
&=&\det(M_{i-1})\det\left( I + \frac{(M_{i-1})^{-1/2}a_ia_i^T
(M_{i-1})^{-1/2} }{(\kappa(a_i,b_i))^2} \right)\nonumber
\end{eqnarray}
For an arbitrary vector $v\in\mathbf{R}^n$, the operator $(I\pm
vv^T)$ has set of eigenvalues $\{1,1,\ldots,1\pm v^Tv\}$.  Thus,
for $i\geq 1$,
\begin{eqnarray}
\det(M_i) &=& \det(M_{i-1})\left( 1+ \frac{a^T_i(M_{i-1})^{-1}a}{(\kappa(a_i,b_i))^2} \right)\nonumber\\
&\geq& \det(M_{i-1})(1+C_2)\nonumber.
\end{eqnarray}
Therefore, we have
\begin{eqnarray}
\det(M_h) &\geq& (\det M_0)(1+C_2)^h\label{AV-eqn45approx}\\
&=&2^n(1+C_2)^h.
\end{eqnarray}
The hypotheses of the theorem state that Case 2 is about to occur.
This means that (\ref{AV-eqn36approx}) and (\ref{Av-eqn37approx})
hold at the current $z$ and for $h\geq 1$. Likewise,
(\ref{AV-eqn38approx}) is true with the current $z$ in place of
$z(s_k)$ and with $h$ in place of $k-1$, i.e., we have
$\xi^T\hessfz\xi\geq\frac{1}{4}\xi^TM_h\xi$ for all $\xi$.  Thus,
\begin{eqnarray}
\det(\hessfz)\geq \frac{1}{4^n}\det(M_h)\nonumber,
\end{eqnarray}
and so
\begin{eqnarray}
\det(\hessfz)\geq \frac{1}{4^n}\cdot 2^n\cdot (1+C_2)^h =
2^{-n}(1+C_2)^h\nonumber,
\end{eqnarray}
which proves the theorem.  Alternatively, taking potential
advantage of (\ref{eqn_ManualBound}),
\begin{eqnarray}
\det(\hessfz)\geq
2^{-n}(2.5)(1+C_2)^{h-1}\label{ineq_TightBoundDetHess}.
\end{eqnarray}
 \qed
\end{proof}

\mylemma{AV-lem19}{Lemma 19 in \cite{AV95}}{For the approximate
analytic center $z$ with $\w\in E(\hessfz,z,\zeta)$, we have
\begin{eqnarray}
P\subset E(\hessfz,z,\vartheta),\nonumber
\end{eqnarray}
where
\begin{eqnarray}
\vartheta &:=& \sqrt{2\left(\frac{14h^2}{(1-\zeta)^2}
+\zeta^2\right)} ,\hspace{5mm}\textrm{if $h>31$.}\nonumber
\end{eqnarray}
}
\begin{proof}
For every $x\in P$,
\begin{eqnarray}
&&(x-z)^T\hessfz (x-z)\nonumber\\
&=&(x-\w+\w-z)^T\hessfz (x-\w+\w-z)\nonumber\\
&=&(x-\w)^T\hessfz (x-\w) + (\w-z)^T\hessfz
(\w-z)+2(x-\w)^T\hessfz (\w-z)\nonumber\\
&\leq& 2(x-\w)^T\hessfz (x-\w) + 2(\w-z)^T\hessfz
(\w-z),\hspace{5mm}[\textrm{since $2a\cdot b\leq a\cdot a+b\cdot b$}]\nonumber\\
&\leq& 2(x-\w)^T\frac{\hessfw}{(1-\zeta)^2}
(x-\w)+2\zeta^2\hspace{5mm}[\textrm{by Lemma \ref{AV-lem3} and
hypothesis}].\nonumber
\end{eqnarray}
The lemma follows by substituting the
conclusion of Lemma \ref{AV-lem9} for $(x-\w)^T{\hessfw}
(x-\w)$.\qed
\end{proof}

\mytheorem{AV-thm14approx}{Approximation version of Theorem 14 in
\cite{AV95}}{There exists a constant $\nu$, independent of $h$,
$n$, $R$, and $\delta$, and there exists a function
$u(n,\delta)\in\Theta(\mathrm{poly}(n,\log(\frac{R}{\delta})))$
such that if $h=\nu n u(n,\delta)$, then the volume of $K_p$ is
sufficiently small so as to assert that $p\in S(K,\delta)$. }
\begin{proof}
The volume of an ellipsoid $E(A,z,r)$ is upper-bounded by
$r^n2^n/\sqrt{\det A}$ \cite{GLS88}.  Thus, from Lemma
\ref{AV-lem19},
\begin{eqnarray}
\mathrm{volume}(P) \leq \mathrm{volume}(E(\hessfz,z,6h)) \leq
\frac{(12h)^n}{\sqrt{\det \hessfz}}.\nonumber
\end{eqnarray}
To prove the theorem, the bound in (\ref{ineq_LBdVolKp}) implies
that it suffices to show that there exists $\nu$ and $u$ such that
$h=\nu n u$ implies
\begin{eqnarray}
\frac{(12\nu n u)^n}{\sqrt{\det \hessfz}} <
\left(\frac{r}{n}\right)^n\label{AV-eqn47approx}.
\end{eqnarray}
Theorem \ref{AV-thm13approx} strengthens this further to
\begin{eqnarray}
\frac{(12\nu n u)^n}{2^{[\log_2(1+C_2)]\nu n u/2-n/2}} &<&\left(\frac{r}{n}\right)^n\label{AV-eqn49approx}\\
\frac{12\nu n u}{2^{[\log_2(1+C_2)]\nu u/2-1/2}}&<&
\frac{r}{n}\label{AV-eqn50approx}\\
\log_2(12\nu n u) - ({{[\log_2(1+C_2)]\nu u/2-1/2}})&<&
\log_2(r/n)\nonumber\\
\log_2(n/r)+\log_2(n)+\log_2(u)+\log_2(12)+1/2 &<& \nu
u\log_2(1+C_2)/2-\log_2(\nu)\nonumber\\
\frac{\log_2(n/r)}{u}+\frac{\log_2(n)}{u}+\frac{\log_2(u)}{u}+\frac{\log_2(12)+1/2}{u}
&<& \nu \log_2(1+C_2)/2\nonumber\\
&&-\frac{\log_2(\nu)}{u}\label{AV-eqn51approx}.
\end{eqnarray}
Setting
\begin{eqnarray}
u:=\log_2(n/r)+\log_2(n)=2\log_2(n)+\log_2(1/r)\label{eqn_defu}
\end{eqnarray}
and assuming $n\geq 2$ gives $u\geq 2$ and hence an upper bound on
the left side of (\ref{AV-eqn51approx}) of
$1+1+1+(\log_2(12)+1/2)/2\leq 5.0424 $.  Thus it suffices to find
$\nu$ such that
\begin{eqnarray}
5.0424<\frac{1}{2}(\nu\log_2(1+C_2)-{\log_2(\nu)})\label{ineq_defnu}.
\end{eqnarray}
Since $C_2$ is a constant, it suffices that $\nu$ be constant.
Later we will see that $r$ is roughly $\delta/R$, thus the theorem
is proven.  The higher the value of $C_2$, the smaller the value
$\nu$ that we need. Note that the constant $\nu$ may be improved
(lowered) with knowledge of $r$ and for specific (larger) values
of $n$.\qed
\end{proof}


\mylemma{AV-lem16}{Lemma 16 in \cite{AV95}}{Let $\zeta<1$. If
$\w\in E(\hessfz,z,\zeta)$, then for all $i$, $1\leq i\leq h$,
\begin{eqnarray}
\sigma_i(\w) \leq \frac{\sigma_i(z)}{(1-\zeta)^4}.
\end{eqnarray}
}
\begin{proof}
From Lemmas \ref{AV-lem3} and \ref{AV-cor4}, we know that for all
$\xi\in\mathbf{R}^n$
\begin{eqnarray}
\frac{1}{(1-\zeta)^2}\xi^T(\hessfz)^{-1}\xi \geq
\xi^T(\hessfw)^{-1}\xi \geq
(1-\zeta)^2\xi^T(\hessfz)^{-1}\xi.\nonumber
\end{eqnarray}
Therefore,
\begin{eqnarray}
\sigma_i(\w) &=& \frac{a^T_i
(\hessfw)^{-1}a_i}{(a_i^T\w-b_i)^2}\leq\frac{1}{(1-\zeta)^2}\frac{a^T_i
(\hessfz)^{-1}a_i}{(a_i^T\w-b_i)^2}\nonumber\\
&=&\frac{1}{(1-\zeta)^2}\frac{a^T_i
(\hessfz)^{-1}a_i/(a_i^Tz-b_i)^2}{[(a_i^T\w-b_i)/(a_i^Tz-b_i)]^2}\nonumber\\
&=&\frac{1}{(1-\zeta)^2}\frac{a^T_i
(\hessfz)^{-1}a_i/(a_i^Tz-b_i)^2}{[1+a_i^T(\w-z)/(a_i^Tz-b_i)]^2}\nonumber\\
&=&\frac{1}{(1-\zeta)^2}\frac{\sigma_i(z)}{[1+a_i^T(\w-z)/(a_i^Tz-b_i)]^2}\label{AV-eqn68}.
\end{eqnarray}
We know from Lemma \ref{AV-KKT} that $\zeta\sqrt{a^T_i\hessfzinv
a_i}\geq |a_i^T(\w-z)|$, and therefore
\begin{eqnarray}
\frac{|a_i^T(\w-z)|}{|a_i^Tz-b_i|}\leq\zeta\frac{\sqrt{a_i^T\hessfzinv
a_i}}{a_i^Tz-b_i}\leq\zeta,\label{AV-eqn69}
\end{eqnarray}
where the second inequality follows since $P\in E(\hessfz,z,1)$
(by Lemma \ref{AV-lem1}) and hence $\sqrt{a_i^T\hessfzinv a_i}\leq
a_i^Tz-b_i$.  It follows that
\begin{eqnarray}
\sigma_i(\w)\leq
\frac{1}{(1-\zeta)^2}\frac{\sigma_i(z)}{(1-\zeta)^2}=\frac{\sigma_i(z)}{(1-\zeta)^4}.\nonumber
\end{eqnarray}
\qed
\end{proof}

\mytheorem{AV-thm11approx}{Approximation version of Theorem 11 in
\cite{AV95}}{There exists a positive constant $\theta$,
independent of $h$, $n$, $R$, and $\delta$, such that after
$\iota$ iterations of the algorithm, $N(\w)\geq\theta \iota$.  The
constant $\theta$ will depend on the parameters of the algorithm.
}
\begin{proof}
The proof is a case analysis following the different cases in the
algorithm.

\vspace{3mm}

\noindent \textit{Case 1, Subcase 1.1}: Let $F_\d(x)$,
$\Psi_\d(x)$, and $N_\d(x)$ denote the functions $F(x)$,
$\Psi(x)$, and $N(x)$ resulting after a hyperplane is dropped.  We
have immediately that
\begin{eqnarray}\label{AV-eqn70}
N_\d(z)\geq N(z)+\ln 2.
\end{eqnarray}
This is because by discarding a hyperplane, we have eliminated a
term from (\ref{AV-eqn14}) that is known to be less than or equal
to $-\ln 2$. In this subcase, however, we must consider the effect
of moving to a new approximate analytic center $z_\d$.  It works
against us that, after the drop, $N_\d(\w)\geq N_\d(\w_\d)$.  Our
goal, however, is to show that the difference is small relative to
the guaranteed $\ln 2$ increase.

We know
\begin{eqnarray}\label{AV-eqn16} (\grad F_\d(\w))^T
=\underbrace{(\grad F(\w))^T}_{=0} +
\frac{a^T_j}{a^T_j\w-b_j}=\frac{a^T_j}{a^T_j\w-b_j}
\end{eqnarray}
and
\begin{eqnarray}\label{AV-eqn17}
\hess F_\d(\w)=\hess F(\w)-\frac{a_ja^T_j}{(a^T_j\w-b_j)^2}.
\end{eqnarray}
Since $\hess F(\w)$ is symmetric positive definite, it has a
square root, and hence
\begin{eqnarray}
\hess F_\d(\w)&=&(\hess F(\w))^\half \left[ I -
\frac{(\hessfw)^{-\half}a_ja_j^T(\hessfw)^{-\half}}{(a^T_j\w-b_j)^2}
\right]\nonumber\\
&&\times(\hess F(\w))^\half.\label{AV-eqn18}
\end{eqnarray}
For an arbitrary vector $v\in\mathbf{R}^n$, the operator
$(I-vv^T)$ has set of eigenvalues $\{1,1,\ldots,1-v^Tv\}$.  If
$v^Tv<1$ then $(I-vv^T)>0$, and for all $\xi\in\mathbf{R}^n$,
\begin{eqnarray}\label{AV-eqn19}
(1-v^Tv)\xi^T\xi\leq \xi^T (1-vv^T)\xi\leq\xi^T\xi.
\end{eqnarray}
We can apply equation (\ref{AV-eqn19}) to the inner matrix in
(\ref{AV-eqn18}) with
\begin{eqnarray}
v:=(\hessfw)^{-\half}a_j/(a_j^T\w-b_j)\nonumber
\end{eqnarray}
and with $\xi:=(\hessfw)^{\half}\chi$ to conclude that for every
$\chi\in\mathbf{R}^n$,
\begin{eqnarray}
(1-\sigma_j(\w))\chi^T\hessfw\chi \leq \chi^T\hess F_\d(\w)\chi
\leq \chi^T\hessfw\chi
\end{eqnarray}
and thus by Corollary \ref{AV-cor4}, for every
$\chi\in\mathbf{R}^n$
\begin{eqnarray}
\frac{1}{1-\sigma_j(\w)}\chi^T(\hessfw)^{-1}\chi \geq \chi^T(\hess
F_\d(\w))^{-1}\chi\geq \chi^T(\hessfw)^{-1}\chi.
\end{eqnarray}
The substitution $\chi:=\hess F_\d(\w)$ gives by equation
(\ref{AV-eqn16})
\begin{eqnarray}\nonumber
\frac{\sigma_j(\w)}{1-\sigma_j(\w)}&\geq&\Psi_\d(\w)\geq
\sigma_j(\w).
\end{eqnarray}
Since in this Subcase 1.1 we have $\sigma_j(z)<\sigma_0$, Lemma
\ref{AV-lem16} gives
\begin{eqnarray}
\sigma_j(\w)\leq
\frac{\sigma_j(z)}{(1-\zeta)^4}<\frac{\sigma_0}{(1-\zeta)^4}.\label{AV-eqn71}
\end{eqnarray}
We then have
\begin{eqnarray}
\Psi_\d(\w)\leq
\frac{\sigma_j(\w)}{1-\sigma_j(\w)}<\frac{\sigma_0/(1-\zeta)^4}{1-\sigma_0/(1-\zeta)^4}\nonumber,
\end{eqnarray}
or,
\begin{eqnarray}
\lambda_\d(\w):=\sqrt{\Psi_\d(\w)}
<C_3:=\sqrt{\frac{\sigma_0/(1-\zeta_0)^4}{1-\sigma_0/(1-\zeta_0)^4}}\label{AV-eqn72_defC3}.
\end{eqnarray}
It thus follows from Lemma \ref{AV-lem6} that
\begin{eqnarray}\label{AV-eqn73_defC4}
F_\d(\w)-F_\d(\w_\d)< C_4:=\half
q_{C_3}^2\left(\frac{1+q_{C_3}}{1-q_{C_3}}\right),
\end{eqnarray}
and consequently $N_\d(\w)-N_\d(\w_\d)<C_4$.  Now,
\begin{eqnarray}
N_\d(\w_\d)-N(\w) &=& (N_\d(\w_\d)-N_\d(\w))+(N_\d(\w)-N_\d(z))\nonumber\\
&& + (N_\d(z)-N(z))+(N(z)-N(\w))\nonumber\\
&=&(N_\d(\w_\d)-N_\d(\w))+\left(N(\w)-N(z)+\ln\left(\frac{a_i^T\w-b_i}{a_i^Tz-b_i}\right)\right)\nonumber\\
&& + (N_\d(z)-N(z))+(N(z)-N(\w))\nonumber\\
&=& (N_\d(\w_\d)-N_\d(\w)) + (N_\d(z)-N(z)) +\ln\left(\frac{a_i^T\w-b_i}{a_i^Tz-b_i}\right)\nonumber\\
&\geq& -C_4 +\ln 2 +\ln\left(1+\frac{a_i^T(\w-z)}{a_i^Tz-b_i}\right)\nonumber\\
&\geq& -C_4 +\ln 2 +\ln\left(1-\frac{|a_i^T(\w-z)|}{|a_i^Tz-b_i|}\right)\nonumber\\
&>& -C_4 +\ln 2 +\ln\left(1-\zeta\right)\nonumber\hspace{5mm}\textrm{[see (\ref{AV-eqn69})]}\\
&>&-C_4 +\ln
2-\zeta-4\zeta^2\nonumber\hspace{5mm}\textrm{[assuming
$\zeta<\frac{1}{2}$]}\\
&>&-C_4 +0.615\label{AV-eqn74}\hspace{5mm}\textrm{[assuming
$\zeta<\frac{1}{16}$].}
\end{eqnarray}
Thus, as long as $C_4<0.615$, the theorem is proven in Subcase
1.1.

\vspace{3mm}

\noindent \textit{Case 1, Subcase 1.2}:  The only action in this
subcase is a change in $\kappa(a_i,b_i)$ for some hyperplane. This
change does not affect $F$ or the analytic center $\w$.  One of
the $h$ terms in (\ref{AV-eqn14}) was previously less than or
equal to $-\ln 2$, and now it becomes $0$. Let $N_{\mathrm{new}}$
be the new function $N$ with the newly reset $\kappa(a_i,b_i)$. We
have $N_\mathrm{new}(z)-N(z)\geq \ln 2$. Noting that,
\begin{eqnarray}
N_{\mathrm{new}}(\w)-N(\w) &=&
(N_{\mathrm{new}}(\w)-N_{\mathrm{new}}(z))+(N_{\mathrm{new}}(z)-N(z))+(N(z)-N(\w))\nonumber\\
&=&
(F(\w)-F(z))+(N_{\mathrm{new}}(z)-N(z))+(F(z)-F(\w))\nonumber\\
&=&N_{\mathrm{new}}(z)-N(z)\nonumber\\
&\geq& \ln 2,\label{AV-eqn75}
\end{eqnarray}
proves the theorem in Subcase 1.2.

\vspace{3mm}

\noindent \textit{Case 2}:  Let $F_\a(x)$, $\Psi_\a(x)$, and
$N_\a(x)$ be the functions $F(x)$, $\Psi(x)$, and $N(x)$ resulting
from the addition of a hyperplane in Case 2.  For notational
simplicity, define $H:=(\hessfw)^{1/2}$, $d:=||H(\w_\a-\w)||_2$.
Define the function $x(t):= \w + t(\w_\a-\w)$. We have
\begin{eqnarray}
||H^{-1}\grad F(\w_\a)||_2 &=& ||H^{-1}\grad F(\w_\a)-H^{-1}\grad F(\w)||_2 \nonumber \\
&=& ||\int_{0}^{1}H^{-1}\hess F(x(t))(w_\a-\w)dt||_2\nonumber \\
&=& ||\int_{0}^{1}[H^{-1}\hess F(x(t))H^{-1}]H(w_\a-\w)dt||_2\nonumber \\
&=& ||\int_{0}^{1}[I+\Delta(t)]H(w_\a-\w)dt||_2\nonumber \\
&\leq& \int_{0}^{1}||[I+\Delta(t)]H(w_\a-\w)||_2dt\nonumber \\
&\leq& \max_t ||I+\Delta(t)||_2||H(w_\a-\w)||_2\nonumber \\
&=& \max_t ||I+\Delta(t)||_2d.\label{AV-eqn23}
\end{eqnarray}
$I+\Delta(t)$ is simply notation for the identity matrix plus some
perturbation of the identity matrix; notice that $H^{-1}\hess
F(x(0))H^{-1}=I$.  We need to examine the effects of the
perturbation.  By definition of $d$, $\w_\a\in E(\hessfw, \w,d)$.

We will want to invoke Lemma \ref{AV-lem3} shortly.  Let us first
dispense with the case $d\geq 1/4$.  Join $\w_\a$ and $\w$ by a
line segment and let $x'$ denote the point on that segment such
that $(x'-\w)^T\hessfw (x'-\w)=(1/4)^2$.  Using the Taylor
approximation, convexity, and the fact that $\w$ minimises $F$, we
conclude
\begin{eqnarray}
F(\w_\a)-F(\w) &\geq& F(x')-F(\w) \nonumber\\
&=& \grad F(\w)^T(x'-\w)+\frac{1}{2}(x'-\w)^T\hessfw
(x'-\w)+\mathrm{Error}\nonumber\\
&=& 0 + \frac{1}{2}\left(\frac{1}{4}\right)^2+\mathrm{Error}\nonumber\\
&\geq&\frac{1}{32} -
\frac{(1/4)^3}{3(1-1/4)}\nonumber\hspace{5mm}\textrm{[by Lemma
\ref{AV-lem5}]}\nonumber\\
&>&0.02,\nonumber
\end{eqnarray}
and so $N(\w_\a)-N(\w)>0.02$.  However,
\begin{eqnarray}
N_\a(\w_\a)-N(\w_\a)&=&
-\ln\left(\frac{a^T\w_\a-\beta}{\kappa(a,\beta)}\right)=-\ln\left(\frac{a^T\w_\a-\beta}{c^Tz_\a-\beta}\right)\nonumber\\
&>&-\ln\left(1+\frac{|a^T(\w_\a-z_\a)|}{|a^Tz_\a-\beta|}\right)>-\ln(1+\zeta)\nonumber\hspace{5mm}\textrm{[by
(\ref{AV-eqn69})]}\\
&>& -\zeta,\nonumber
\end{eqnarray}
and so
\begin{eqnarray}
N_\a(\w_\a)-N(\w)=N_\a(\w_\a)-N(\w_\a)+N(\w_\a)-N(\w)>0.02-\zeta
\end{eqnarray}
for the case in which we assume $d\geq 1/4$.  Recall $\zeta$,
which measures the quality of the approximation of $z$ to $\w$, is
under our control. Up until this point, we have assumed
$\zeta<1/16=0.0625$; now, we must assume that $\zeta<0.02$.

Now consider the case $d<1/4$.  Since $\w_\a\in E(\hessfw,\w,d)$
and $d<1$, it follows by Lemma \ref{AV-lem3} that
\begin{eqnarray}
\xi^T(I+\Delta(t))\xi &=& \xi^TH^{-1}\hess
F(x(t))H^{-1}\xi\nonumber\\
&\leq& \frac{1}{(1-d)^2}\xi^TH^{-1}\hess
F(\w)H^{-1}\xi,\nonumber\hspace{5mm}\textrm{for every $t\in[0,1]$}\\
&\leq& \frac{1}{(1-d)^2}\xi^TI\xi,\nonumber
\end{eqnarray}
for every $\xi$ in $\mathbf{R}^n$.  Therefore, $\max_t
||I+\Delta(t)||_2\leq 1/(1-d)^2$.  Returning to (\ref{AV-eqn23}),
we may now conclude further that
\begin{eqnarray}
||H^{-1}\grad F(\w_\a)||_2\leq d/(1-d)^2.\label{AV-eqn25}
\end{eqnarray}
Consider a different perspective on $||H^{-1}\grad F(\w_\a)||_2$.
We know
\begin{eqnarray}
\grad F(\w_\a)=\grad
F_\a(\w_\a)+\frac{a^T}{a^T\w_\a-\beta}=\frac{a^T}{a^T\w_\a-\beta}\label{AV-eqn26}
\end{eqnarray}
where, recall, we select the appropriate $\beta$ when Case 2
occurs.  Now,
\begin{eqnarray}
a^T\w_\a-\beta &=& a^Tz-\beta+a^T(\w-z)+a^T(\w_\a-\w)\nonumber\\
&\leq& \underbrace{\frac{1}{\gamma}\sqrt{a^T\hessfzinv
a}}_{\textrm{by def'n of $\gamma$}}+\underbrace{\zeta\sqrt{a^T\hessfzinv a}}_{\textrm{since $\w\in E(\hessfz,z,\zeta)$}}\nonumber\\
&&+\underbrace{d\sqrt{a^T\hessfwinv a}}_{\textrm{since $\w_\a\in E(\hessfw,\w,d)$}}\nonumber\\
&\leq&\frac{1}{\gamma}\frac{\sqrt{a^T\hessfwinv a}}{1-\zeta} +
\zeta \frac{\sqrt{a^T\hessfwinv a}}{1-\zeta} +d\sqrt{a^T\hessfwinv
a}\nonumber
\end{eqnarray}
where the last inequality follows from two applications of Lemma
\ref{AV-lem3}.  Using (\ref{AV-eqn26}),
\begin{eqnarray}
||H^{-1}\grad F(\w_\a)||_2 &=&
\frac{||H^{-1}a||_2}{a^T\w_\a-\beta}=\frac{\sqrt{a^T(H^{-1})^TH^{-1}a}}{a^T\w_\a-\beta}=\frac{\sqrt{a^T
H^{-2}a}}{a^T\w_\a-\beta}\nonumber\\
&=&\frac{\sqrt{a^T \hessfwinv a}}{a^T\w_\a-\beta}\nonumber\\
&\geq& \frac{1}{1/\gamma(1-\zeta) + \zeta/(1-\zeta) +
d}.\label{AV-eqn76}
\end{eqnarray}
Combining (\ref{AV-eqn25}) and (\ref{AV-eqn76}), we conclude that
\begin{eqnarray}
d\geq
\frac{1-\zeta}{(1/\gamma)+2+\zeta}=C_5:=\frac{1-\zeta_0}{(1/\gamma_0)+2+\zeta_0}.\
\end{eqnarray}
Now invoke Lemma \ref{AV-lem5} again to get
\begin{eqnarray}
N(\w_\a)-N(\w)=F(\w_\a)-F(\w)\geq C_6:=
\frac{1}{2}(C_5)^2-\frac{(C_5)^3}{3(1-C_5)}.\label{AV-eqn31approx}
\end{eqnarray}
Thus, as long as $C_6>0$, the theorem is proven. \qed
\end{proof}

\mytheorem{AV-thm15approx}{Approximation version of Theorem 15 in
\cite{AV95}}{If the algorithm does not first find a separating
hyperplane or halt by Stopping Condition 1, then, within
$O(nu\log(nuR/\delta))$ iterations, Stopping Condition 2 must be
met. If Stopping Condition 2 is met, then the set $K_p$ is
negligibly small and the algorithm may return ``$p\in
S(K,\delta)$''.}
\begin{proof}
By Lemma \ref{AV-lem10}, a stopping condition that says ``the
width of $P$ is too small to contain $K_p$'' is
\begin{eqnarray}
2r > [\min_i\{a_i^T\w-b_i\}](3h+4).\label{ineq_IdealStopCond2}
\end{eqnarray}
Because we do not have our hands on $\w$ during the algorithm,
condition (\ref{ineq_IdealStopCond2}) is not feasible to check
directly. Instead, we will use Theorem \ref{AV-thm11approx} and
show that if $F(\w)$ is larger than a certain value $\mathcal{F}$,
then Stopping Condition 2 is satisfied, which, in turn, implies
that (\ref{ineq_IdealStopCond2}) is satisfied.

First, we need an upper and lower bound on $a_i^Tz-b_i$ relative
to $a_i^T\w-b_i$. Suppose that $z$ is an approximation to $\w$
with $\w\in E(\hessfz,z,\zeta)$. For any index $i$, $1\leq i\leq
h$,
\begin{eqnarray}
|\ln(a_i^T\w-b_i)-\ln(a_i^Tz-b_i)| &=&
|\ln(1+(a_i^T(\w-z))/(a_i^Tz-b_i))|\nonumber\\
&=& |\ln(1+t)|,\hspace{5mm}\textrm{[where $|t|<\zeta$; see
(\ref{AV-eqn69})]}\nonumber\\
&<&-\ln(1-\zeta)\nonumber.
\end{eqnarray}
This implies
\begin{eqnarray}
(1-\zeta)(a_i^T\w-b_i)<a_i^Tz-b_i<(a_i^T\w-b_i)/(1-\zeta)\label{ineq_BoundOnaiTz_bi}.
\end{eqnarray}

Assuming $\zeta<\zeta_0$, starting with
(\ref{ineq_IdealStopCond2}) we have the series of implications, as
promised above:
\begin{eqnarray}
\begin{array}{lrcl}
& 2r &>& [\min_i\{a_i^T\w-b_i\}](3h+4)\nonumber\\
\Leftarrow\hspace{3mm}\hspace{3mm}&2r &>&  \frac{[\min_i\{a_i^Tz-b_i\}]}{1-\zeta_0}(3h+4)\hspace{3mm}\textrm{[by (\ref{ineq_BoundOnaiTz_bi})]}\nonumber\\
\Leftarrow\hspace{3mm}&  2r &>& \frac{[\min_i\{a_i^T\w-b_i\}]}{(1-\zeta_0)^2}(3h+4)\hspace{3mm}\textrm{[by (\ref{ineq_BoundOnaiTz_bi})]}\nonumber\\
\Leftrightarrow\hspace{3mm}\textrm{for some $j$:}&\hspace{3mm}  2r &>& \frac{a_j^T\w-b_j}{(1-\zeta_0)^2}(3h+4)\nonumber\\
\Leftrightarrow\hspace{3mm}\textrm{for some $j$:}&\hspace{3mm}  \frac{2r(1-\zeta_0)^2}{3h+4} &>& a_j^T\w-b_j\nonumber\\
\Leftrightarrow\hspace{3mm}\textrm{for some $j$:}&\hspace{3mm}  \ln \left(\frac{3h+4}{2r(1-\zeta_0)^2}\right) &<& -\ln(a_j^T\w-b_j)\nonumber\\
\Leftarrow\hspace{3mm}\textrm{for some $j$:}&\hspace{3mm}  \frac{\nu n u}{h}\ln\left(\frac{3\nu n u+4}{2r(1-\zeta_0)^2}\right) &<& -\ln(a_j^T\w-b_j)\hspace{3mm}\textrm{[since $h\leq \nu n u$]}\nonumber\\
\Leftarrow\hspace{3mm}&  F(\w)+\ln(1-\w^T\w)&>& \nu n u\ln\left(\frac{3\nu n u+4}{2r(1-\zeta_0)^2}\right)\nonumber\\
\Leftrightarrow\hspace{3mm}&  F(\w)&>& \nu n u\ln\left(\frac{3\nu n u+4}{2r(1-\zeta_0)^2}\right)-\ln(1-\w^T\w)\nonumber\\
\Leftarrow\hspace{3mm}&  F(\w)&>& \mathcal{F}:=\nu n
u\ln\left(\frac{3\nu n
u+4}{2r(1-\zeta_0)^2}\right)+\ln\left(\frac{2+\nu n
u}{2}\right),\nonumber
\end{array}
\end{eqnarray}
where the second line is Stopping Condition 2, and the last line
follows from (\ref{ineq_UpperBdw}).

It remains to show that $F(\w)$ reaches the value $\mathcal{F}$
within $O(nu^2)$ iterations. Each $\kappa(a_i,b_i)$ value is the
distance of the approximate analytic center, at some iteration,
from the $i$th hyperplane. Consider the first distance to a
hyperplane, set when the hyperplane is introduced in Case 2. By
selection of $\beta$,
\begin{eqnarray}
a^Tz-\beta = \gamma_0^{-1}\sqrt{a^T\hessfzinv
a}=\gamma_0^{-1}\max_{y\in
E(\hessfz,z,1)}a^T(y-z)\leq\gamma_0^{-1}\label{AV-eqn87}
\end{eqnarray}
since $E(\hessfz,z,1)\subset P\subset B_n$.  In subsequent
iterations, $\w$ may drift farther from the hyperplane
$(a,\beta)$.  At worst, it can drift to the edge of the unit
hyperball $B_n$.  Therefore, we may safely say that
\begin{eqnarray}
\kappa(a_i,b_i) \leq \gamma_0^{-1}+2,\hspace{5mm}\textrm{for all
$i$.}\label{AV-eqn88}
\end{eqnarray}
By Theorem \ref{AV-thm11approx}, after $\iota$ iterations,
$N(\w)\geq \theta \iota$ for some $\theta>0$.  That is:
\begin{eqnarray}
-\sum_{i=1}^{h}\ln(a_i^T\w-b_i)-\ln(1-\w^T\w)\geq\theta\iota-\sum_{i=1}^{h}\ln\kappa(a_i,b_i)\label{AV-eqn54}
\end{eqnarray}
after $\iota$ iterations.  From (\ref{AV-eqn54}) and
(\ref{AV-eqn88}), we get
\begin{eqnarray}
F(\w)\geq\theta\iota-h\ln(\gamma_0^{-1}+2).\label{AV-eqn55}
\end{eqnarray}
Thus, $F$ is lower-bounded by a function that grows linearly with
$\iota$.  Therefore, $F(\w)>\mathcal{F}$ if the number of
iterations $\iota$ satisfies
\begin{eqnarray}
\theta\iota-\nu n u\ln(\gamma_0^{-1}+2) > \nu n
u\ln\left(\frac{3\nu n
u+4}{2r(1-\zeta_0)^2}\right)+\ln\left(\frac{2+\nu n
u}{2}\right)\nonumber
\end{eqnarray}
or, equivalently,
\begin{eqnarray}
\iota > \frac{\nu n u\ln\left(\frac{3\nu n
u+4}{2r(1-\zeta_0)^2}\right)+\ln\left(\frac{2+\nu n
u}{2}\right)+\nu n u\ln(\gamma_0^{-1}+2)}{\theta}.\nonumber
\end{eqnarray}
Again, since we will see that $r\approx \delta/R$, $\iota \in
O(nu\log(nuR/\delta))$ iterations suffices.\qed
\end{proof}

\mytheorem{AV-thm20}{Theorem 20 in \cite{AV95}}{There exists some
constant $C_\d$ such that any time a hyperplane is discarded in
Subcase 1.1, $F_\d(z)-F_\d(\w_\d)\leq K_1$. Likewise, there exists
some constant $C_\a$ such that any time a hyperplane is added in
Case 2, $F_\a(z)-F_\a(\w_\a)\leq K_2$. }
\begin{proof}
We have already seen in (\ref{AV-eqn73_defC4}) that
\begin{eqnarray}
F_\a(z)-F_\a(\w_\a)<C_4.\label{AV-eqn91}
\end{eqnarray}
By definition of the approximation, we know that $\w\in
E(\hessfz,z,\zeta)$.  But $(\w-z)^T\hessfdz (\w-z)\leq
(\w-z)^T\hessfz (\w-z)$, and thus $\w\in E(\hessfdz,z,\zeta)$. Let
$(a_j,b_j)$ denote the discarded hyperplane. It follows from Lemma
\ref{AV-lem5} that
\begin{eqnarray}
F_\d(z)-F_\d(\w) &=& \grad F_\d
(\w)^T(z-\w)+\frac{1}{2}(z-\w)^T\hessfdw(z-\w)
+\mathrm{Error}\nonumber\\
&=&\frac{a_j^T(z-\w)}{a_j^T\w-b_j}+\frac{1}{2}(z-\w)^T\hessfdw(z-\w)
+\mathrm{Error}\hspace{5mm}\textrm{[see
(\ref{AV-eqn16})].}\nonumber
\end{eqnarray}
An argument essentially identical to the argument used to prove
Lemma \ref{AV-lem17}(b) shows that
\begin{eqnarray}
|(a_j^T(z-\w))/(a_j^T\w-b_j)|\leq
\frac{\zeta\sqrt{\sigma_j(z)}}{1-\zeta\sqrt{\sigma_j(z)}}.\label{AV-eqn93}
\end{eqnarray}
Therefore,
\begin{eqnarray}
&&\frac{a_j^T(z-\w)}{a_j^T\w-b_j}+\frac{1}{2}(z-\w)^T\hessfdw
(z-\w)+\mathrm{Error}\nonumber\\
&\leq&\frac{\zeta\sqrt{\sigma_j(z)}}{1-\zeta\sqrt{\sigma_j(z)}}+\frac{1}{2}(z-\w)^T\hessfdw
(z-\w)+\mathrm{Error}\nonumber\\
&\leq&\frac{\zeta\sqrt{\sigma_j(z)}}{1-\zeta\sqrt{\sigma_j(z)}}+\frac{1}{2}(z-\w)^T\hessfw
(z-\w)+\mathrm{Error}\label{AV-eqn94}\\
&\leq&\frac{\zeta\sqrt{\sigma_j(z)}}{1-\zeta\sqrt{\sigma_j(z)}}+\frac{1}{2}(z-\w)^T\frac{\hessfz}{(1-\zeta)^2}
(z-\w)+\mathrm{Error}\nonumber\hspace{5mm}\textrm{by Lemma \ref{AV-lem3}}\\
&\leq&\frac{\zeta\sqrt{\sigma_j(z)}}{1-\zeta\sqrt{\sigma_j(z)}}+\frac{1}{2}\frac{\zeta^2}{(1-\zeta)^2}+\mathrm{Error}\nonumber\\
&\leq&\frac{\zeta\sqrt{\sigma_0}}{1-\zeta\sqrt{\sigma_0}}+\frac{1}{2}\frac{\zeta^2}{(1-\zeta)^2}+\frac{\zeta^3}{3(1-\zeta)}.\label{AV-eqn95}
\end{eqnarray}
So we have
\begin{eqnarray}
F_\d(z)-F_\d(\w)\leq
\frac{\zeta\sqrt{\sigma_0}}{1-\zeta\sqrt{\sigma_0}}+\frac{1}{2}\frac{\zeta^2}{(1-\zeta)^2}+\frac{\zeta^3}{3(1-\zeta)}.\label{AV-eqn96}
\end{eqnarray}
Together, (\ref{AV-eqn91}) and (\ref{AV-eqn96}) imply
\begin{eqnarray}
F_\d(z)-F_\d(\w_\d) &=& F_\d(z)
-F_\d(\w)+F_\d(\w)-F_\d(\w_\d)\nonumber\\
&\leq&
\frac{\zeta\sqrt{\sigma_0}}{1-\zeta\sqrt{\sigma_0}}+\frac{1}{2}\frac{\zeta^2}{(1-\zeta)^2}+\frac{\zeta^3}{3(1-\zeta)}+C_4\nonumber\\
&\leq& C_\d:=
\frac{\zeta_0\sqrt{\sigma_0}}{1-\zeta_0\sqrt{\sigma_0}}+\frac{1}{2}\frac{\zeta_0^2}{(1-\zeta_0)^2}+\frac{\zeta_0^3}{3(1-\zeta_0)}+C_4.\nonumber
\end{eqnarray}
This obviously provides a constant upper bound $C_\d$ for Subcase
1.1.

The proof for Case 2 is harder.  Rewrite $F_\a(z)-F_\a(\w_a)$ as
\begin{eqnarray}
F_\a(z)-F_\a(\w_a) =
F_\a(z)-F_\a(\w)+F_\a(\w)-F_\a(\w_a).\nonumber
\end{eqnarray}
We first work on a bound for $F_\a(z)-F_\a(\w)$.  We have
\begin{eqnarray}
(z-\w)^T\hessfaz (z-\w) &=& (z-\w)^T\hessfz
(z-\w)+\left(\frac{a^T(z-\w)}{a^Tz-\beta}\right)^2\nonumber\\
&\leq& \zeta^2 + \zeta^2\gamma^2 \hspace{5mm}\textrm{[by Lemma
\ref{AV-lem17}(a)]}\nonumber\\
&=& (1+\gamma^2)\zeta^2.\nonumber
\end{eqnarray}
It follows that $\w\in E(\hessfaz,z,\sqrt{1+\gamma^2}\zeta)$.
Lemma \ref{AV-lem3} then implies that
\begin{eqnarray}
(\w-z)^T\hessfaw (\w-z) \leq \frac{(\w-z)^T\hessfaz
(\w-z)}{(1-\sqrt{1+\gamma^2}\zeta)^2}  \leq
\left(\frac{\sqrt{1+\gamma^2}\zeta}{1-\sqrt{1+\gamma^2}\zeta}\right)^2
\nonumber
\end{eqnarray}
which says $z\in E(\hessfaw,\w,C_7)$, where
$C_7:=\sqrt{1+\gamma_0^2}\zeta_0/(1-\sqrt{1+\gamma_0^2}\zeta_0)$.
The second-degree Taylor approximation now gives
\begin{eqnarray}
&& F_\a(z)-F_\a(\w)\nonumber\\
&=& \grad F_\a(\w)^T(z-\w)+\frac{1}{2}(z-\w)^T\hessfaw (z-\w) +
\mathrm{Error}\nonumber\\
&=& \grad F_\a(\w)^T(z-\w)+\frac{1}{2}(z-\w)^T\hessfw (z-\w)
+\frac{1}{2}\left(\frac{a^T(z-\w)}{a^T\w-\beta}\right)^2+
\mathrm{Error}\nonumber\\
&=& \frac{-a^T(z-\w)}{a^T\w-\beta}+\frac{1}{2}(z-\w)^T\hessfw
(z-\w) +\frac{1}{2}\left(\frac{a^T(z-\w)}{a^T\w-\beta}\right)^2+
\mathrm{Error}\nonumber\\
&\leq&
\underbrace{\frac{\zeta\gamma}{1-\zeta\gamma}}_{\textrm{Lemma
\ref{AV-lem17}}} +
\underbrace{\frac{1}{2}\left(\frac{\zeta}{1-\zeta}\right)^2}_{\textrm{
(\ref{AV-eqn65}) and Lemma \ref{AV-lem3}}} +
\underbrace{\frac{1}{2}\left(\frac{\zeta\gamma}{1-\zeta\gamma}\right)^2}_{\textrm{Lemma
\ref{AV-lem17}}}+\underbrace{\frac{C_7^3}{3(1-C_7)}}_{\textrm{Lemma
\ref{AV-lem5}}}\nonumber\\
&\leq& \frac{\zeta_0\gamma_0}{1-\zeta_0\gamma_0}+
\frac{1}{2}\left(\frac{\zeta_0}{1-\zeta_0}\right)^2 +
\frac{1}{2}\left(\frac{\zeta_0\gamma_0}{1-\zeta_0\gamma_0}\right)^2+\frac{C_7^3}{3(1-C_7)},
\label{AV-eqn97}
\end{eqnarray}
where we note that (\ref{AV-eqn65}) and Lemma \ref{AV-lem3} imply
$z\in E(\hessfw,\w,\frac{\zeta}{1-\zeta})$ to get the second term
in the second-last line above.  We clearly have established that
$F_\a(z)-F_\a(\w)$ is bounded above by a constant.

We saw in (\ref{AV-eqn81}) that, in Case 2, $\w_\a\in
E(\hessfaw,\w,q_{\tilde{\gamma}})$.  Lemma \ref{AV-lem3} now
implies
\begin{eqnarray}
(\w_\a-\w)^T\hessfawa (\w_\a-\w) \leq
(\w_\a-\w)^T\frac{\hessfaw}{(1-q_{\tilde{\gamma}})^2} (\w_\a-\w)
\leq
\frac{q^2_{\tilde{\gamma}}}{(1-q_{\tilde{\gamma}})^2}\nonumber
\end{eqnarray}
which implies
\begin{eqnarray}
\w\in E(\hessfawa, \w_\a,
\frac{q_{\tilde{\gamma}}}{1-q_{\tilde{\gamma}}}).\label{AV-eqn98}
\end{eqnarray}
A second-degree Taylor expansion and Lemma \ref{AV-lem5} give
\begin{eqnarray}
F_\a(\w)-F_\a(\w_\a) &=& \grad F_\a
(\w_\a)^T(\w-\w_\a)+\frac{1}{2}(\w-\w_\a)^T\hessfawa (\w-\w_\a)+
\mathrm{Error}\nonumber\\
&\leq& 0
+\frac{1}{2}\left(\frac{q_{\tilde{\gamma}}}{1-q_{\tilde{\gamma}}}\right)^2+\frac{(q_{\tilde{\gamma}}/(1-q_{\tilde{\gamma}}))^3}{3(1-q_{\tilde{\gamma}}/(1-q_{\tilde{\gamma}}))}\nonumber\\
&\leq&\frac{1}{2}\left(\frac{q_{\tilde{\gamma_0}}}{1-q_{\tilde{\gamma_0}}}\right)^2+\frac{(q_{\tilde{\gamma_0}}/(1-q_{\tilde{\gamma_0}}))^3}{3(1-q_{\tilde{\gamma_0}}/(1-q_{\tilde{\gamma_0}}))}
.\label{AV-eqn99}
\end{eqnarray}
We put together the results of (\ref{AV-eqn97}) and
(\ref{AV-eqn99}) to get
\begin{eqnarray}
F_\a(z)- F_\a(\w_\a) \leq C_\a,\nonumber
\end{eqnarray}
where
\begin{eqnarray}
C_\a &:=&\frac{\zeta_0\gamma_0}{1-\zeta_0\gamma_0}+
\frac{1}{2}\left(\frac{\zeta_0}{1-\zeta_0}\right)^2 +
\frac{1}{2}\left(\frac{\zeta_0\gamma_0}{1-\zeta_0\gamma_0}\right)^2+\frac{C_7^3}{3(1-C_7)}\nonumber\\
&&+\frac{1}{2}\left(\frac{q_{\tilde{\gamma_0}}}{1-q_{\tilde{\gamma_0}}}\right)^2+\frac{(q_{\tilde{\gamma_0}}/(1-q_{\tilde{\gamma_0}}))^3}{3(1-q_{\tilde{\gamma_0}}/(1-q_{\tilde{\gamma_0}}))}\nonumber
\end{eqnarray}
We have established that a constant upper bound $C_\a$ exists for
$F_\a(z)-F_\a(\w_\a)$.\qed
\end{proof}

\end{appendix}



\end{document}